\pdfoutput=1
\documentclass[11pt,twoside,a4paper,cmspaper,final,collab]{cms-tdr}
\def\svnVersion{c3804c2}\def\svnDate{2022/02/17}\def\cmsCernNoTag{CERN-EP-2021-210}\def\cmsCernDate{\today}\def\cmsMessage{Published in the Journal of High Energy Physics as \href{http://dx.doi.org/10.1007/JHEP02(2022)107}{\doi{10.1007/JHEP02(2022)107}.}}
\begin{document}\cmsNoteHeader{TOP-20-010}

\newcommand{\flavour}{flavor\xspace}
\newcommand{\colour}{color\xspace}
\newcommand{\coloured}{colored\xspace}
\newcommand{\centre}{center\xspace}
\newcommand{\grey}{gray\xspace}
\newcommand{\labelled}{labeled\xspace}
\newcommand{\modelled}{modeled\xspace}

\newcommand{\THxsec}{{\ensuremath{94.2^{+1.9}_{-1.8}\,\text{(scale)} \pm 2.5\,\text{(PDF)}\unit{fb}}}\xspace}
\newcommand{\THxsecFourF}{{\ensuremath{73.6 \pm 6.2\,\text{(scale)} \pm 0.4\,\text{(PDF)}\unit{fb}}}\xspace}
\newcommand{\EXss}{{\ensuremath{0.933_{-0.077}^{+0.080}\stat {}_{-0.064}^{+0.078}\syst}}\xspace}
\newcommand{\EXxsec}{{\ensuremath{87.9_{-7.3}^{+7.5}\stat {}_{-6.0}^{+7.3}\syst \unit{fb}}}\xspace}
\newcommand{\EXxsecTZ}{{\ensuremath{62.2_{-5.7}^{+5.9}\stat {}_{-3.7}^{+4.4}\syst \unit{fb}}}\xspace}
\newcommand{\EXxsecTbarZ}{{\ensuremath{26.1_{-4.6}^{+4.8}\stat {}_{-2.8}^{+3.0}\syst \unit{fb}}}\xspace}
\newcommand{\EXssTZ}{{\ensuremath{1.02_{-0.09}^{+0.10}\stat {}_{-0.06}^{+0.07}\syst}}\xspace}
\newcommand{\EXssTbarZ}{{\ensuremath{0.79_{-0.14}^{+0.15}\stat {}_{-0.08}^{+0.09}\syst}}\xspace}
\newcommand{\Rvalue}{{\ensuremath{2.37_{-0.42}^{+0.56}\stat {}_{-0.13}^{+0.27}\syst}}\xspace}
\newcommand{\Avalue}{{\ensuremath{0.54 \pm 0.16\stat \pm 0.06\syst}}\xspace}
\newcommand{\il}{{138\fbinv}\xspace}

\newcommand{\stZq}{{\ensuremath{\sigma_{\tZq}}}\xspace}
\newcommand{\stZqSM}{{\ensuremath{\sigma_{\tZq}^\mathrm{SM}}}\xspace}
\newcommand{\chit}{{\ensuremath{\chi^2}}\xspace}
\newcommand{\PZSys}{{\ensuremath{\PQt + \PZ}}\xspace}
\newcommand{\yexpi}{\ensuremath{y^\text{exp}_i}\xspace}
\newcommand{\yobsi}{\ensuremath{y^\text{obs}_i}\xspace}

\newcommand{\pp}{\ensuremath{\Pp\Pp}\xspace}
\newcommand{\tZq}{\ensuremath{\PQt\PZ\PQq}\xspace}
\newcommand{\tbarZq}{\ensuremath{\PAQt\PZ\PQq}\xspace}
\newcommand{\tplusZq}{\ensuremath{\tZq (\ell^+_{\PQt})}\xspace}
\newcommand{\tminusZq}{\ensuremath{\tbarZq (\ell^-_{\PQt})}\xspace}
\newcommand{\tWZ}{\ensuremath{\PQt\PW\PZ}\xspace}
\newcommand{\tHq}{\ensuremath{\PQt\PH\PQq}\xspace}
\newcommand{\tHW}{\ensuremath{\PQt\PH\PW}\xspace}
\newcommand{\tg}{\ensuremath{\PQt\PGg}\xspace}
\newcommand{\ttg}{\ensuremath{\ttbar\PGg}\xspace}
\newcommand{\Zg}{\ensuremath{\PZ\PGg}\xspace}
\newcommand{\Wg}{\ensuremath{\PW\PGg}\xspace}
\newcommand{\ttH}{\ensuremath{\ttbar\PH}\xspace}
\newcommand{\ttZ}{\ensuremath{\ttbar\PZ}\xspace}
\newcommand{\ttW}{\ensuremath{\ttbar\PW}\xspace}
\newcommand{\ttVV}{\ensuremath{\ttbar\PV\PV}\xspace}
\newcommand{\ttVH}{\ensuremath{\ttbar\PV\PH}\xspace}
\newcommand{\ttHH}{\ensuremath{\ttbar\PH\PH}\xspace}
\newcommand{\tttt}{\ensuremath{\ttbar\ttbar}\xspace}
\newcommand{\ttX}{\ensuremath{\PQt(\PAQt)\PX}\xspace}
\newcommand{\VV}{\ensuremath{\PV\PV}\xspace}
\newcommand{\WZ}{\ensuremath{\PW\PZ}\xspace}
\newcommand{\ZZ}{\ensuremath{\PZ\PZ}\xspace}
\newcommand{\VVV}{\ensuremath{\PV\PV\PV}\xspace}
\newcommand{\ZGS}{\ensuremath{\PZ/\PGg^{*}}\xspace}

\newcommand*{\DeepJet}{\textsc{DeepJet}\xspace}
\newcommand*{\TMVA}{\textsc{tvma}\xspace}
\newcommand*{\Tensorflow}{\textsc{tensorflow}\xspace}
\newcommand*{\NNPDFThreeOne}{NNPDF3.1\xspace}
\newcommand*{\NNPDFThreeZero}{NNPDF3.0\xspace}

\newcommand{\qVecStar}{\ensuremath{\vec{p}({\PQq'}^{\star})}\xspace}
\newcommand{\lVecStar}{\ensuremath{\vec{p}(\Pell_{\PQt}^{\star})}\xspace}
\newcommand{\asym}{\ensuremath{A_{\Pell}}\xspace}
\newcommand{\apow}{\ensuremath{a_{\Pell}}\xspace}

\newcommand{\toplep}{\ensuremath{\Pell_{\PQt}}\xspace}
\newcommand{\toplepptvec}{\ensuremath{\ptvec(\Pell_{\PQt})}\xspace}
\newcommand{\zpt}{\ensuremath{\pt(\PZ)}\xspace}
\newcommand{\toppt}{\ensuremath{\pt(\PQt)}\xspace}
\newcommand{\topleppt}{\ensuremath{\pt(\Pell_{\PQt})}\xspace}
\newcommand{\mtz}{\ensuremath{m(\PQt,\PZ)}\xspace}
\newcommand{\mlll}{\ensuremath{m(3\Pell)}\xspace}
\newcommand{\DPhi}{\ensuremath{\Delta \phi}\xspace}
\newcommand{\DEta}{\ensuremath{\Delta \eta}\xspace}
\newcommand{\delphill}{\ensuremath{\DPhi(\Pell,\Pell')}\xspace}
\newcommand{\jprimpt}{\ensuremath{\pt(\mathrm{j}')}\xspace}
\newcommand{\jprimeta}{\ensuremath{\abs{\eta(\mathrm{j}')}}\xspace}
\newcommand{\costheta}{\ensuremath{\cos(\theta^{\star}_{\text{pol}})}\xspace}
\newcommand{\abseta}{\ensuremath{\abs{\eta}}\xspace}
\newcommand{\pz}{\ensuremath{p_\PZ}\xspace}
\newcommand{\mll}{\ensuremath{m_{\Pell\Pell'}}\xspace}
\newcommand{\sigmatop}{\ensuremath{\sigma_{\tZq (\ell^+_\PQt)}}\xspace}
\newcommand{\sigmaantitop}{\ensuremath{\sigma_{\tbarZq (\ell^-_\PQt)}}\xspace}
\newcommand{\Rtz}{\ensuremath{R}\xspace}
\newcommand{\mtw}{\ensuremath{\mT^\PW}\xspace}
\newcommand{\leptonMVA}{\ensuremath{\text{lepton MVA}}\xspace}

\newlength\cmsTabSkip\setlength{\cmsTabSkip}{1ex}

\cmsNoteHeader{TOP-20-010}

\title{Inclusive and differential cross section measurements of single top quark production in association with a \texorpdfstring{\PZ}{Z} boson in proton-proton collisions at \texorpdfstring{$\sqrt{s} = 13\TeV$}{sqrt(s) = 13 TeV}}

\author*[cern]{The CMS Collaboration}

\date{\today}

\abstract{
Inclusive and differential cross sections of single top quark production in association with a \PZ boson are measured in proton-proton collisions at a \centre-of-mass energy of 13\TeV with a data sample corresponding to an integrated luminosity of 138\fbinv recorded by the CMS experiment. Events are selected based on the presence of three leptons, electrons or muons, associated with leptonic \PZ boson and top quark decays. The measurement yields an inclusive cross section of \EXxsec for a dilepton invariant mass greater than 30\GeV, in agreement with standard model (SM) calculations and represents the most precise determination to date. The ratio between the cross sections for the top quark and the top antiquark production in association with a \PZ boson is measured as \Rvalue. Differential measurements at parton and particle levels are performed for the first time. Several kinematic observables are considered to study the modeling of the process. Results are compared to theoretical predictions with different assumptions on the source of the initial-state \PQb quark and found to be in agreement, within the uncertainties. Additionally, the spin asymmetry, which is sensitive to the top quark polarization, is determined from the differential distribution of the polarization angle at parton level to be \Avalue, in agreement with SM predictions.
}

\hypersetup{
pdfauthor={CMS Collaboration},
pdftitle={Inclusive and differential cross section measurements of single top quark production in association with a Z boson in proton-proton collisions at sqrt(s) = 13 TeV},
pdfsubject={CMS},
pdfkeywords={CMS, SM, single top quark, Z boson, cross section, differential}
}

\maketitle

\clearpage

\section{Introduction}
\label{sec:intro}
\begin{figure}[htb!]
\centering
\setlength\intextsep{18pt}
\includegraphics[width=0.25\textwidth]{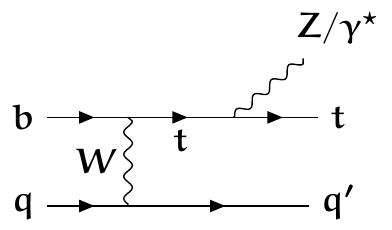}
\hspace{1.0cm}
\includegraphics[width=0.25\textwidth]{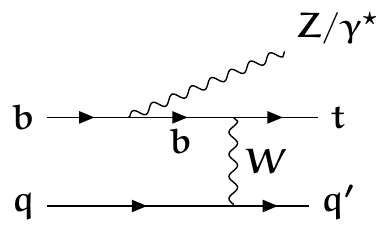}
\hspace{1.0cm}
\includegraphics[width=0.25\textwidth]{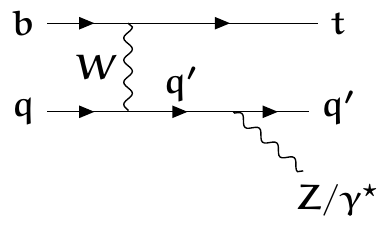}\\
\vspace{1.0cm}
\includegraphics[width=0.25\textwidth]{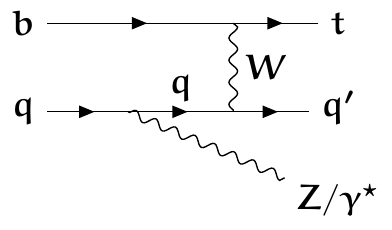}
\hspace{1.0cm}
\includegraphics[width=0.25\textwidth]{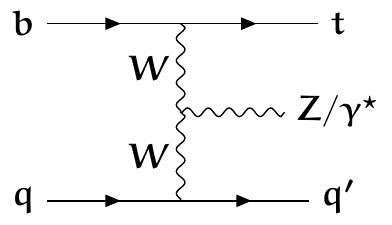}
\hspace{1.0cm}
\includegraphics[width=0.25\textwidth]{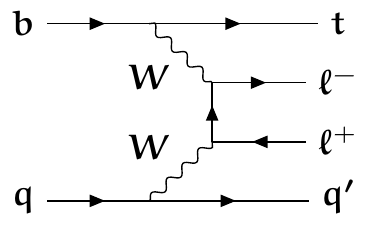}
\caption{Representative LO Feynman diagrams for the \tZq production process.
The production mechanism of nonresonant lepton pairs (lower right) is included in the signal definition to correctly account for interference effects.
}
\label{fig:diagramtzq}
\end{figure}

The electroweak production of a top quark or antiquark in association
with a \PZ boson, the \tZq process, was recently observed in proton-proton (\pp) collisions
at a \centre-of-mass energy of 13\TeV at the
CERN LHC by both the CMS and ATLAS experiments~\cite{Sirunyan:2018zgs,Aad:2020wog}.
The process has unique features that make it a suitable probe for several
interactions in the standard model (SM) of particle physics.
Figure~\ref{fig:diagramtzq} shows representative leading-order (LO) Feynman diagrams of \tZq, where $\ell$ stands for an electron or muon,
including also off-shell photons ($\PGg^{*}$) and the possibility of nonresonant dilepton emission to correctly account for interference effects.
Throughout the text, unless stated otherwise,
\tZq stands collectively for the top quark and antiquark production, including nonresonant dilepton emission.
Because of the pure electroweak nature of \tZq production, corrections to the cross section
arising from quantum chromodynamics (QCD) are typically small.
As a result, the study of {\PQt}{\PZ}, {\PQt}{\PW}{\PQb}, and {\PW}{\PW}{\PZ}
couplings in \tZq production is not primarily affected by QCD uncertainties~\cite{Degrande:2018fog}.
This makes an analysis of \tZq production advantageous in comparison to the associated production
of a top quark-antiquark pair (\ttbar) and a \PZ boson (\ttZ), where the \ttbar is produced via a QCD interaction.

The top quark is strongly polarized in this process because of its electroweak production mechanism.
Measurement of the top quark polarization in the \tZq process
provides complementary information to the existing studies of
the top quark electroweak interactions~\cite{jhep04_2016_023, Khachatryan_2016, jhep04_2017_124, Aaboud_2017, Sirunyan:2019hqb}. Furthermore, the \tZq process offers the possibility
of measuring the top quark and antiquark production cross sections separately, as well as their ratio.
These measurements yield potential sensitivity to different parameterizations of the parton distribution functions (PDFs) of the proton.

Previous measurements of the inclusive \tZq cross section \stZq in leptonic final states have reached a precision of about 15\%~\cite{Sirunyan:2018zgs, Aad:2020wog, Aaboud:2017ylb,Sirunyan:2017nbr} and are in agreement with the SM predictions at next-to-LO (NLO),
$\stZqSM=\THxsec$, for a dilepton invariant mass greater than 30\GeV~\cite{Sirunyan:2017nbr}.
The systematic uncertainty associated with the energy scale used in the calculations arises from variations of the factorization and renormalization scales. The calculation is performed in the five-\flavour
scheme (5FS), where the \PQb quark content of the proton is described by the appropriate PDF.
No differential measurement of the \tZq process has been reported so far.

This paper presents the most precise measurement of \stZq to date,
as well as the first measurement of differential cross sections for the \tZq process.
Data from \pp collisions at $\sqrt{s} = 13\TeV$ collected by the CMS experiment, corresponding to an integrated luminosity of \il~\cite{lumipaper}, are used in this analysis.
The improved precision on \stZq in this measurement compared to the previous results~\cite{Aad:2020wog,Sirunyan:2018zgs,Aaboud:2017ylb,Sirunyan:2017nbr} is due to the larger data sample, an optimized lepton identification, and the use of various control regions in data~\cite{Sirunyan:2018zgs}.
The inclusive measurement is extended to report the first separate determination of the {\PZ} boson associated production cross sections of the top quark and antiquark, as well as their ratio.
The studies are performed in final states with three leptons (electrons or muons), including also a small contribution from sequential $\tau$ lepton decays.
Two selected leptons of same \flavour and opposite charge are assumed to come from the \PZ boson decay, while the third lepton is associated with the leptonic decay of the \PW boson produced in the top quark decay.
Both the inclusive and differential cross section measurements heavily rely on multivariate classifiers to separate
the \tZq signal from various background processes including \ttZ.
The results are obtained by performing maximum likelihood fits on distributions that are obtained from the responses of the classifiers. Tabulated results are provided in HEPData~\cite{hepdata}.

The differential distributions are extracted at both parton and particle levels using a likelihood-based unfolding (as detailed in, \eg, Ref.~\cite{Sirunyan_2019}).
Measured observables at parton level are the transverse momenta, \pt, of the top quark, the \PZ boson, and the lepton from the top quark decay, together with the invariant masses of the three leptons and the \PZSys system. The azimuthal angular distance between the two leptons from the \PZ boson decay, as well as the cosine of the top quark polarization angle, are also measured. The differential measurement of the top quark polarization angle is used to determine the top quark spin asymmetry.
Additionally, the \pt and absolute pseudorapidity, \abseta, of the jet corresponding to the light-flavor quark that recoils against the virtual \PW boson (${\PQq}'$ in Fig.~\ref{fig:diagramtzq}), denoted as the recoiling jet or j', are measured at the particle level.
Results are compared with predictions using the four-\flavour scheme (4FS), where the incoming \PQb quark is produced in the gluon-splitting process, and the 5FS.

The paper is organized as follows: the CMS detector is briefly introduced in
Section~\ref{sec:detector}. Section~\ref{sec:samples} is devoted to the data and simulated samples,
and the identification and selection requirements applied to the reconstructed objects.
A description of the event selection and reconstruction is presented
in Section~\ref{sec:selection}, while the estimation of the backgrounds is discussed
in Section~\ref{sec:backgrounds}. Discussion of systematic uncertainties
affecting the presented measurements follows in Section~\ref{sec:systematics}.
Sections~\ref{sec:inclusive} and~\ref{sec:differential} are dedicated
to the description of the methodology and the obtained results
relevant to the inclusive and differential measurements, respectively.
Finally, the paper is summarized in Section~\ref{sec:summary}.

\section{The CMS detector, data, and simulated samples}
\label{sec:detector}
The central feature of the CMS apparatus~\cite{CMS_detector}
is a superconducting solenoid of 6\unit{m} internal diameter,
providing a magnetic field of 3.8\unit{T}. Silicon pixel and
strip trackers, a lead tungstate crystal electromagnetic
calorimeter (ECAL), and a brass and scintillator hadron calorimeter,
each composed of a barrel and two endcap sections, reside within the solenoid.
Forward calorimeters extend the $\eta$ coverage provided by the
barrel and endcap detectors. Muons are detected in gas-ionization detectors
embedded in the steel flux-return yoke outside the solenoid.

The data events used in the analysis correspond to
the \pp collisions recorded by the CMS experiment in 2016--2018.
Events are required to pass several selection criteria defined at trigger level including the presence of either one, two, or three leptons (electrons or muons)~\cite{Khachatryan:2016bia}. The combination of these triggers yields a trigger selection efficiency close to 100\% in the full phase space relevant to the presented study.
In order to compare the recorded and selected data with SM predictions, dedicated sets of simulated samples are employed, with consistent modeling of the running conditions for each data-taking year.

\begin{figure}[htb!]
\centering
\includegraphics[width=0.4\textwidth]{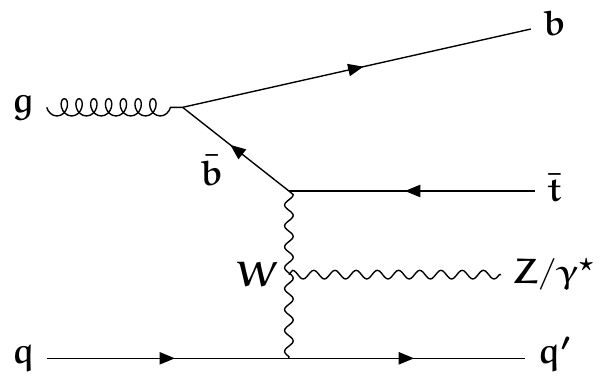}
\hspace{1.0cm}
\includegraphics[width=0.4\textwidth]{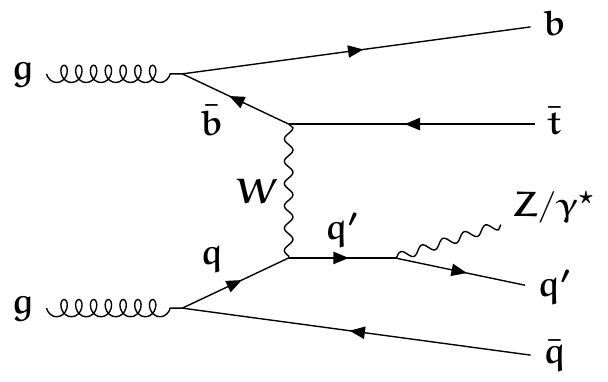}
\caption{An example of the LO (left) and NLO (right) Feynman diagrams for the \tZq
production process in the 4FS.
The \ZGS interference term is included in the MC event
simulation. In the case of the NLO generation, the
{\Pg}{\Pg}- and \qqbar-initiated processes are possible, with an additional quark or gluon present in the final state.}
\label{fig:diagramtzq2}
\end{figure}

The \tZq process requires the presence of a bottom quark
in the initial state.
This can be described using the 5FS as shown in Fig.~\ref{fig:diagramtzq}, where the \PQb quark production depends on the proton PDF.
The Monte Carlo (MC) event simulation produced in the 5FS is preferable in the calculation of the total production cross sections~\cite{pagani2020nlo}.
The modeling of the kinematic properties of the particles in the final state is, however, expected to be more precise in the 4FS, where the \PQb quark is explicitly required to be associated with the gluon-splitting process at the matrix element (ME) level~\cite{pagani2020nlo}. This directly leads to the presence of a second \PQb quark that is produced with relatively small \pt. On the other hand, the additional vertex in the 4FS ME leads to an increased uncertainty related to the renormalization and factorization scales used in the calculation.
Examples of LO and NLO Feynman diagrams corresponding to the \tZq production in
the 4FS are shown in Fig.~\ref{fig:diagramtzq2}.
In the extraction procedure applied to the \tZq events, the prediction of the signal process is based on the 4FS calculations and is normalized to the production cross section obtained in the 5FS.
An alternative \tZq signal sample is generated in the 5FS
with the same generator as used in the simulation of the default \tZq sample,
and is compared to the unfolded results at the parton and particle levels.

Using \MGvATNLO (v2.4.2)~\cite{MadGraph, Frederix2012},
the \tZq signal events are generated at NLO precision in perturbative QCD,
such that processes initiated by gluon ({\Pg}{\Pg}) and quark (\qqbar) pairs are
included and the radiation of an additional gluon is allowed. The nonresonant dilepton
production and \ZGS interference is also included
in the simulation. The same ME generator is used to simulate the
dominant background processes: associated production of a \PW and \PZ boson (\WZ),
\ttbar production in association with a \PZ
(\ttZ) or \PW (\ttW) boson,
production of a photon in association with a \PZ
(\Zg) or \PW (\Wg) boson or with a top quark (\tg),
production of three electroweak gauge bosons (\VVV),
and production of four top quarks (\tttt).
Other background processes, which are simulated at LO
with \MADGRAPH~\cite{MadGraph}, include single top quark production
in association with a Higgs
boson (\tHq), with a Higgs boson and an additional \PW boson (\tHW),
and with a \PW and \PZ boson (\tWZ). Additional background
processes considered include \ttbar production in association with a photon (\ttg),
with two electroweak gauge bosons (\ttVV), with one electroweak gauge
boson and a Higgs boson (\ttVH), and with two Higgs bosons (\ttHH).
For processes with the associated production of two \PZ bosons
(\ZZ), as well as \ttbar production in association with a Higgs boson (\ttH),
the \POWHEG v2~\cite{Nason:2004rx, Frixione:2007vw, Alioli:2010xd, Hartanto_2015}
generator is used at NLO in QCD.
The \MCFM~\cite{Campbell_2010} generator (v7.0.1) is used for the MC event
simulation at LO for the gluon-initiated \ZZ production (${\Pg}{\Pg} \to \ZZ$).
In the measurement of the lepton misidentification rate (see Section~\ref{sec:backgrounds}),
simulated samples with the Drell--Yan (DY) and \ttbar production processes are used.
The processes are simulated at NLO with \MGvATNLO and \POWHEG~v2, respectively.

Simulated events are processed with \PYTHIA (v8.2)~\cite{Sjostrand:2014zea}
to model the fragmentation and the parton shower. The FxFx~\cite{Frederix2012} merging
scheme is used to avoid double counting associated with the MC event
simulation in the same phase space due to the ME generation of extra partons at NLO with \MGvATNLO.
A set of CP5 tuning parameters~\cite{TuneCP5} is used for the parton shower, hadronization, and
modeling of the underlying event in the 2017 and 2018 MC samples,
as well as of the \tZq, \ttZ, \ttW, \ttH, \ttg, \Zg, and \tttt MC
samples in all three years.
The CUETP8M1~\cite{TuneCUETP8M1_1, TuneCUETP8M1_2}, CUETP8M2,
and CUETP8M2T4~\cite{TuneCUETP8M2} tunes are used
for other 2016 MC samples.
The \tZq and \ttZ MC samples in all three years, as well as
MC samples for 2017 and 2018 data, are generated with the \NNPDFThreeOne~\cite{NNPDF31} PDF set
(with next-to-NLO precision in perturbative QCD for \tZq).
Other MC samples for 2016 are generated with the \NNPDFThreeZero~\cite{NNPDF30} PDF set.
The effects of additional \pp collisions attributed to the same or adjacent bunch crossings (pileup)~\cite{cms_pileup_mitigation},
are simulated with \PYTHIA.
The simulated events are reweighted according to the distribution of the number of interactions in each bunch
crossing corresponding to a total inelastic \pp cross section of 69.2\unit{mb}~\cite{Sirunyan_2018}.
The simulation of the CMS detector is performed with \GEANTfour~\cite{GEANT4}.

\section{Reconstruction and identification of physics objects}
\label{sec:samples}
To reconstruct the physics objects described below, the same algorithms are applied to simulated events and data.
The particle-flow (PF) algorithm~\cite{CMS-PRF-14-001} is used to reconstruct and identify photons, electrons, muons, and charged and neutral hadrons in an event, with an optimized combination of information from the various elements of the CMS detector.
The missing transverse momentum vector \ptvecmiss is computed as the negative vector \ptvec sum of all the PF candidates in an event~\cite{Sirunyan:2019kia}.

The candidate vertex with the largest value of summed $\pt^2$ of all physics objects assigned to this vertex is taken to be the primary vertex (PV) of the \pp interaction.

Jets are reconstructed by clustering the PF candidates using the anti-\kt algorithm~\cite{Cacciari:2008gp, Cacciari:2011ma} with a distance parameter of $R = 0.4$.
Charged particles identified as originating from pileup
interactions are discarded and an offset correction is applied to correct for remaining contributions.
Jet energy corrections are derived from simulation to bring
the measured response of jets to that of particle-level jets on average.
In situ measurements of the momentum balance in dijet,
photon+jet, {\PZ}+jet, and multijet
events are used to account for any residual differences in the
jet energy scale between data and simulation~\cite{Khachatryan:2016kdb}.
The jet energy resolution in simulation is corrected to match the one observed
in data. Additional selection criteria are applied to each jet
to remove jets potentially dominated by anomalous contributions
from various subdetector components, or misreconstruction.
Jets are required to have $\pt > 25\GeV$, $\abseta < 5$, and be separated from any identified
lepton by $\DR = \sqrt{\smash[b]{(\DEta)^2 + (\DPhi)^2}} > 0.4 $,
where \DEta and \DPhi are the pseudorapidity and
azimuthal angular separation between the jet and
the lepton, respectively. The relatively loose
selection criterion applied to the jet \abseta is necessary
for reconstructing the light quark jet in the \tZq process,
which is predominantly produced in the forward region
of the detector (see Fig.~\ref{fig:diagramtzq}).
Jets that are reconstructed within the acceptance of the CMS pixel
detector ($\abseta < 2.4$ for 2016, $\abseta < 2.5$ for
2017 and 2018) are denoted as central jets.

Using the \DeepJet algorithm~\cite{BTV-16-002, Bols:2020bkb, CMS-DP-2018-058},
central jets containing \PQb hadrons are identified as \PQb-tagged jets.
The \PQb tagging requirement used in the analysis corresponds to a \PQb quark jet selection efficiency of about 85\% for jets with $\pt > 30\GeV$ as estimated in simulated \ttbar events. 
An associated misidentification rate of 1\% for jets arising from \PQu, \PQd, or \PQs quarks and gluons, and 15\% for jets arising from \PQc quarks is obtained for those events.

The electron momentum is estimated by combining the energy measurement
in the ECAL with the momentum measurement in the tracker. The momentum
resolution for electrons with $\pt \approx 45\GeV$ from
$\PZ \to \Pe\Pem$ decays is within 1.7--4.5\%. The resolution
is generally better in the barrel than in the endcap region, and depends
on the bremsstrahlung energy emitted by the electron as it
traverses the material in front of the ECAL~\cite{Khachatryan:2015hwa}.
Electrons are selected within $\abseta < 2.5$.

Muons are reconstructed within $\abseta < 2.4$ using drift tubes,
cathode strip chambers, and resistive plate chambers.
Association of muon objects to reconstructed tracks that are measured in the silicon tracker
yields the relative \pt resolution of 1\% in the barrel and 3\% in the endcaps
for muons with \pt up to 100\GeV, and
of better than 7\% in the barrel for muons with \pt up to 1\TeV~\cite{Sirunyan:2018}.

Leptons originating from decays of electroweak bosons are referred to as ``prompt'',
while those originating from hadron decays, as well as misidentified leptons from jets or hadrons, are collectively referred to as ``nonprompt''.
A strong separation between prompt and nonprompt leptons is obtained by
using a set of discriminating variables based on the reconstructed properties of leptons and jets.
A relative isolation variable is defined as the scalar \pt sum of all
PF objects inside a cone of $\DR = 0.3$ around the direction of the lepton, excluding the lepton itself,
divided by the \pt of the lepton~\cite{relIso_0,relIso_1}. A relative isolation parameter
computed with a cone size that decreases for higher lepton \pt values is
also used.
The isolation variables are corrected for pileup effects.
The reconstructed transverse and longitudinal impact
parameters, as well as the signed impact
parameter significance, of the tracks associated with the leptons,
computed with respect to the PV position, are used to determine the consistency of the leptons originating from the PV.
A number of variables discriminating between prompt and nonprompt leptons
use information about the reconstructed jet with the smallest $\DR$ with respect to the identified lepton, requiring $\DR < 0.4$.
This jet is used to compute the number of charged particles
matched to the jet, the ratio of the jet \pt to the lepton \pt,
the lepton momentum projected on the
transverse plane to the reconstructed jet direction, as well as the
output discriminator value of the \DeepJet \PQb tagging
algorithm. In addition, the muon segment
compatibility criteria~\cite{Sirunyan:2018} are used for selected muons, while
the output discriminator value of the electron identification
algorithm is used for electrons~\cite{Khachatryan:2015hwa}.

The aforementioned variables are combined into a multivariate analysis
(MVA) based discriminant (\leptonMVA), which is trained and evaluated with the \TMVA
package~\cite{tmva2007}.
A boosted decision tree (BDT) algorithm is trained on a large sample of simulated prompt leptons originating from the \tZq, \ttZ,
and \ttW processes, as well as nonprompt leptons taken from simulated \ttbar
events.
The requirement on the \leptonMVA value corresponds to a prompt lepton selection efficiency of about 95\%, while rejecting 98\% of the nonprompt leptons, as evaluated from MC simulation for leptons with $\pt > 25\GeV$.
The \leptonMVA and its
training discussed here are an extension and reoptimization of a similar
MVA used in the first observation of the \tZq process by CMS~\cite{Sirunyan:2018zgs}.

Leptons that pass the requirement on the \leptonMVA are labeled as ``tight'' leptons and are selected for further analysis. Leptons that fail this requirement are subjected to additional criteria, including requirements on the relative lepton isolation and the \DeepJet discriminator value of the jet that is closest to the lepton.
Leptons that are either tight or satisfy those additional criteria are labeled as ``loose'' leptons and are used in the estimation of the nonprompt-lepton background from control samples in data (as discussed in Section~\ref{sec:backgrounds}).

\section{Event reconstruction and signal selection}
\label{sec:selection}
Selected events are required to contain exactly
three tight leptons and at least two jets, of which at least one is \PQb tagged. The three leptons, ordered according to their \pt,
must have a \pt of at least 25, 15, and 10\GeV, respectively.
Two of the three leptons are required to form a pair with electric charge of opposite sign and same lepton \flavour (OSSF).
Furthermore, the invariant mass of the OSSF pair must
be compatible with the \PZ boson mass within 15\GeV.
In case of an ambiguity, the OSSF pair with the mass closest to the \PZ boson mass is chosen.

Events that satisfy the aforementioned conditions
define the signal region.
For \tZq events with three prompt leptons the selection efficiency is about 20\%.
For the inclusive and differential measurements the
signal region is furthermore divided into subregions based
on the number of jets and \PQb-tagged jets.
In order to study the background prediction, we define
dedicated control regions that are complementary to the signal region.
They are discussed in detail in Section~\ref{sec:backgrounds}.

A good discrimination between the \tZq process and various
backgrounds contributing to the signal
region is achieved by using MVA techniques.
In the final step of the measurement, the output score
of an MVA is used in maximum likelihood fits to extract
the inclusive and differential \tZq cross sections. A full
event reconstruction, described below, is performed to obtain a set of additional
variables used as inputs for the MVA to improve the separation between signal and background
events. Identified physics objects are used to compute
several observables for the differential
cross section measurement.

The four-momentum of the neutrino in the decay of the \PW boson originating from the top quark is
reconstructed similarly to Ref.~\cite{Chatrchyan_2011}.
First, the lepton that is not associated with the OSSF lepton pair,
denoted as \toplep, is assigned to the \PW boson.
Then, a \PW boson mass constraint on the system of the \ptvecmiss and \toplepptvec is imposed.
This leads to two solutions, or in some cases one solution, for the neutrino four-momentum.
The top quark candidate is reconstructed by combining the four-momenta of the
neutrino solution(s), the \toplep, and a \PQb-tagged jet.
In the case of an ambiguity, the combination that gives the top quark candidate mass closest to the value of 172.5\GeV is chosen.
A particular feature of the \tZq process is the recoiling jet that is often radiated in the forward detector region.
This jet is identified with an efficiency of about 86\% by selecting the jet
with the highest \pt, excluding \PQb-tagged jets.

\begin{figure}[p!]
  \centering
  \includegraphics[width=0.45\textwidth]{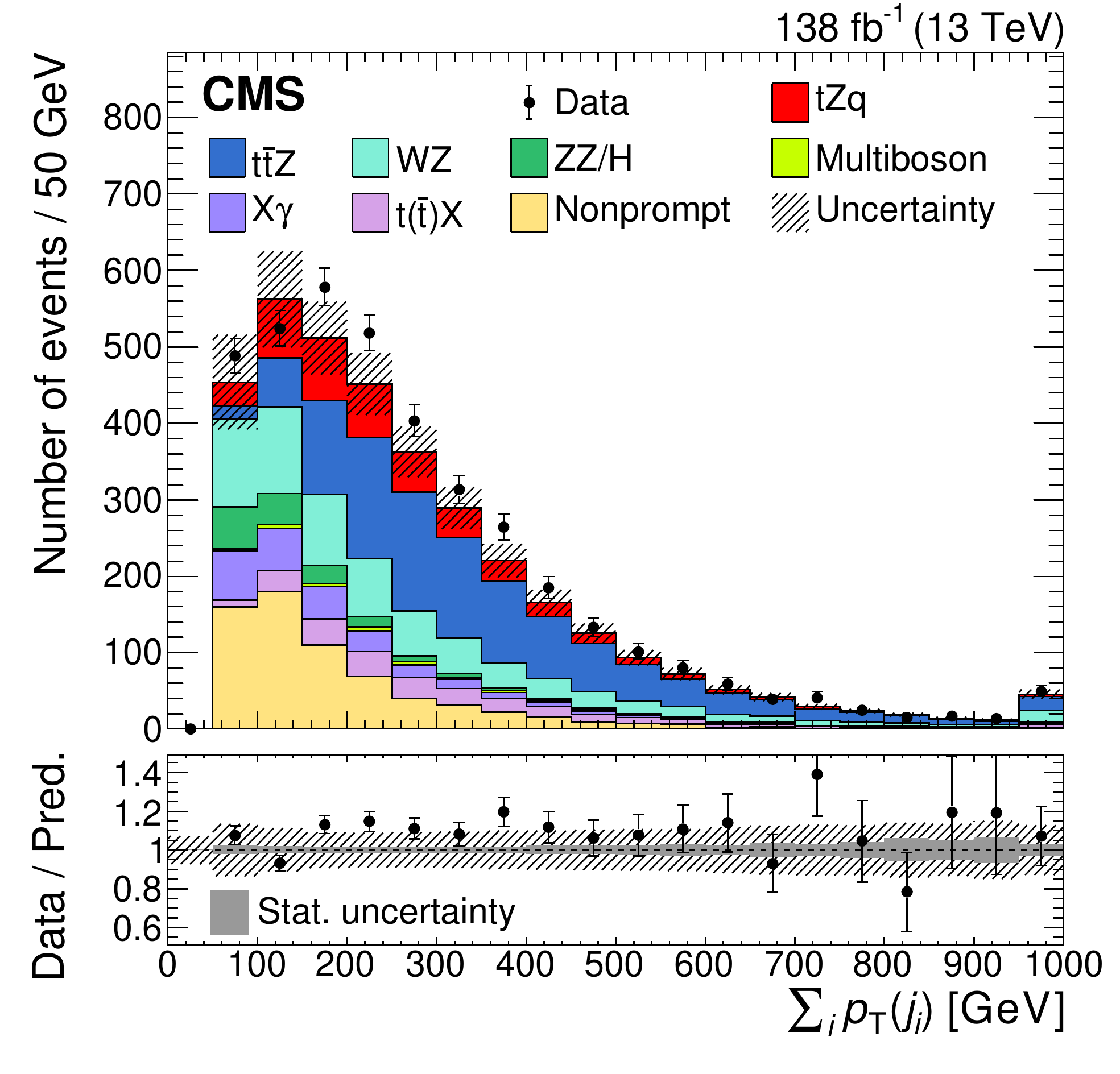}
  \includegraphics[width=0.45\textwidth]{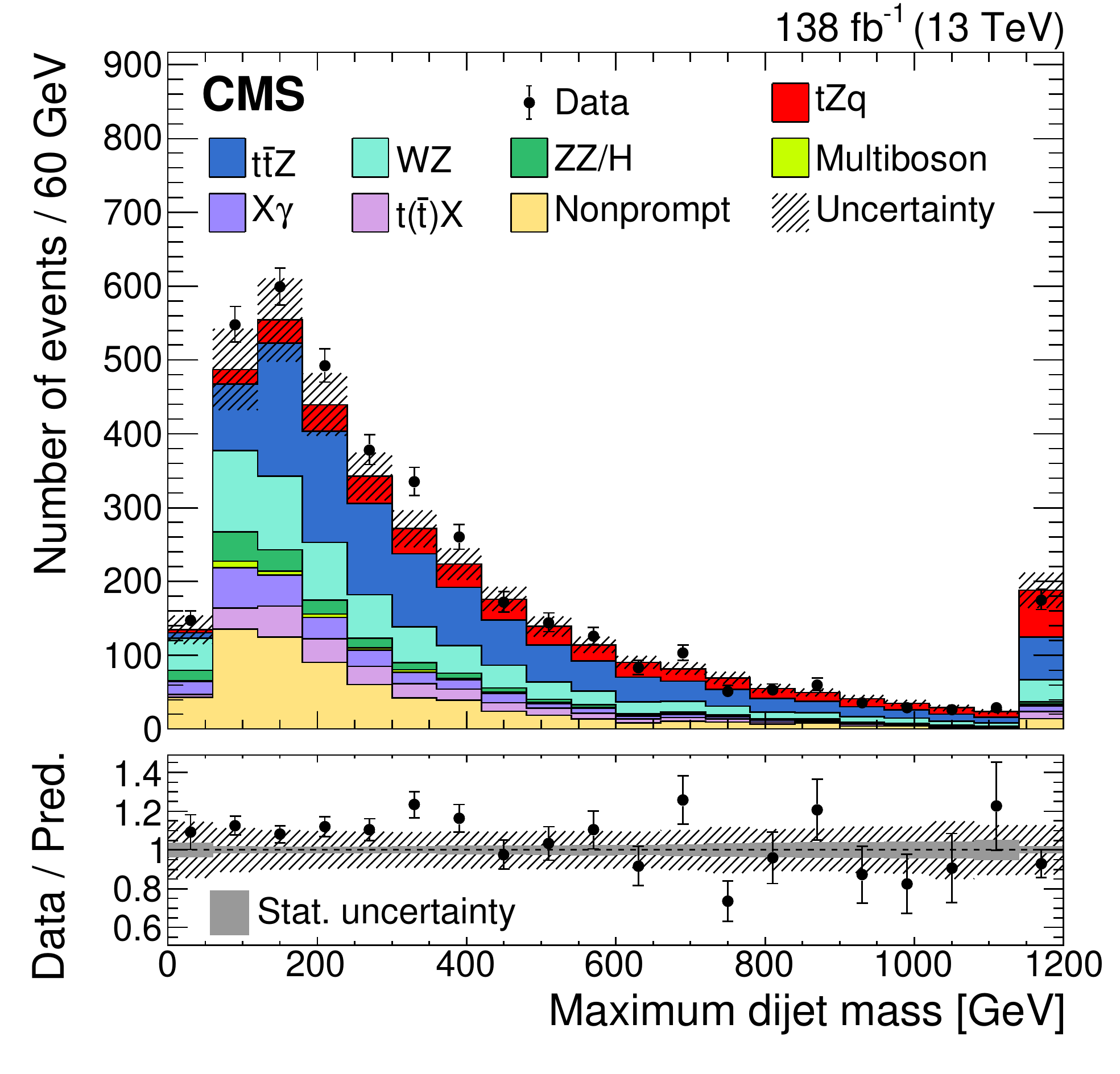} \\
  \includegraphics[width=0.45\textwidth]{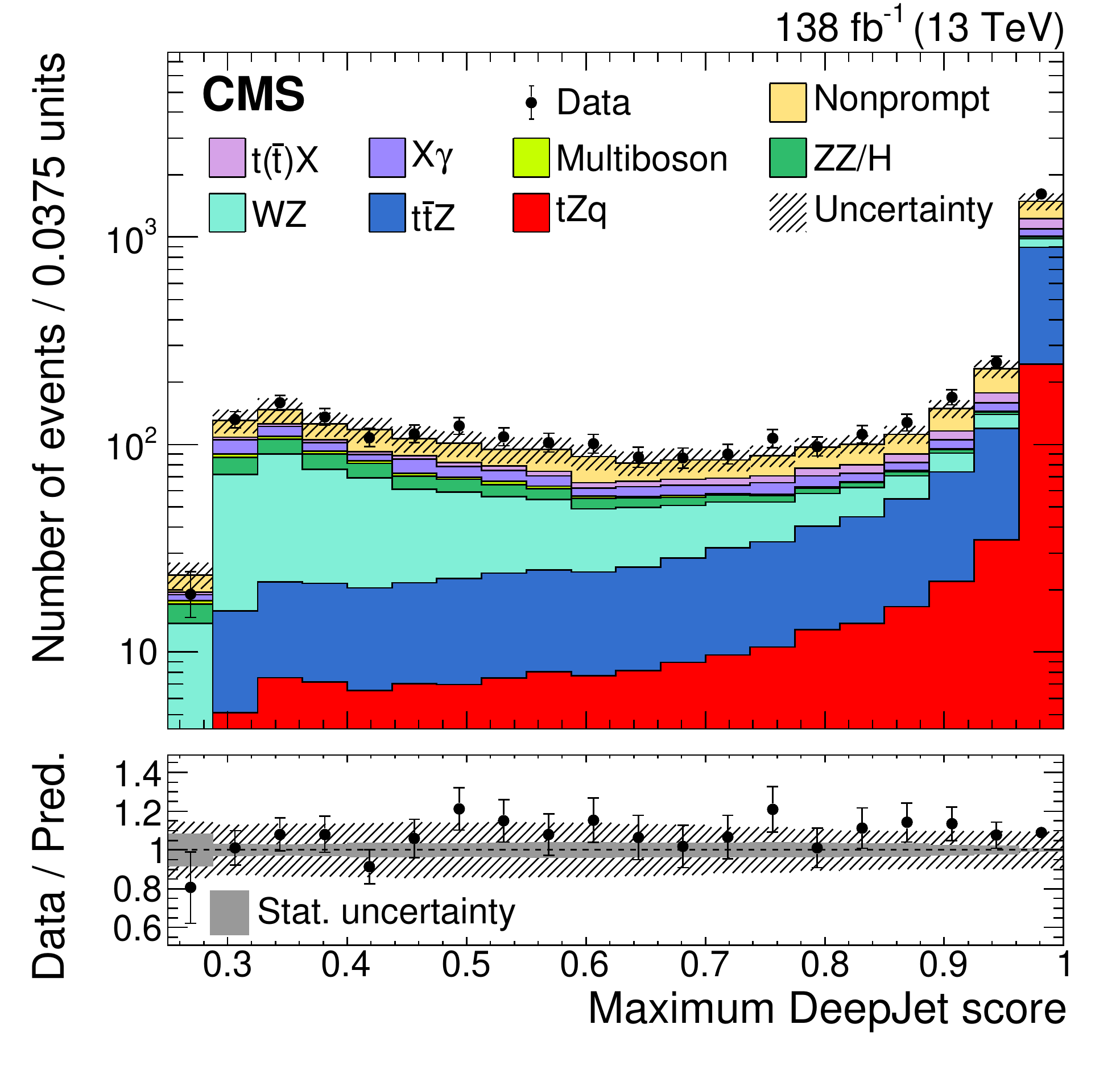}
  \includegraphics[width=0.45\textwidth]{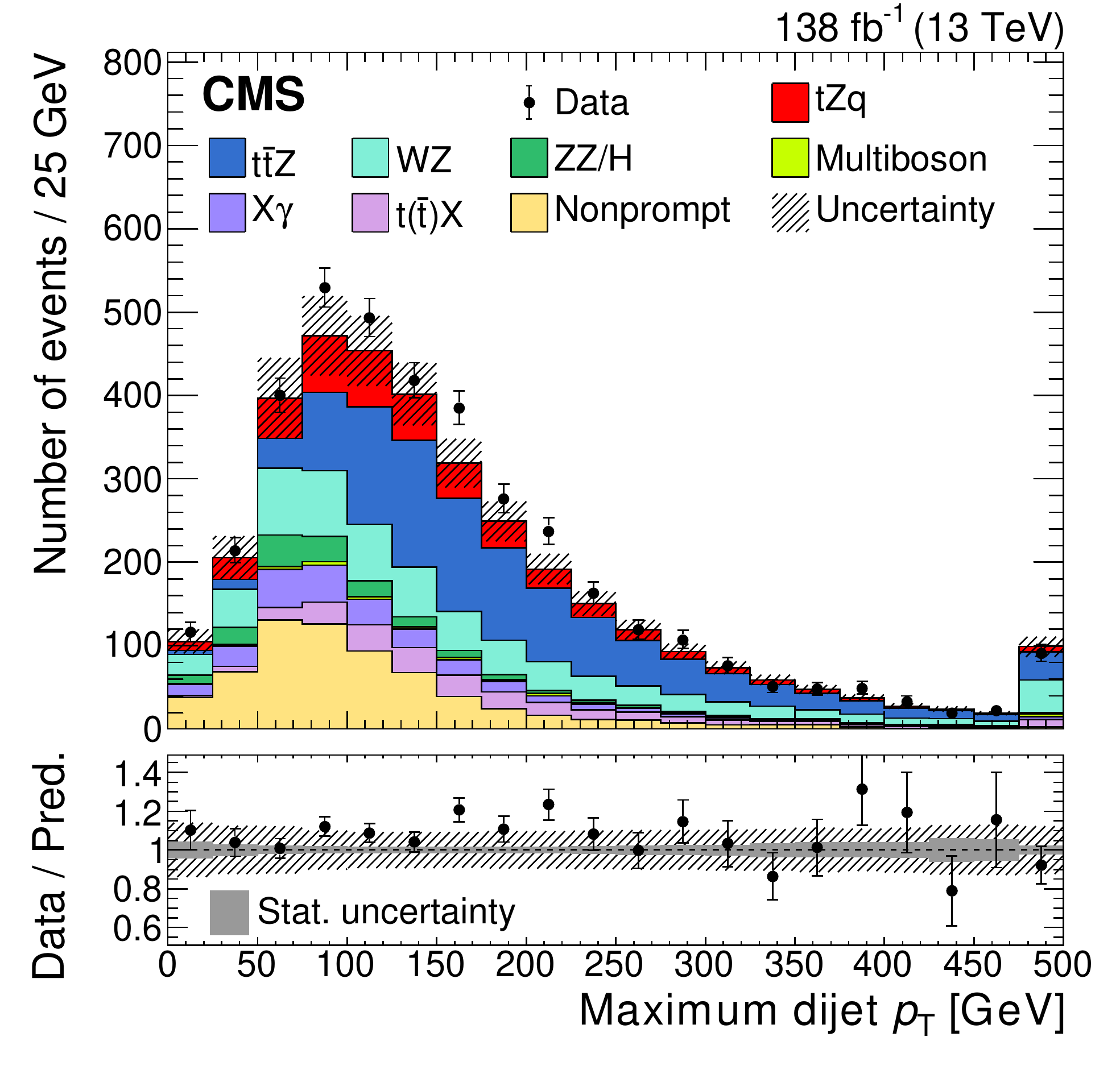} \\
  \includegraphics[width=0.45\textwidth]{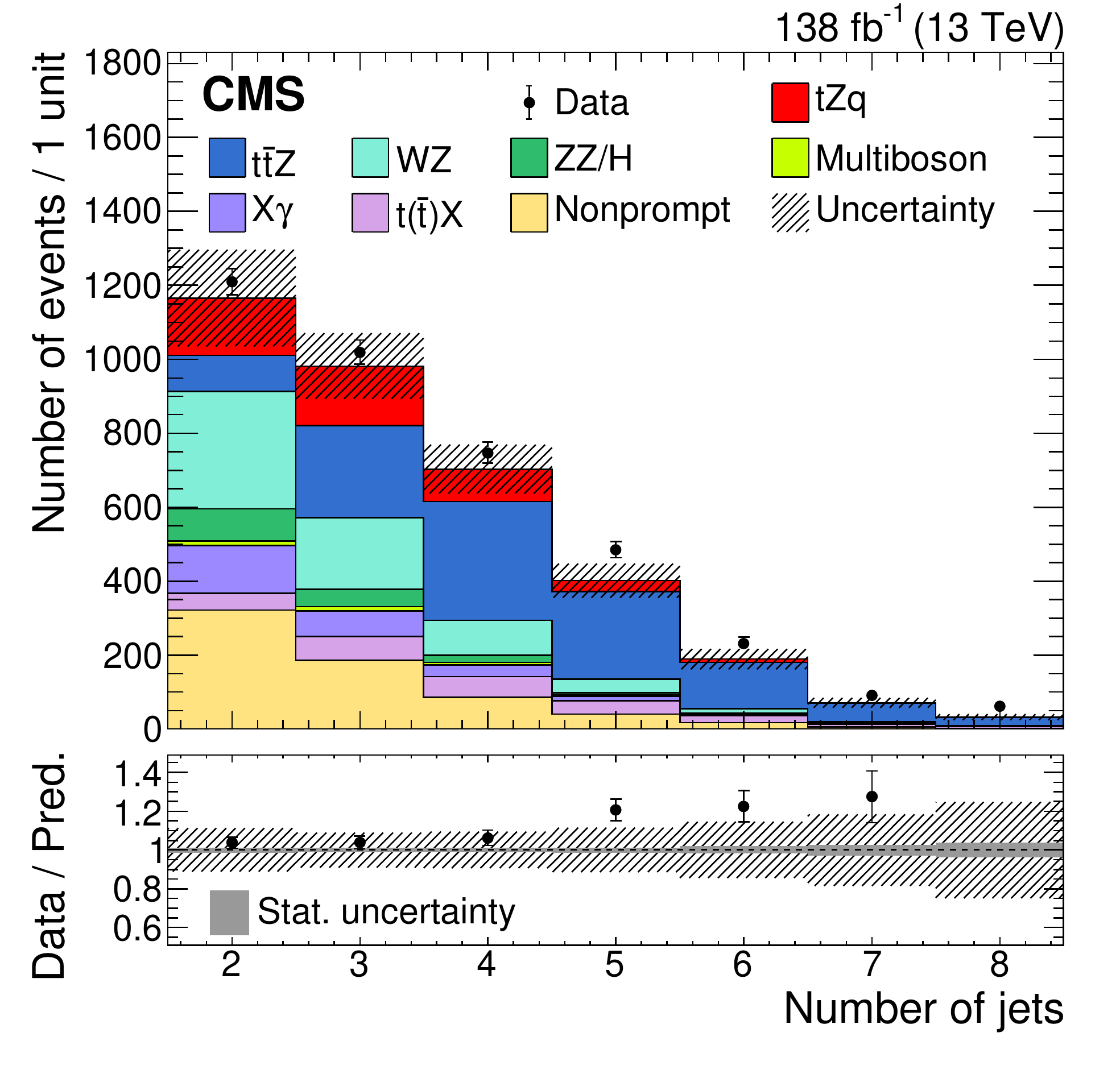}
  \includegraphics[width=0.45\textwidth]{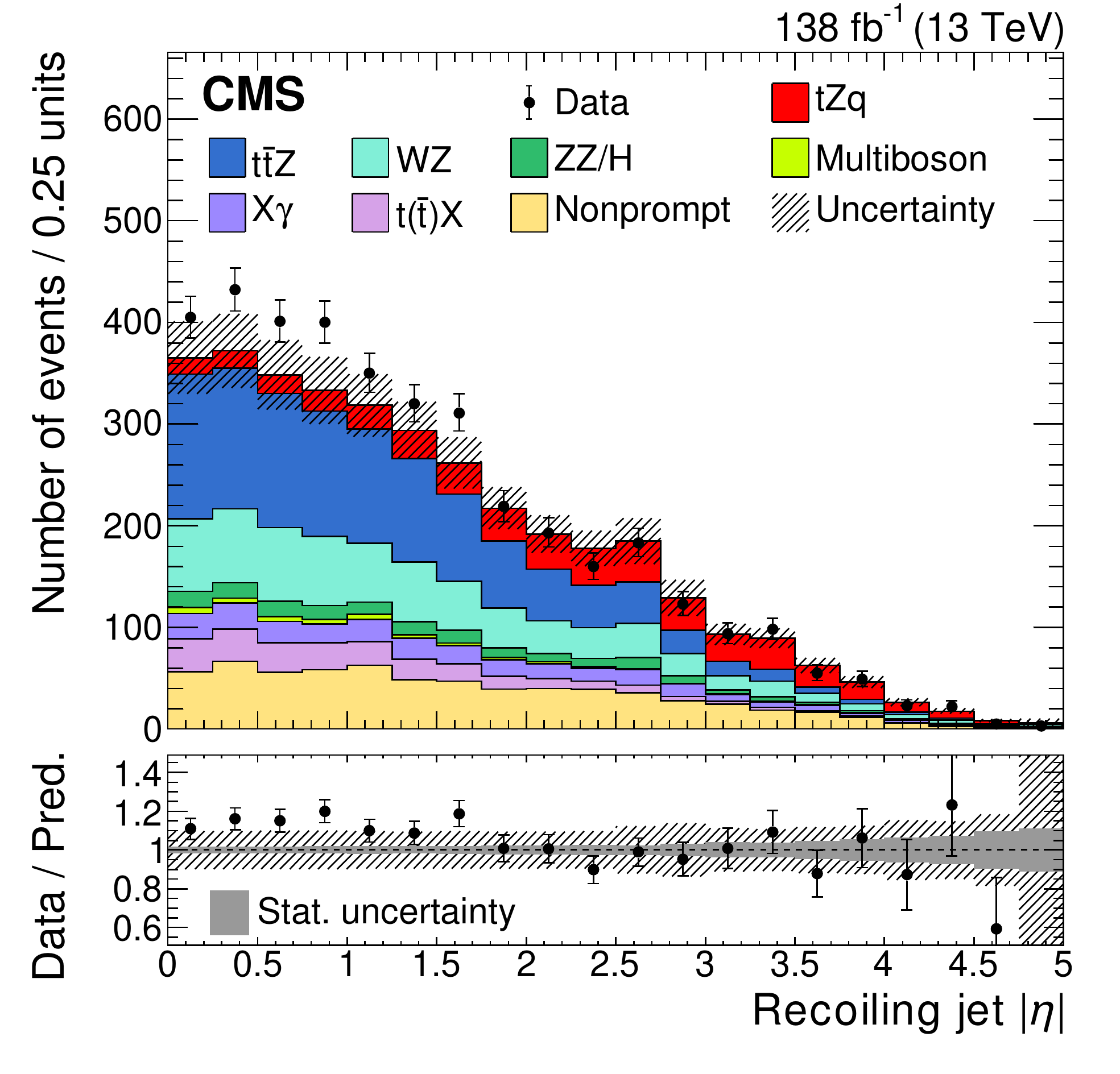}
\caption{Distributions of the most
powerful discriminating variables in the signal region for the data (points) and predictions (colored histograms),
including the scalar \pt sum of all jets (upper left),
the maximum invariant mass of any two-jet system (upper right),
the maximum \DeepJet score of any jet (middle left),
the maximum \pt value of any two-jet system (middle right),
the number of jets in the event (lower left),
and the \abseta of the recoiling jet (lower right).
The last bins include the overflows.
The lower panels show the ratio of the data to the sum of the predictions. The vertical lines on the data points represent the statistical uncertainty in the data; the shaded area corresponds to the total uncertainty in the prediction; the \grey area in the ratio indicates the uncertainty related to the limited statistical precision in the prediction.}
\label{fig:SRFeatures}
\end{figure}

The most powerful discriminating variables used in the event classification
are obtained from the event reconstruction
and correspond to the scalar \pt sum of all jets,
the maximum invariant mass and maximum \pt of any two-jet system,
the maximum \DeepJet score of any jet,
the number of jets in the event,
and the \abseta of the recoiling jet.
Predicted distributions of these variables compared to data are shown in Fig.~\ref{fig:SRFeatures}.

Other discriminating variables are the reconstructed top quark mass,
the invariant mass of the OSSF lepton pair,
the angle between j' and the \PQb-tagged jet associated with the top quark decay,
the number of \PQb-tagged jets,
the scalar sum of the \pt of all selected leptons and \ptmiss, and
\ptmiss itself.
The transverse mass of the reconstructed \PW boson (\mtw) is also included in the event classification and is defined as:
\begin{equation}
    \mtw = \sqrt{2 \ptmiss \topleppt \left[ 1 - \cos \DPhi \right]},
\end{equation}
where \DPhi is the difference in azimuthal angle between \toplepptvec and \ptvecmiss.

The top quark polarization is linked to the polarization of the lepton from its decay
and can be measured with respect to the direction of the recoiling quark.
The cosine of the top quark polarization angle \costheta is defined
similarly to Ref.~\cite{Sirunyan:2019hqb} as:
\begin{equation}
    \costheta = \frac{\qVecStar \cdot \lVecStar)}{\abs{\qVecStar} \abs{\lVecStar}},
\end{equation}
where $\qVecStar$ and $\lVecStar)$
are the three-momenta of the recoiling quark and the lepton from
the top quark decay, respectively.
The asterisks indicate that the three-momenta are measured in the top quark candidate rest frame.
The polarization $P$ of the top quark is related to
the spin asymmetry as $\asym = \frac{1}{2} P \apow$,
where \apow refers to the spin-analyzing power
of the lepton associated with the top quark decay
and is equal to unity in LO calculations~\cite{Je_abek_1994, Aguilar_Saavedra_2010}.
The spin asymmetry \asym is related to the differential cross section as a function of \costheta by:
\begin{equation}
    \label{eq:spin_asymmetry}
    \frac{\rd\sigma}{\rd\costheta} = \stZq \left(\frac{1}{2} + \asym \costheta \right).
\end{equation}

\section{Background determination}
\label{sec:backgrounds}
Several background contributions to the signal region are studied, divided into two main categories.
The first contains processes that include three genuine prompt leptons.
Events in the second category contain at least one nonprompt lepton,
and therefore enter the signal region by virtue of imperfect nonprompt-lepton rejection.
Background contributions from the first category are \modelled using the MC simulations,
whereas the backgrounds from the second are estimated using a technique based on control samples in data.

\begin{figure}[htb!]
	\centering
	\includegraphics[width=0.45\textwidth]{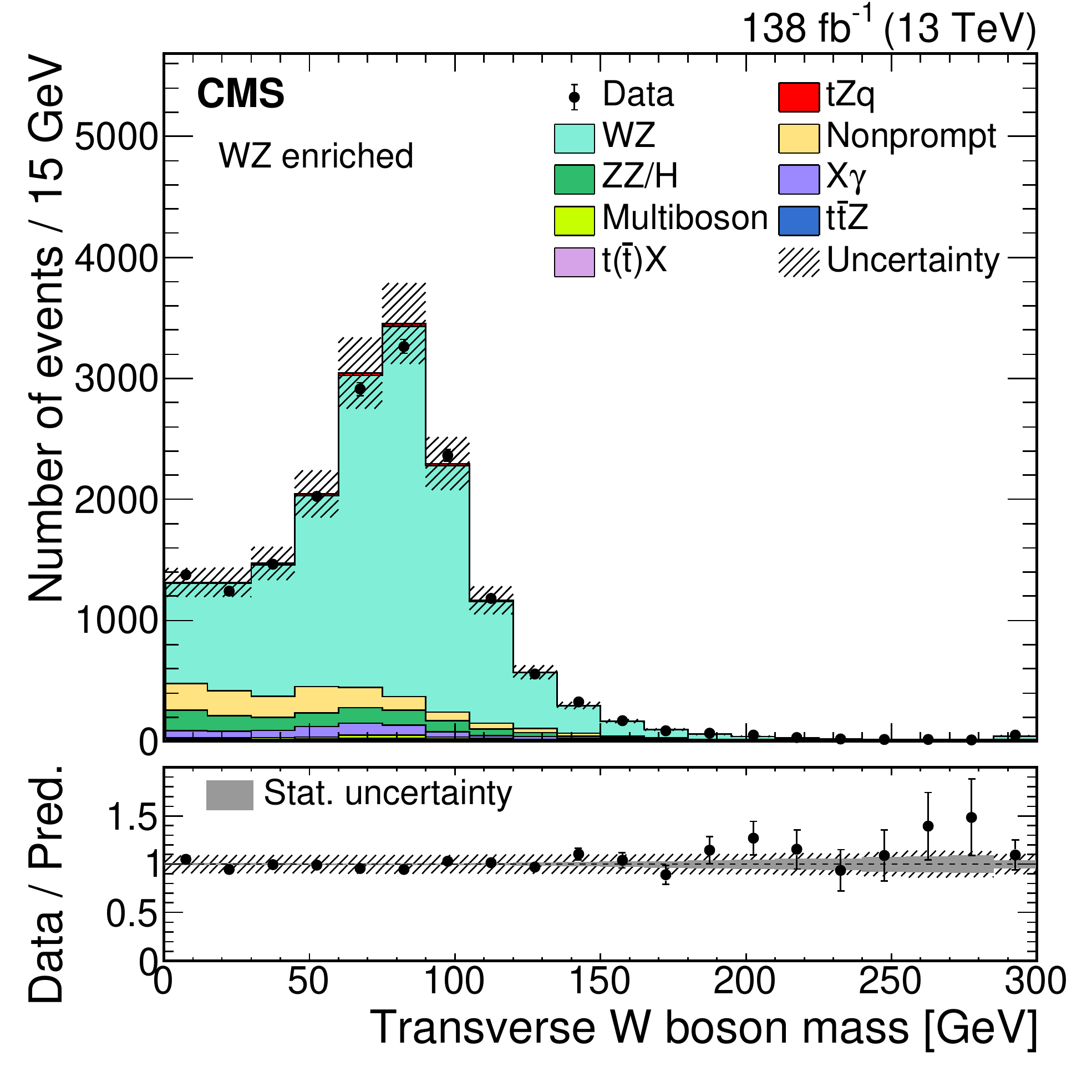}
	\includegraphics[width=0.45\textwidth]{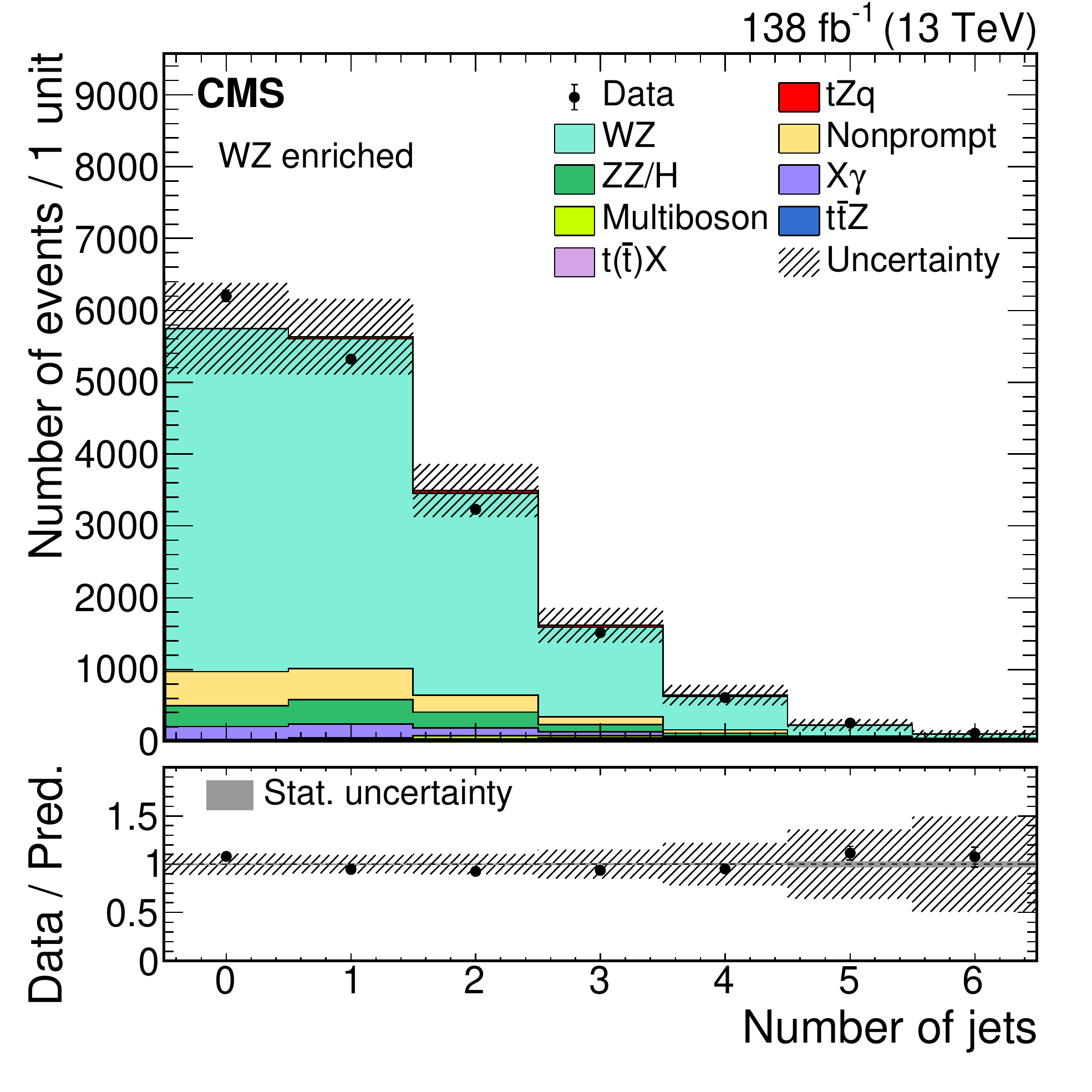} \\
    \includegraphics[width=0.45\textwidth]{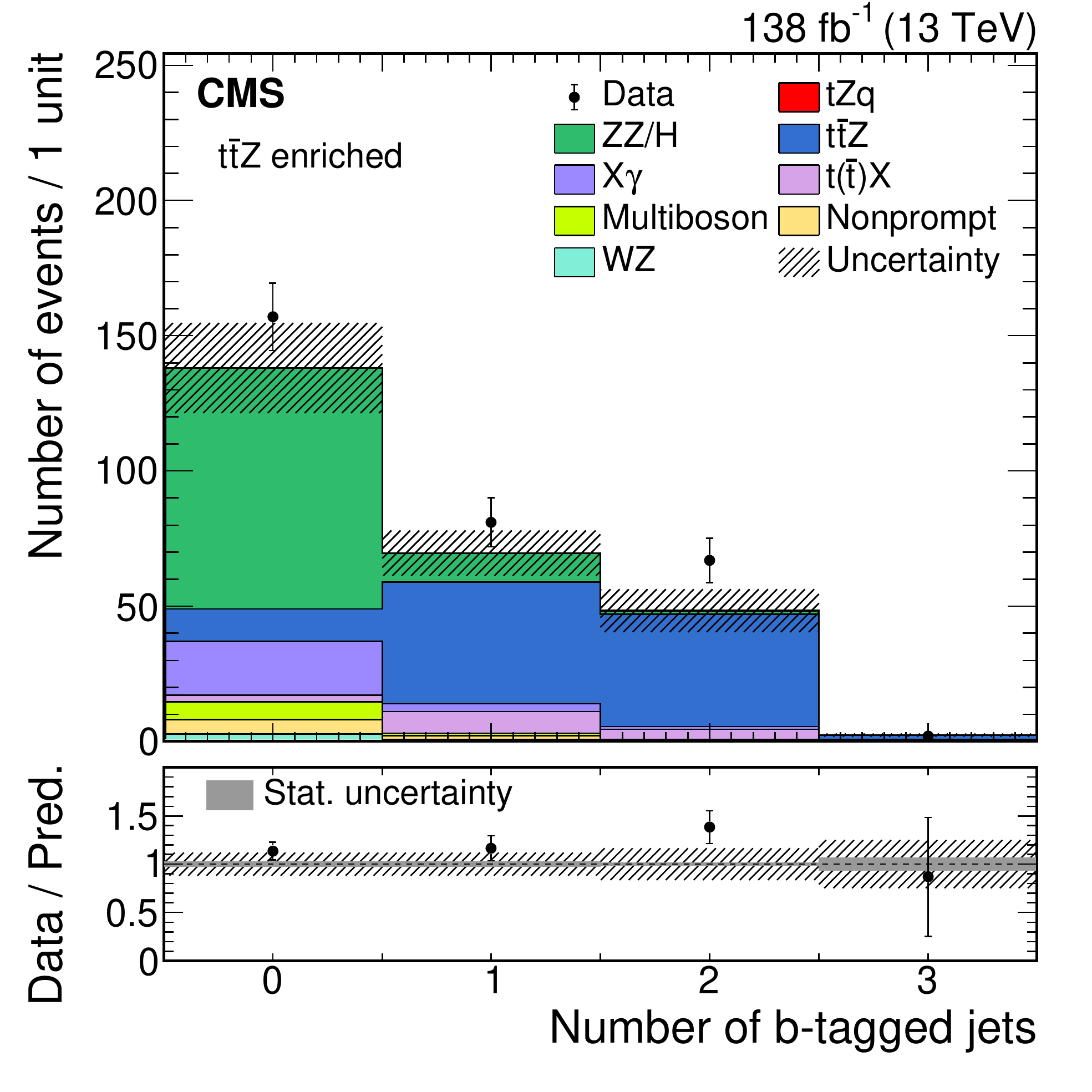}
    \includegraphics[width=0.45\textwidth]{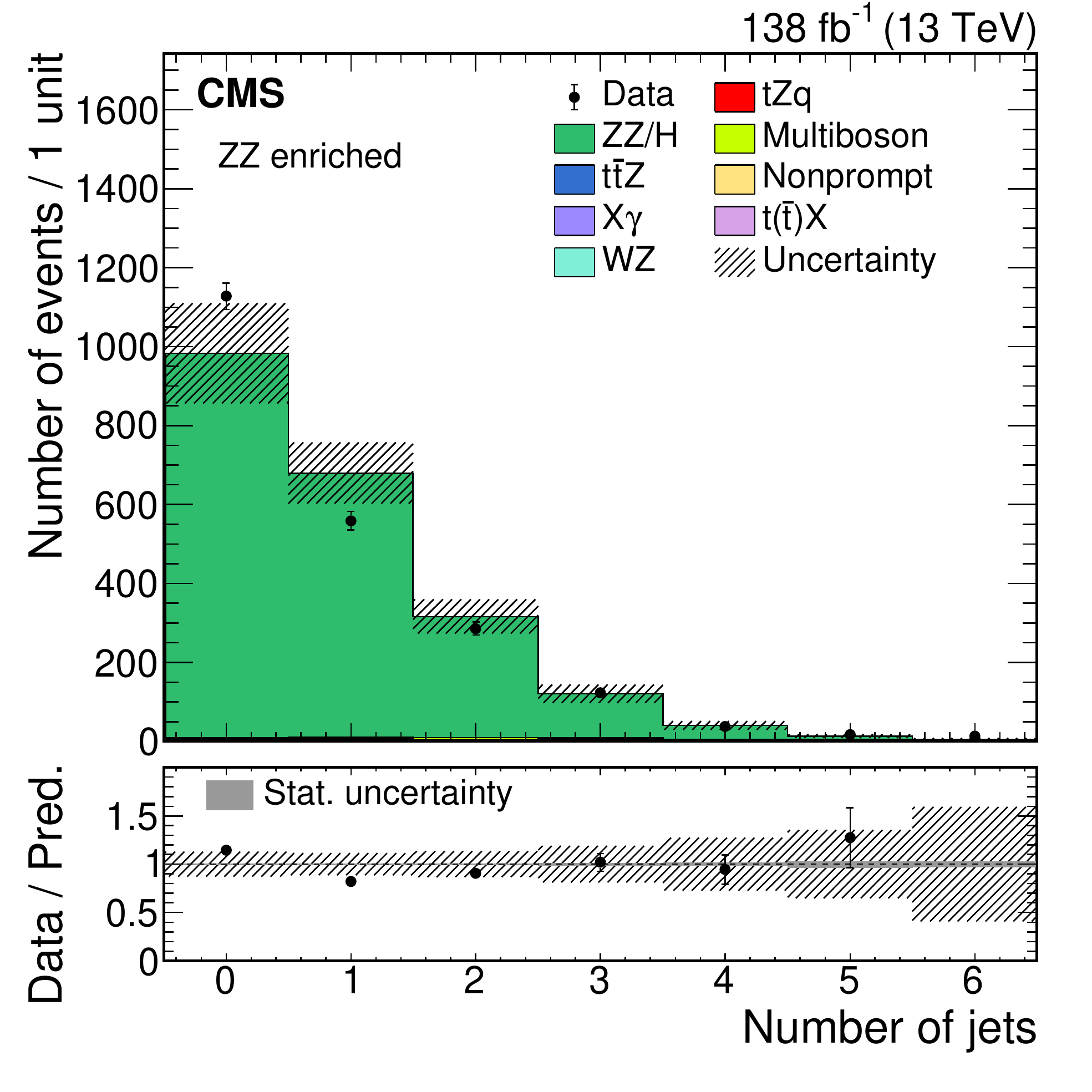}\\
	\caption{
    Distributions of the transverse \PW boson mass (upper left)
    and the number of selected jets (upper right) for the \WZ-enriched control region,
    the number of \PQb-tagged jets (lower left) in the \ttZ-enriched control region,
    and the number of jets (lower right) in the \ZZ-enriched control region for the data (points) and predictions (colored histograms).
    The lower panels show the ratio of the data to the sum of the predictions. The vertical lines on the data points represent the statistical uncertainty in the data; the shaded area corresponds to the total uncertainty in the prediction; the \grey area in the ratio indicates the uncertainty related to the limited statistical precision in the prediction.}
	\label{fig:controlregion}
\end{figure}

The production of \WZ bosons is an important source of background events, especially for events with
a small number of reconstructed jets or \PQb-tagged jets.
The inclusive production cross section of this process is both predicted and measured with high precision~\cite{SirunyanWZ:2019}.
In order to validate the predictions obtained for \WZ production with
additional jets, a dedicated data control region is defined with similar
lepton identification requirements as used in the signal region,
but vetoing events containing a \PQb-tagged jet.
Additionally, $\ptmiss > 50\GeV$ is required, accounting for the reconstructed missing momentum originating due to the neutrino coming from the \PW boson decay.
Figure~\ref{fig:controlregion} shows the predicted jet multiplicity
and \mtw distributions compared to data in this control region.
Good agreement in the overall normalization
and shape of the presented distributions is observed.
In the signal region, about 30\% of the simulated \WZ events
have a jet containing a bottom quark.
The other 70\% enter the signal region because of misidentification in the heavy-flavor jet tagging algorithm, with jets originating from light quarks and \PQc quarks each contributing about half.
The modeling of the \WZ process with \PQb quarks
is subject to an uncertainty that is not constrained in the control
region because there is a negligible fraction of events with an
additional \PQb quark.
A dedicated study of this uncertainty was performed in DY
events~\cite{SirunyanTTZ:2020}, resulting in an additional uncertainty of 20\%
assigned to the normalization of \WZ selected events containing an
additional \PQb quark in the signal region.

The dominant background in the subregions with a large
number of jets or \PQb-tagged jets comes from the \ttZ process.
The modeling and normalization of \ttZ is validated in two distinct ways.
In the signal region, good separation of this process from \tZq is achieved with the MVA technique.
Hence, the signal region contains an implicit control region for \ttZ at low MVA discriminant values.
Additionally, a dedicated control region is defined that requires
an event to contain four leptons, with an OSSF pair
compatible with the \PZ boson mass.
If a second OSSF pair is present, it is required
not to be compatible with the \PZ boson mass (within 15\GeV) in
order to suppress the contribution from the \ZZ process.
The distributions of the number of \PQb-tagged jets in data and the prediction for this \ttZ-enriched control region are shown in the lower-left plot of Fig.~\ref{fig:controlregion}. 
The data exceed the prediction especially for the bin with two \PQb-tagged jets where the \ttZ contribution is large. This observed underprediction is consistent with previous measurements~\cite{SirunyanTTZ:2020}.

The background from \ZZ events in the final states involving
three leptons consists of events where both \PZ bosons decay
leptonically, but one of the leptons is not reconstructed
or does not satisfy the lepton selection requirements.
The \ZZ control region requires the presence of four leptons that are used
to form two OSSF pairs, both of them compatible with the \PZ boson
mass within 15\GeV.
The distributions of the number of jets in data and the prediction for this \ZZ-enriched control region are shown in the lower-right plot of Fig.~\ref{fig:controlregion}.

\begin{figure}[htb!]
	\centering
	\includegraphics[width=0.45\textwidth]{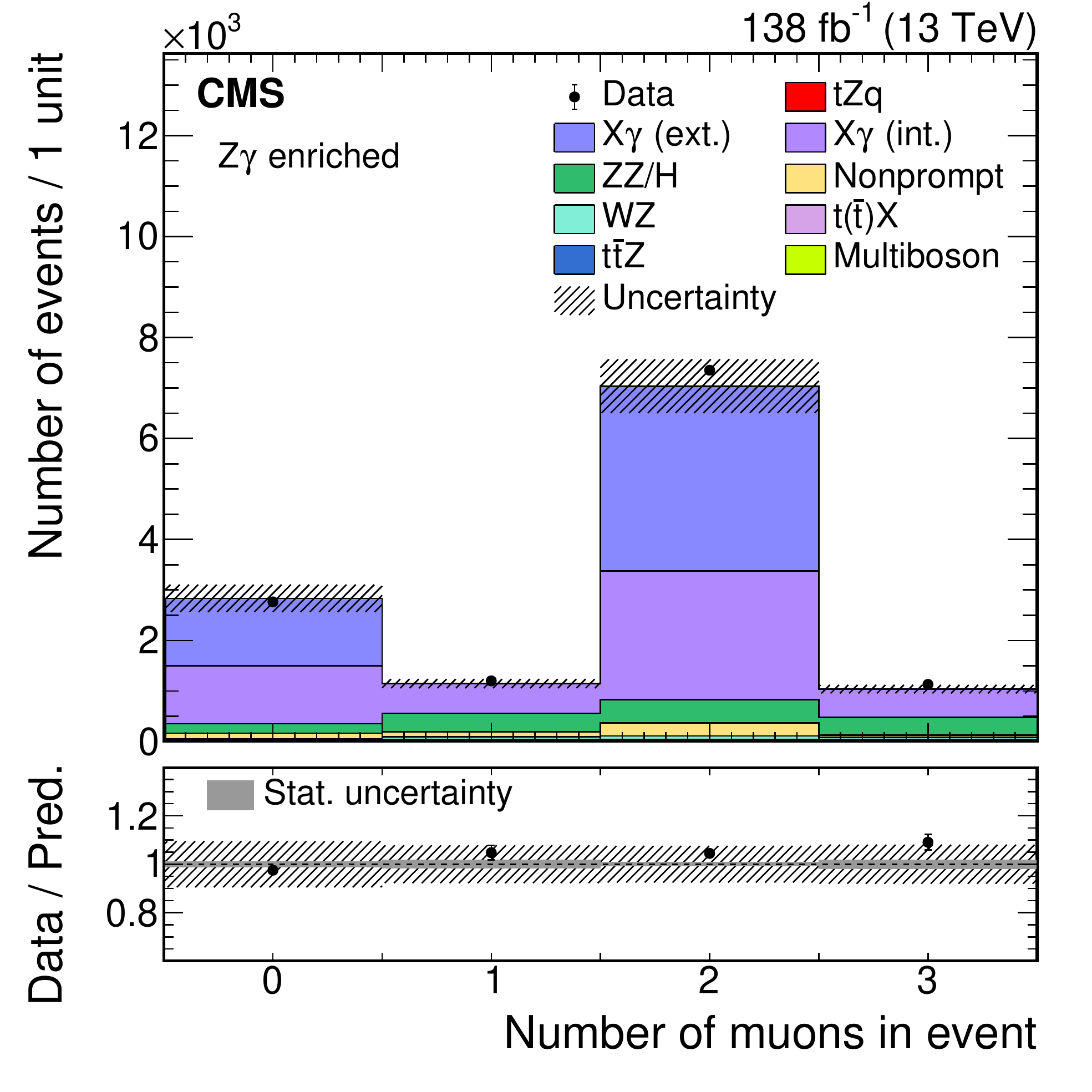}
	\includegraphics[width=0.45\textwidth]{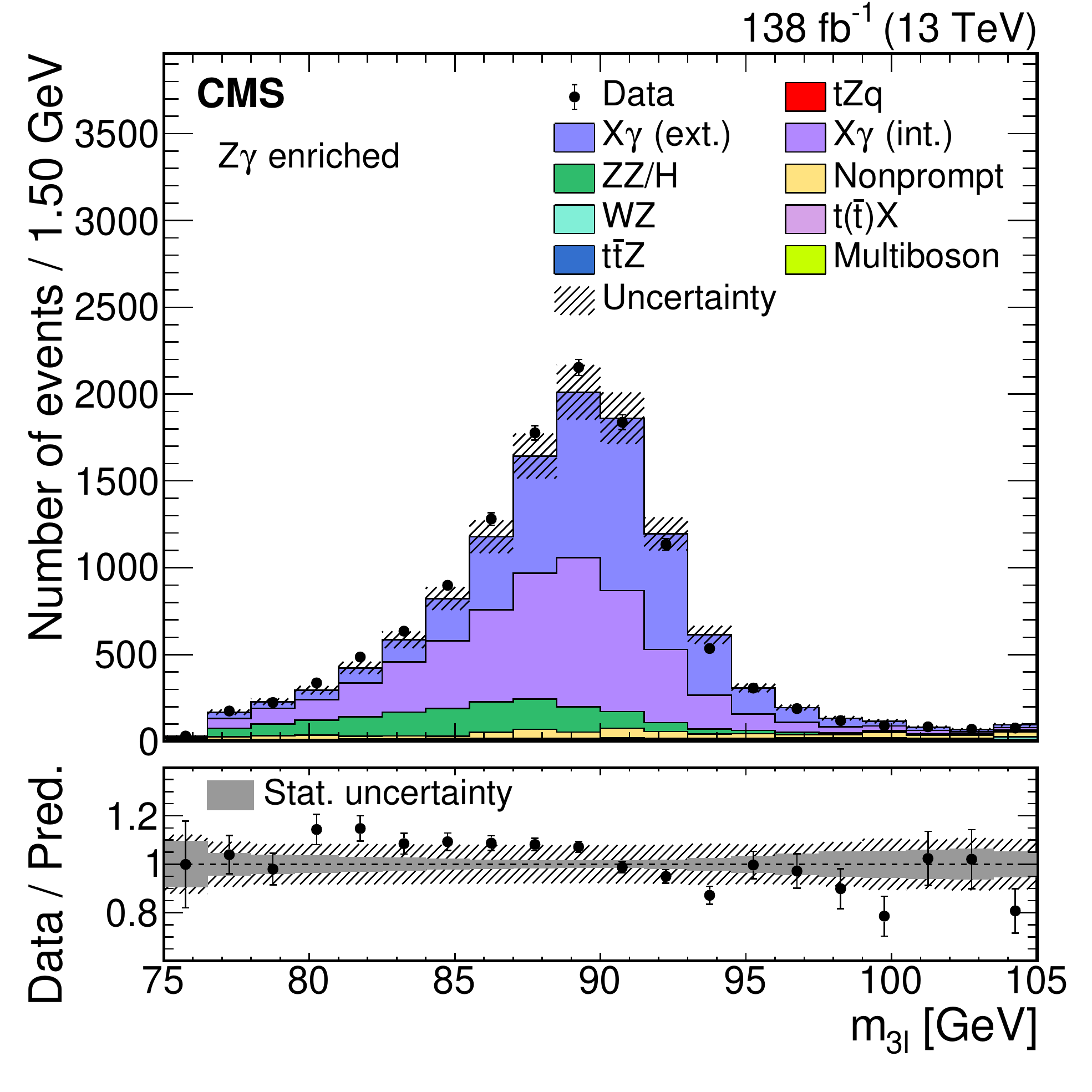}
	\caption{Distributions in the \Zg-enriched
	control region of the number of selected muons (left) and the invariant mass of the three-lepton system (right) for the data (points) and predictions (colored histograms). The contributions in the simulation from ``external'' photon conversions, where a real photon converts into a pair of (mostly) electrons from its interaction in the detector material, and so-called ``internal'' conversions, where a virtual photon decays into a pair of leptons, are shown separately. The lower panels show the ratio of the data to the sum of the predictions. The vertical lines on the data points represent the statistical uncertainty in the data; the shaded area corresponds to the total uncertainty in the prediction; the \grey area in the ratio indicates the uncertainty related to the limited statistical precision in the prediction.}
	\label{fig:zgcontrolregion}
\end{figure}

The \Zg process can represent a background to the signal region via conversion of the photon to an electron-positron pair in the detector material. In such a process, the converted photon may transfer a large part of its momentum to one of the two leptons. This leads to the production of one lepton of sufficient \pt to pass the selection criteria, with the other lepton failing those requirements.
The leptonic decay of the \PZ boson yields the additional two leptons needed to satisfy the three-lepton selection.
The selected events for the \Zg control region must contain three tight leptons whose
combined invariant mass must be compatible with the \PZ boson mass within 15\GeV, whereas any pair of leptons is required to fail this invariant mass constraint. With this selection, a pure \Zg region is obtained, allowing the validation of the modeling of photon conversions in the detector material.
Figure~\ref{fig:zgcontrolregion} displays the distributions for the number of selected muons in an event (left) and the three-lepton invariant mass (right) from the data and prediction for the \Zg control region.
The plots show the contributions from ``external'' photon conversions, where a real photon converts into a pair of (mostly) electrons from its interaction in the detector material, and so-called ``internal'' conversions, where a virtual photon decays into a pair of leptons. Also note that in all figures, \Zg is the major contribution to the background category labeled `X\PGg', with only minor contributions from other processes involving photons.

Other sources associated with the prompt-lepton backgrounds lead to much smaller contributions and are also estimated from simulation. They mainly include \tWZ, \ttH, and \ttW events, as well as the production of three massive electroweak bosons (\VVV).

The second major category of background includes those containing at least one nonprompt lepton. These arise from either \ttbar dileptonic events or DY production with an additional selected lepton that is either misidentified from a jet or a genuine lepton from hadron decay. This contribution is estimated from data using the so-called ``tight-to-loose ratio'' method~\cite{relIso_1}. The key feature of this method is the measurement of the probability for a nonprompt lepton satisfying the loose-quality definition to pass the tight-selection criteria. This probability, the ``misidentification rate'', is measured in a kinematic region enriched in QCD multijet events. In order to estimate the nonprompt-lepton background contribution to the signal region, the measured misidentification rate is used to compute a transfer factor, which is applied to events in a region with similar selection criteria as the signal region, except that at least one of the three loose leptons is not identified as a tight lepton.

In the first step of this procedure, the method is validated using simulated samples.
The misidentification rate is measured in simulated QCD multijet events and applied to simulated \ttbar and DY events.
A good description of nonprompt leptons in simulation is obtained in terms of all kinematic variables used in the multivariate classifier for signal extraction, as well as the classifier output score itself (shown in Appendix~\ref{sec:appendix}). This indicates that the method can be used to predict the nonprompt-lepton kinematic properties in \ttbar and DY events with misidentification rates measured in QCD multijet events.

Next, the misidentification rate is measured in a multijet-enriched data sample.
The event selection criteria in this measurement aim at selecting events containing nonprompt or misidentified leptons.
Events are required to have exactly one loose lepton, at least one jet
with $\pt > 30\GeV$ that does not overlap with the lepton within $\DR = 0.7$,
and $\ptmiss < 20\GeV$, in order to suppress contamination
from processes containing prompt leptons.

The misidentification rate is defined as the ratio of the number of events with a nonprompt lepton passing the tight selection to the number of events with a nonprompt lepton passing the loose selection.
Prompt-lepton contamination from electroweak processes is subtracted using simulation.
The misidentification rate is calculated separately for electrons and muons, and binned in \pt and \abseta to take into account the kinematic properties of nonprompt leptons passing the tight-lepton definition.

\begin{figure}[htb!]
	\centering
	\includegraphics[width=0.45\textwidth]{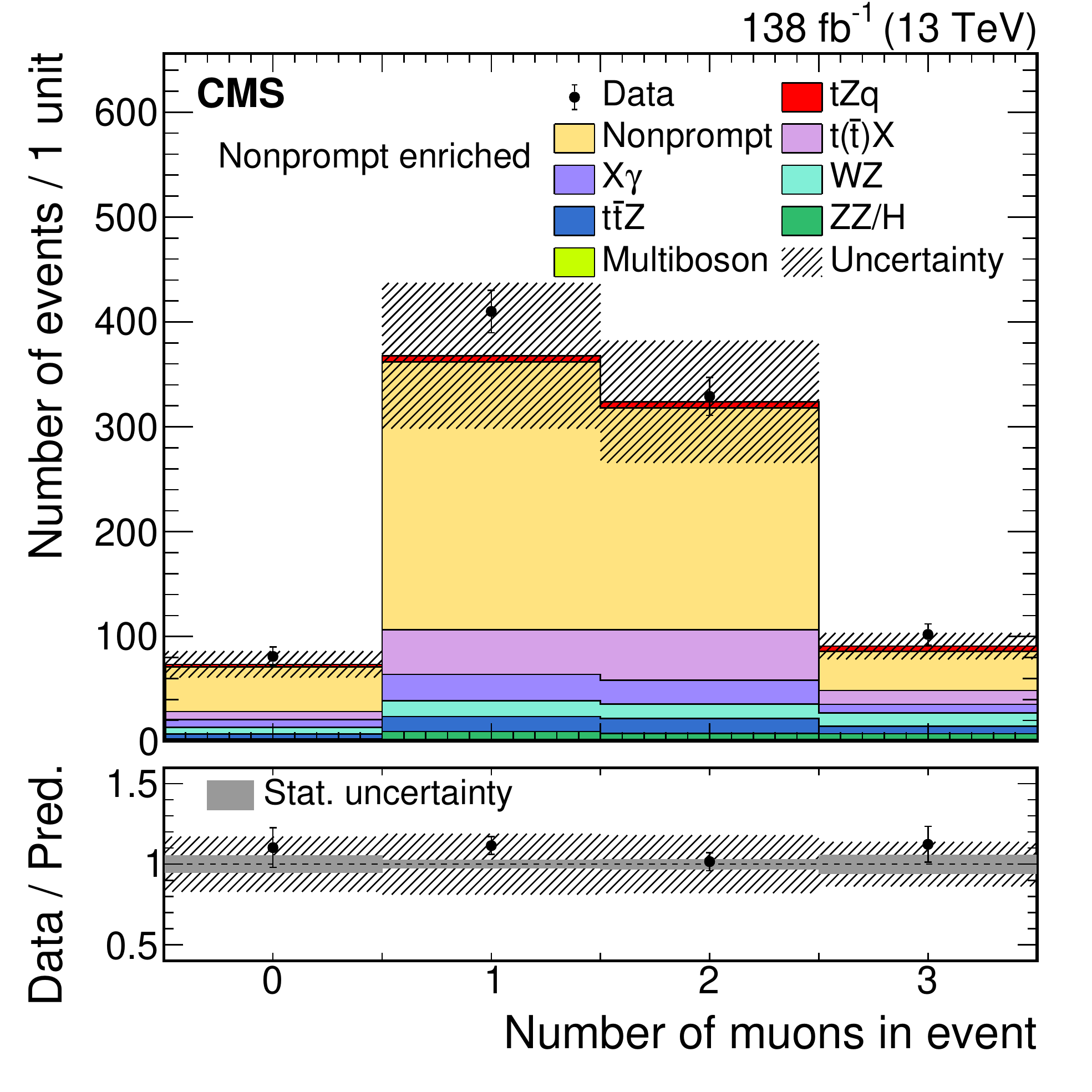}
	\includegraphics[width=0.45\textwidth]{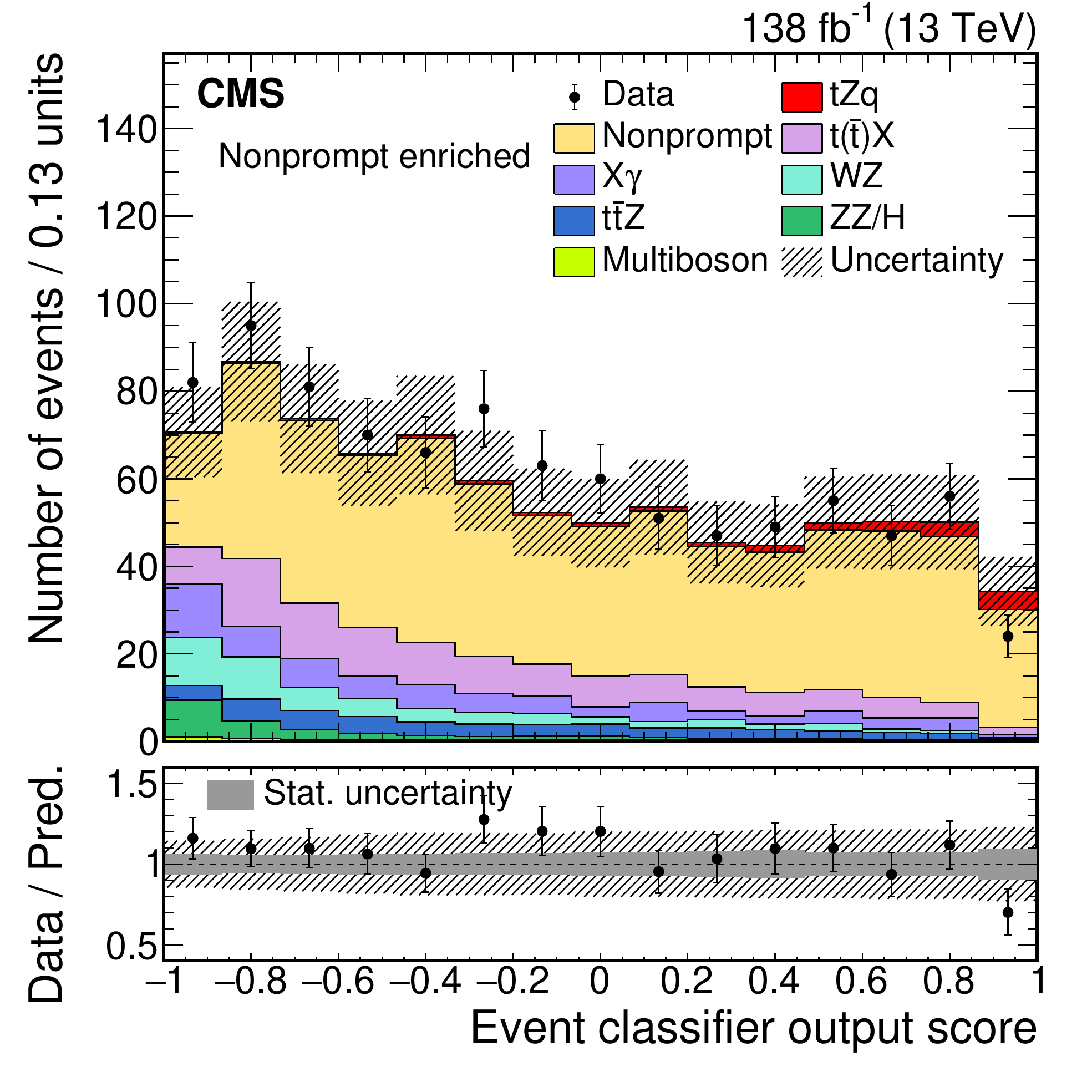}
	\caption{Distributions of the number of selected muons per event (left) and the output score of the multivariate classifier used for the signal extraction in the inclusive cross section measurement (right) in a control region enriched with nonprompt leptons for data (points) and predictions (colored histograms). The lower panels show the ratio of the data to the sum of the predictions. The vertical lines on the data points represent the statistical uncertainty in the data; the shaded area corresponds to the total uncertainty in the prediction; the \grey area in the ratio indicates the uncertainty related to the limited statistical precision in the prediction.}
	\label{fig:npcontrolregion}
\end{figure}

Finally, the method is validated in data using a dedicated nonprompt-lepton control region. This region is defined similarly to the signal region, with the
exception that either an OSSF pair is vetoed or the invariant mass of the OSSF pair is required to be incompatible with the \PZ boson mass. The nonprompt-lepton control region is defined for events with exactly one \PQb-tagged jet and two or three jets, as these represent the events in the signal region for which the nonprompt-lepton background is sizable. The result of this validation is shown in Fig.~\ref{fig:npcontrolregion}, where the contribution labeled ``nonprompt'' is obtained using the approach described above. There is good agreement between the data and the nonprompt-lepton background prediction obtained from the control region in data.

\section{Systematic uncertainties}
\label{sec:systematics}
The inclusive and differential measurements are affected by similar sources of experimental and theoretical uncertainties. The measurements follow the same approach to assessing the systematic uncertainties, modulo some small differences that are motivated by different fitting procedures.

The trigger efficiency is the probability of events that pass
the analysis selection criteria to satisfy any
of the trigger selection requirements.
This efficiency is measured in a data sample where events are required
to pass a set of reference triggers uncorrelated with any
of the trigger requirements used in this analysis.
An efficiency consistent with 100\% is measured in both simulation and data, and therefore no correction is applied.
Driven by the limited statistical precision of the trigger efficiency measurement, a systematic uncertainty of 2\% is assigned to cover potential differences between data and simulation.
This systematic uncertainty affects all processes equally, but because of its statistical origin, it is considered to be uncorrelated across the data-taking years.

The integrated luminosity measured in each data-taking period is used to normalize the predictions obtained from simulation with an associated systematic uncertainty of 1.2--2.5\%, which is partially correlated across the data-taking years~\cite{lumipaper,CMS-PAS-LUM-17-004,CMS-PAS-LUM-18-002}.

The pileup uncertainty alters the distribution of the number of \pp collisions per bunch crossing.
It is estimated by varying the total \pp inelastic cross section in all simulated samples by $\pm$4.6\%~\cite{Sirunyan_2018}, affecting both the shape of the distributions that are fit in the signal extraction and the normalization of the predictions.
This source of uncertainty is fully correlated across the data-taking years and processes.

In the 2016 and 2017 data sets, a too-early response of triggers
related to the ECAL led to the mistaken selection of data events from the previous bunch crossing.
In order to account for this effect, the simulated events are reweighted as a function of the \pt and $\eta$ of selected jets and photons with the corresponding uncertainties.
This source of uncertainty is fully correlated across all 2016 and 2017 simulated samples, but absent for the 2018 simulation.

Several uncertainties arise from the reconstruction and identification of various physics objects.
Data-to-simulation scale factors are derived to correct the efficiencies of prompt-lepton reconstruction and identification, as well as the \PQb tagging efficiency of reconstructed jets.
The scale factors are varied within the associated uncertainties, affecting both the shape and normalization of the derived predictions in simulation.
The uncertainties in the scale factors are split into a statistical part originating from the finite statistical precision of the methods used to obtain them, and a systematic part originating from the methodology itself. The first part is considered to be uncorrelated between the data-taking years, while the latter part is treated as fully correlated.

The four-momentum of each selected jet is varied to
account for the uncertainty in the jet energy resolution
and the jet energy scale~\cite{Khachatryan:2016kdb}.
These variations are consistently propagated to \ptvecmiss and the \PQb tagging efficiency scale factors, and are considered to be partially correlated across all data-taking years and processes. An additional uncertainty, related to the unclustered energy, is taken into account by varying all unclustered contributions to \ptvecmiss within their respective resolutions and propagating these changes to \ptvecmiss. This uncertainty is considered to be uncorrelated among the data-taking years.

The tight-to-loose ratio method that is used for the estimation of the nonprompt-lepton background
from control samples in data was verified and no bias
in the shapes of various variable distributions
was observed (as shown in Appendix~\ref{sec:appendix} and Fig.~\ref{fig:npcontrolregion}).
Therefore, no shape uncertainty is assigned to the nonprompt-lepton background estimate from data.
Based on the level of agreement in the nonprompt-lepton control regions in data, a conservative normalization uncertainty of 30\% is applied to this process, which covers the discrepancies (as shown in Fig.~\ref{fig:npcontrolregion}). It is considered correlated across the data-taking years.

Theoretical uncertainties include systematic effects associated with the renormalization and factorization scales at ME level, as well as with the PDFs used in the simulation.
The former uncertainties are propagated to the final measurement by varying both scales independently up and down by a factor of two, avoiding the case where one scale is varied up while the other is varied down.
The latter is propagated by reweighting the simulation using the corresponding variations in the NNPDF sets~\cite{NNPDF31,NNPDF30}.
Both types of systematic uncertainty are treated as fully correlated across the data-taking years, but while the PDF uncertainty is also correlated across all processes, the QCD scale uncertainties are considered uncorrelated between QCD- and electroweak-induced processes.
These sources affect both the production cross section and the acceptance of all simulated processes.
The systematic effect in the cross section is not taken into
account for the \ttZ, \WZ, \ZZ, and \Zg processes, and only the acceptance effects are considered.
Instead, a global normalization uncertainty is
assigned to each of these processes.
An uncertainty of 10\% is applied to the \WZ, \ZZ, and \Zg processes, which is larger than the typical uncertainty from dedicated
measurements~\cite{SirunyanWZ:2019,SirunyanZZ:2018}.
This covers any difference in the considered phase
space and is based on the study of control regions.
The \ttZ cross section is measured to a precision
of 8\%~\cite{SirunyanTTZ:2020}. However, there is some tension between the
theoretical prediction and the measurement.
In this analysis, a normalization uncertainty of 15\% is assigned to
the \ttZ process to cover this effect.
The theoretical uncertainties associated with the renormalization scales for the initial- and final-state radiations (ISR and FSR)
are estimated in a similar way, independent of the other scale uncertainties, by varying the corresponding scales up
and down by a factor of two. Both sources of uncertainty are fully correlated across the data-taking years, but while the FSR uncertainty is also correlated across all processes, the ISR uncertainty is treated as uncorrelated between QCD-induced (\ttZ) and electroweak-induced (\tZq) processes.
The uncertainties related to the choice of the color-reconnection model used in the parton shower and
to the underlying event tune are estimated for \tZq and
\ttZ with additional MC samples, produced with different color-reconnection models and varied underlying event tunes, respectively~\cite{TuneCP5,Argyropoulos_2014}. Both are considered correlated across the data-taking years and processes.

In the differential measurement, choices used in estimating the background uncertainties are slightly different with respect to the inclusive measurement to give more freedom to the differential fit.
The theoretical uncertainties in  PDFs, ISR, FSR, and the renormalization and factorization scales, are considered for the \tZq and \ttZ processes.
No a priori normalization uncertainty is applied to \WZ, \ZZ, and \Zg events
since the normalizations are kept freely floating in the fit.
In addition, a normalization uncertainty of 25\% is
applied to the triboson processes~\cite{Observation_VVV}.
For other rare processes involving top quarks, such as \tWZ and \ttbar in association with two additional bosons,
a normalization uncertainty of 50\% is assigned.

The impact of the dominant sources of systematic uncertainties in the inclusive \tZq cross section measurement is discussed in more detail in Section~\ref{sec:inclusive_results}

\section{Inclusive cross section measurement}
\label{sec:inclusive}

\subsection{Signal extraction}
\label{sec:inclusive_extraction}

The measurement of the inclusive \tZq cross section is performed by fitting the distribution of the BDT discriminant used to separate the \tZq signal from the various backgrounds. A binary BDT classifier is trained using the \TMVA~\cite{tmva2007} package. The discriminating input variables are similar to those used in an earlier CMS measurement of the \tZq process~\cite{Sirunyan:2018zgs}. The most powerful ones were discussed in Section~\ref{sec:selection} and shown in Fig.~\ref{fig:SRFeatures}.

The measurement uses a maximum likelihood fit of the predicted signal and background contributions to the data,
binned in the distribution of the BDT discriminant. The likelihood $\mathcal{L}$ to be maximized consists of the product of bin-by-bin Poisson probabilities $\mathcal{P}$ for observing a given number of data events in each bin~\cite{Conway2011}:
\begin{equation}
	\mathcal{L} = \prod_i \mathcal{P}(\yobsi | \yexpi) \prod_k p(\hat{\theta}_k|\theta_k),
	\label{eq:likelihood}
\end{equation}
where \yobsi and \yexpi are the observed and expected numbers of events in the $i$th bin, respectively. The number of expected events in bin $i$ depends on the signal and background predictions as:
\begin{equation}
\yexpi\left(\sigma_{\tZq},\ \vec{w},\ \vec{\theta}\right)\ =\ s_{i}\left(\sigma_{\tZq},\ \vec{\theta}\right)\ +\ \sum_j b_{i,j}\left(w_j,\ \vec{\theta}\right),
\label{eq:expected_events}
\end{equation}
where $s_{i}$ is the expected number of \tZq events in the $i$th bin, which depends on the targeted cross section \stZq and nuisance parameters $\vec{\theta}$ to describe the uncertainties in the prediction.
The variable $b_{i,j}$ denotes the number of expected events from the $j$th background process in the $i$th bin, which depends on its normalization $w_j$ and nuisance parameters $\vec{\theta}$.
The factors $p(\hat{\theta}_k|\theta_k)$ in Eq.~(\ref{eq:likelihood}) represent the probability of obtaining a best fit value $\hat{\theta}_k$ for the $k$th nuisance parameter, given its a priori value $\theta_k$. This probability is log-normally distributed for normalization uncertainties and has a Gaussian density for shape uncertainties.

For the inclusive measurement, the signal region is subdivided into three subregions based on the number of jets and \PQb-tagged jets: exactly 1 \PQb-tagged jet and 2 or 3 jets (dominated by the \WZ and nonprompt backgrounds), 1 \PQb-tagged jet and $\geq$4 jets (dominated by the \ttZ and \WZ backgrounds), and $\geq$2 \PQb-tagged jets (dominated by the \ttZ background).

\begin{figure}[p!]
	\centering
	\includegraphics[width=0.4\textwidth]{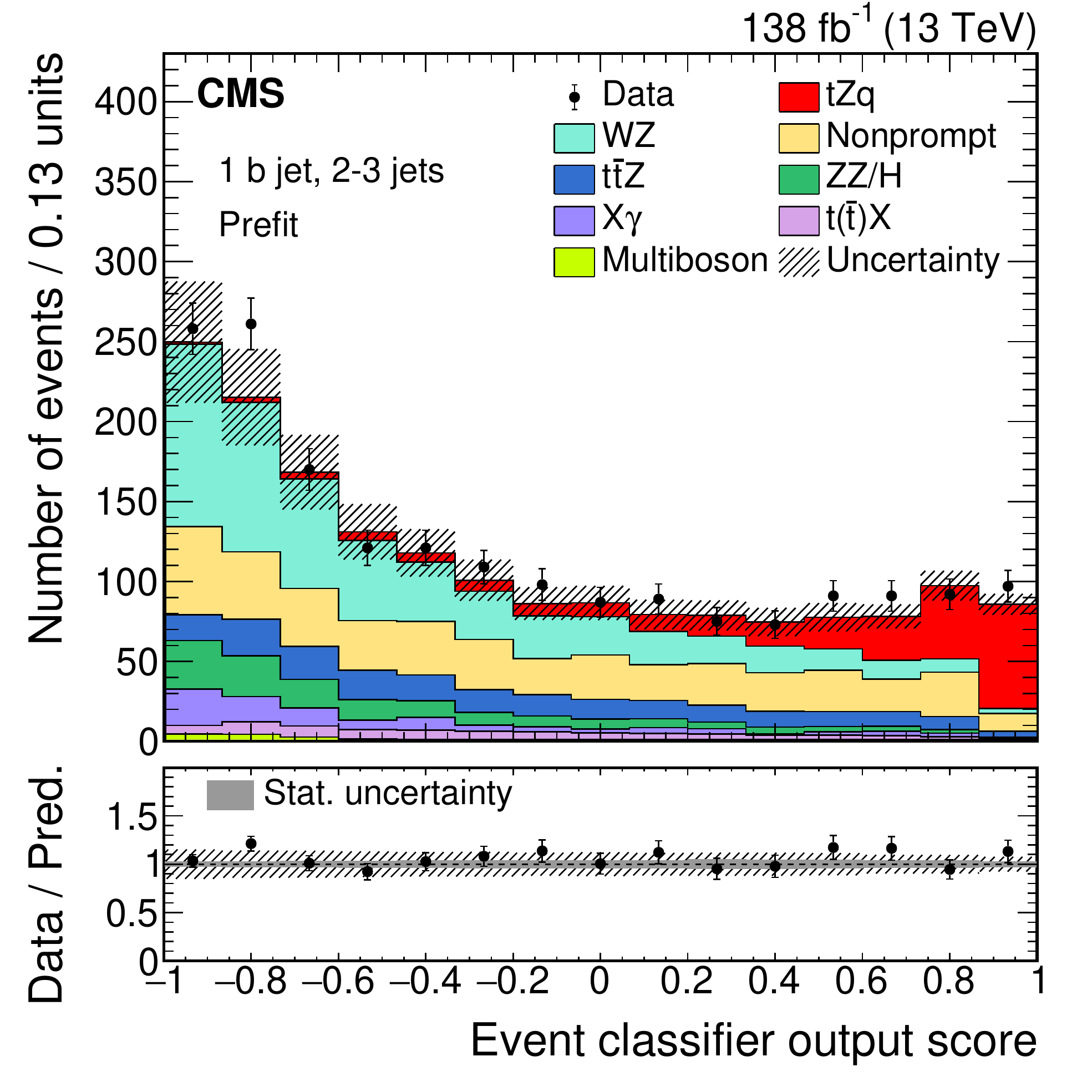}
	\includegraphics[width=0.4\textwidth]{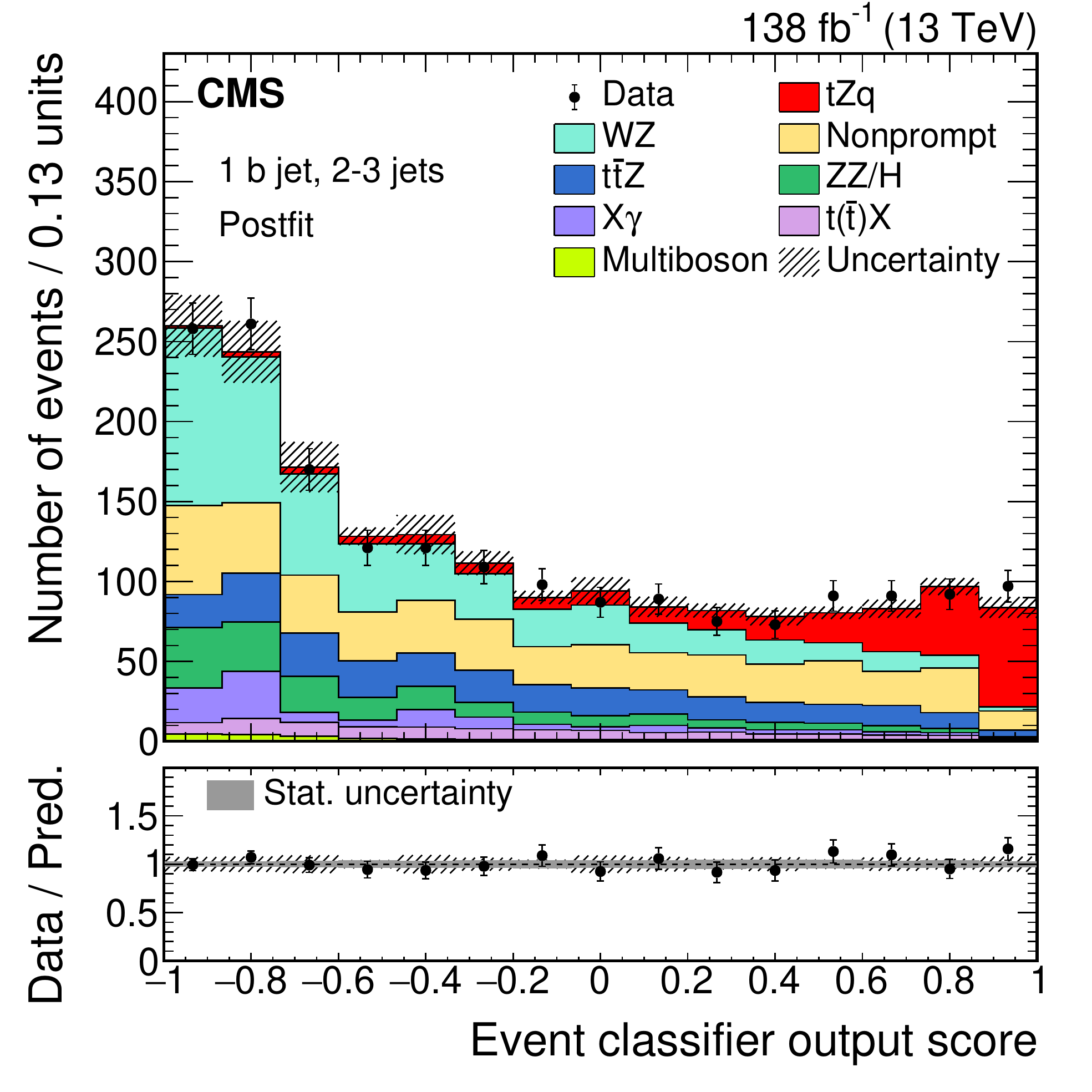} \\
	\includegraphics[width=0.4\textwidth]{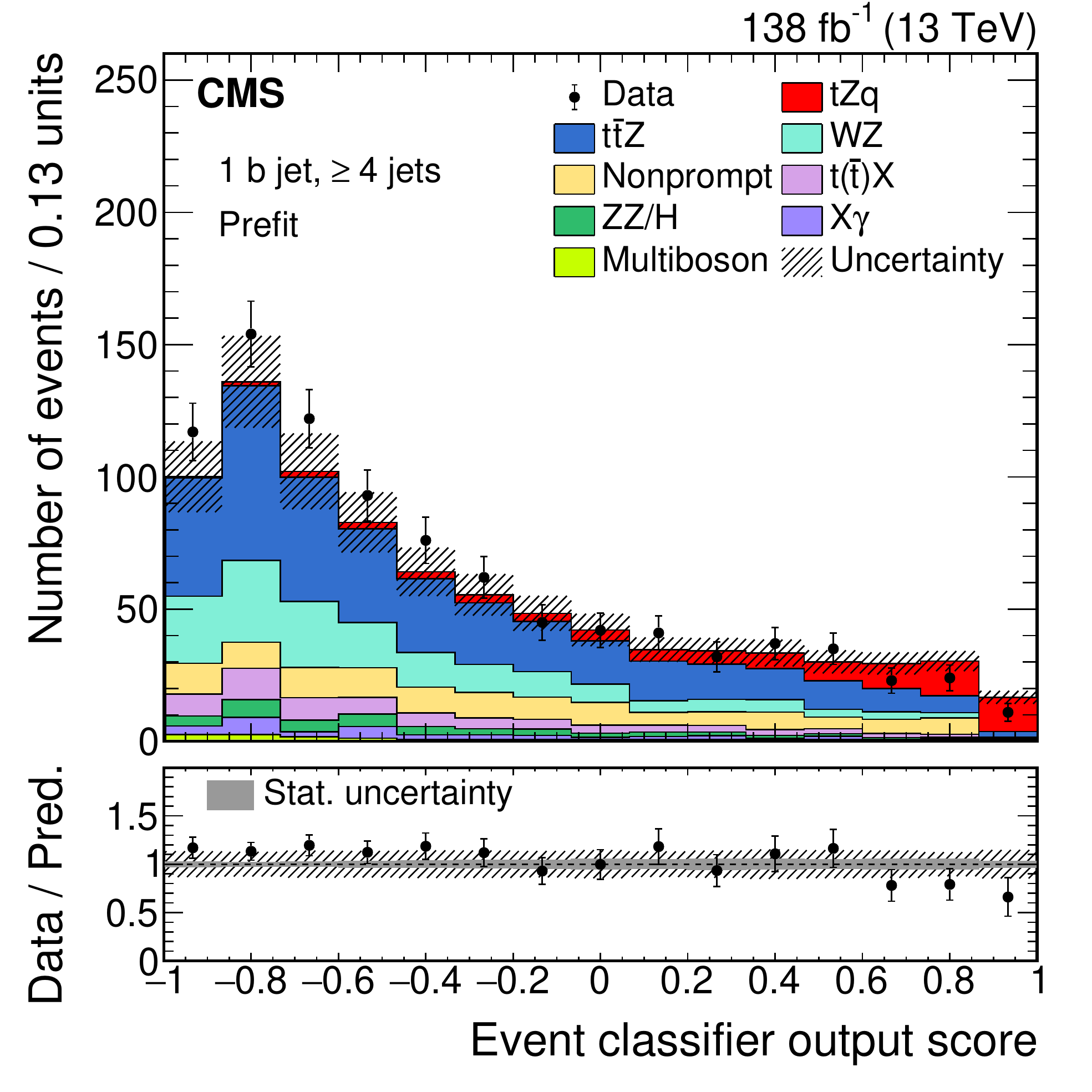}
	\includegraphics[width=0.4\textwidth]{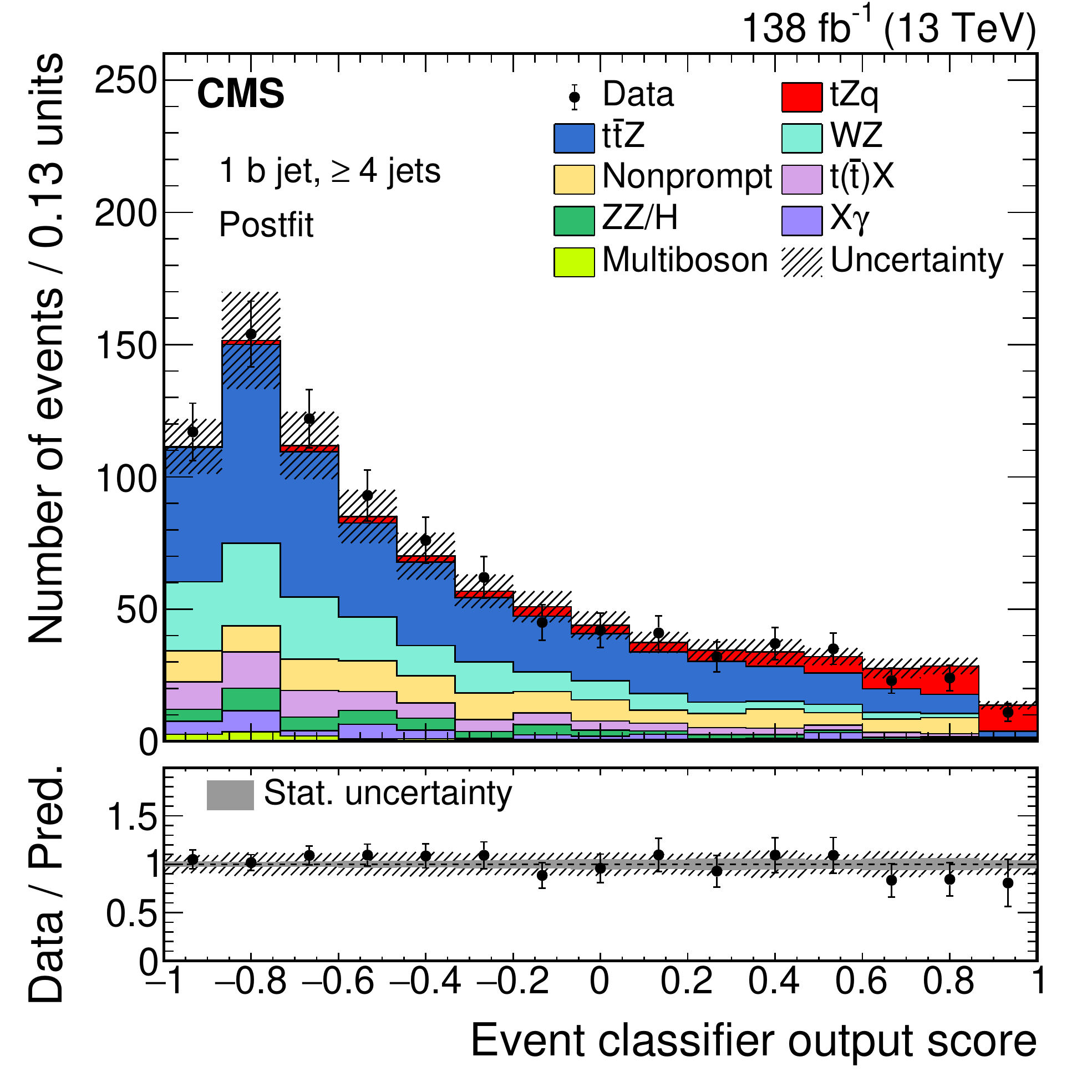} \\
	\includegraphics[width=0.4\textwidth]{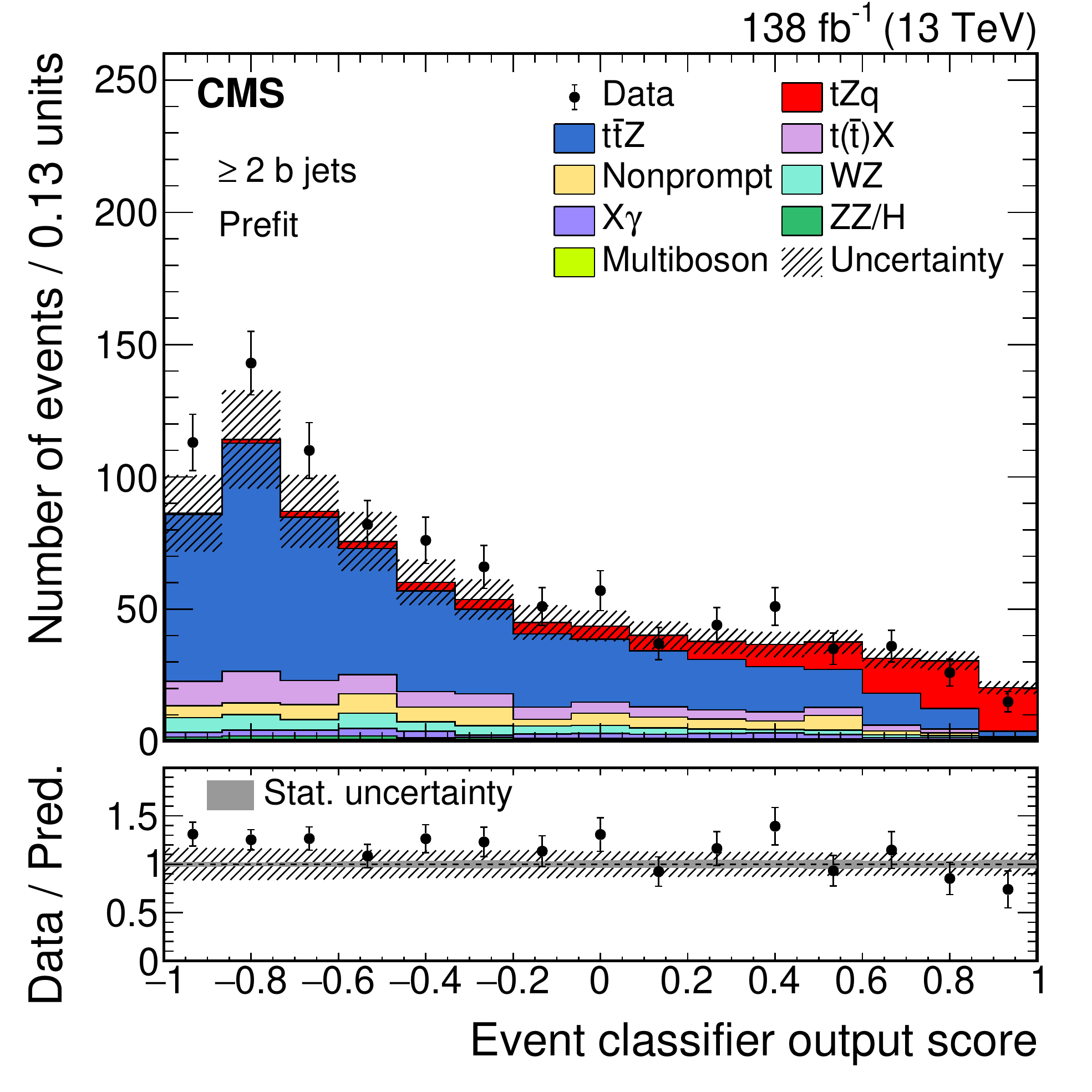}
	\includegraphics[width=0.4\textwidth]{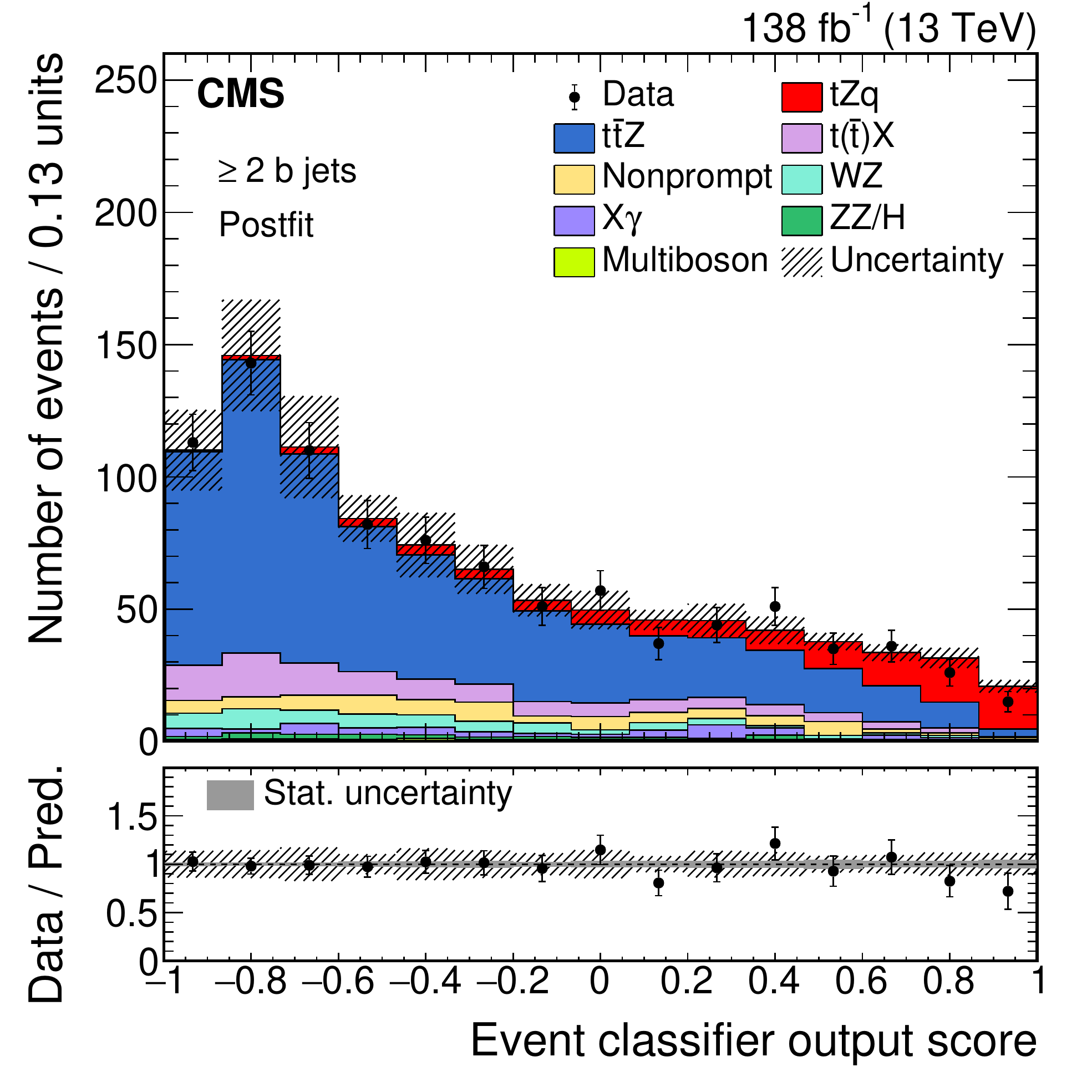}
	\caption{Distributions of the event BDT discriminant in the signal region for data (points) and predictions (colored histograms).
    The results are shown for prefit (left) and postfit (right) distributions in mutually exclusive signal subregions:
    exactly one \PQb-tagged jet, 2--3 jets (upper);
    exactly one \PQb-tagged jet, $\geq$4 jets (middle);
    and $\geq$2 \PQb-tagged jets (lower).
    The lower panels show the ratio of the data to the sum of the predictions. The vertical lines on the data points represent the statistical uncertainty in the data; the shaded area corresponds to the total uncertainty in the prediction; the \grey area in the ratio indicates the uncertainty related to the limited statistical precision in the prediction.}
	\label{fig:inclusive_signalregion_bdt}
\end{figure}

The fit is performed simultaneously for all considered
data-taking years and event categories. The corresponding BDT discriminant distributions are shown in Fig.~\ref{fig:inclusive_signalregion_bdt} for both the prefit (left) and postfit (right) normalizations.
The control regions discussed in Section~\ref{sec:backgrounds} are included in the fit, which allows a better constraint on the relevant systematic uncertainties in the background processes, especially their normalizations. The distributions used in the fit are the transverse \PW boson mass for the \WZ control region, the number of jets for the \ZZ and \Zg control region, the number of \PQb-tagged jets for the \ttZ control region, and finally the total event yield for the nonprompt control region.

All sources of systematic uncertainties discussed in Section~\ref{sec:systematics} are treated as nuisance parameters in the fit, with a consistent treatment of all correlations between various uncertainties.

\subsection{Results}
\label{sec:inclusive_results}

The predicted cross section for
the \tZq process, where the \PZ boson decays to a
pair of electrons, muons, or $\tau$ leptons, is
$ \stZqSM = \THxsec $~\cite{Sirunyan:2017nbr}.
The calculation is performed at NLO in the 5FS and also includes
nonresonant lepton-pair production with $\mll > 30\GeV$.

The signal strength is defined as the ratio of the observed to the predicted \tZq cross sections and is measured in the signal region defined in Sections~\ref{sec:selection} and~\ref{sec:inclusive_extraction}. The result is:
\begin{equation*}
\mu = \frac{\stZq}{\stZqSM} = \EXss,
\end{equation*}

Using the predicted cross section mentioned above, this signal strength corresponds to the measured cross section of
\begin{equation*}
\stZq = \EXxsec.
\end{equation*}

Combining the statistical and systematic uncertainties in quadrature, the measured \tZq cross section has a precision of 11\%, which is an improvement over the previous measurements of this process~\cite{Sirunyan:2018zgs,Aad:2020wog}. This is partly due to the smaller integrated luminosity of 77\fbinv in earlier measurements.
Another improvement comes from a broader definition of the signal by including events with two or more b-tagged jets, and events with at least four selected jets.
The latter subregion, whose distributions are shown in the middle plots of Fig.~\ref{fig:inclusive_signalregion_bdt}, provides an important contribution to the improved sensitivity of this measurement.
Additional gain comes from the improved performance of the \leptonMVA and the looser selection criteria applied to this discriminant. The working point used in this analysis has a signal efficiency and background rejection of about 95\% and 98\%, respectively (as discussed in Section~\ref{sec:samples}), whereas for the typical tighter working point, these numbers are 90\% and 99\%, respectively.
The loosening of the \leptonMVA selection criteria allowed better constraints on the relevant systematic uncertainties in the nonprompt-lepton background prediction using the dedicated control regions.

\begin{figure}[htb!]
	\centering
	\includegraphics[width=0.8\textwidth]{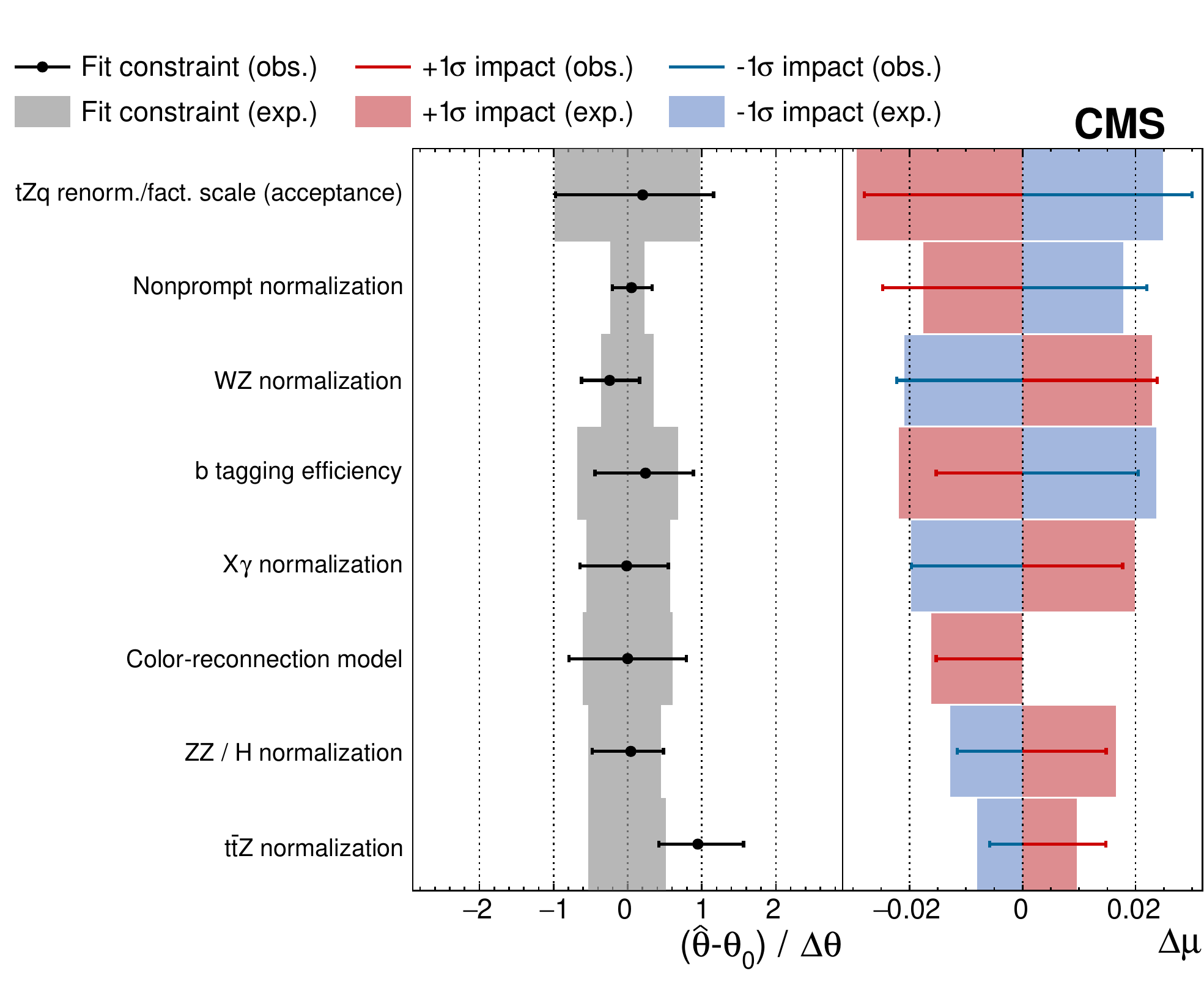}
	\caption{Summary of the dominant systematic uncertainties affecting the inclusive \tZq cross section measurement. The left column lists the sources of systematic uncertainty, treated as nuisance parameters in the fit, in order of importance. In the middle column, the black points with the horizontal bars show for each source the difference between the observed best fit value ($\hat{\theta}$)  and the nominal value ($\theta_0$), divided by the expected standard deviation ($\Delta \theta$). The right column plots the change in the \tZq signal strength $\mu$ if a nuisance parameter is varied one standard deviation up (red), or down (blue). The \grey, red, and blue bands display the same quantity as their corresponding markers, but using a simulated data set where all nuisance parameters are set to their expected values.}
	\label{fig:inclusive_impacts}
\end{figure}

The dominant systematic uncertainties affecting the
inclusive cross section measurement are shown in Fig.~\ref{fig:inclusive_impacts}.
These uncertainties include the effects on acceptance from varying the renormalization
and factorization scales associated with the \tZq and \ttZ processes, the normalization uncertainties in all the considered processes discussed in Section~\ref{sec:systematics}, and several other sources related
to the \PQb tagging efficiency correction, \colour reconnection, and parton showering.
The impact of the \colour-reconnection model choice on the \tZq signal strength is one-sided since it results from using an alternative model as opposed to a model parameter varied up and down.
All sources of systematic uncertainty not shown in Fig.~\ref{fig:inclusive_impacts} have smaller impacts than the dominant ones discussed here.
The best fit values of the nuisance parameters are all within one standard deviation of their expected values, and the measured impacts on the signal strength are generally in agreement with the corresponding expected values.
The measured normalization of \ttZ events is shifted by about one standard deviation with respect to the theoretical prediction, and indicates an underprediction of this background. This is expected and is consistent with the previously published results~\cite{SirunyanTTZ:2020}. 
The increased contribution from \ttZ events after the fit results in a good agreement of prediction to data in both the \ttZ-dominated signal subregion (lower plots in Fig.~\ref{fig:inclusive_signalregion_bdt}) and the \ttZ control region with four leptons.

\begin{figure}[htb!]
	\centering
	\includegraphics[width=0.45\textwidth]{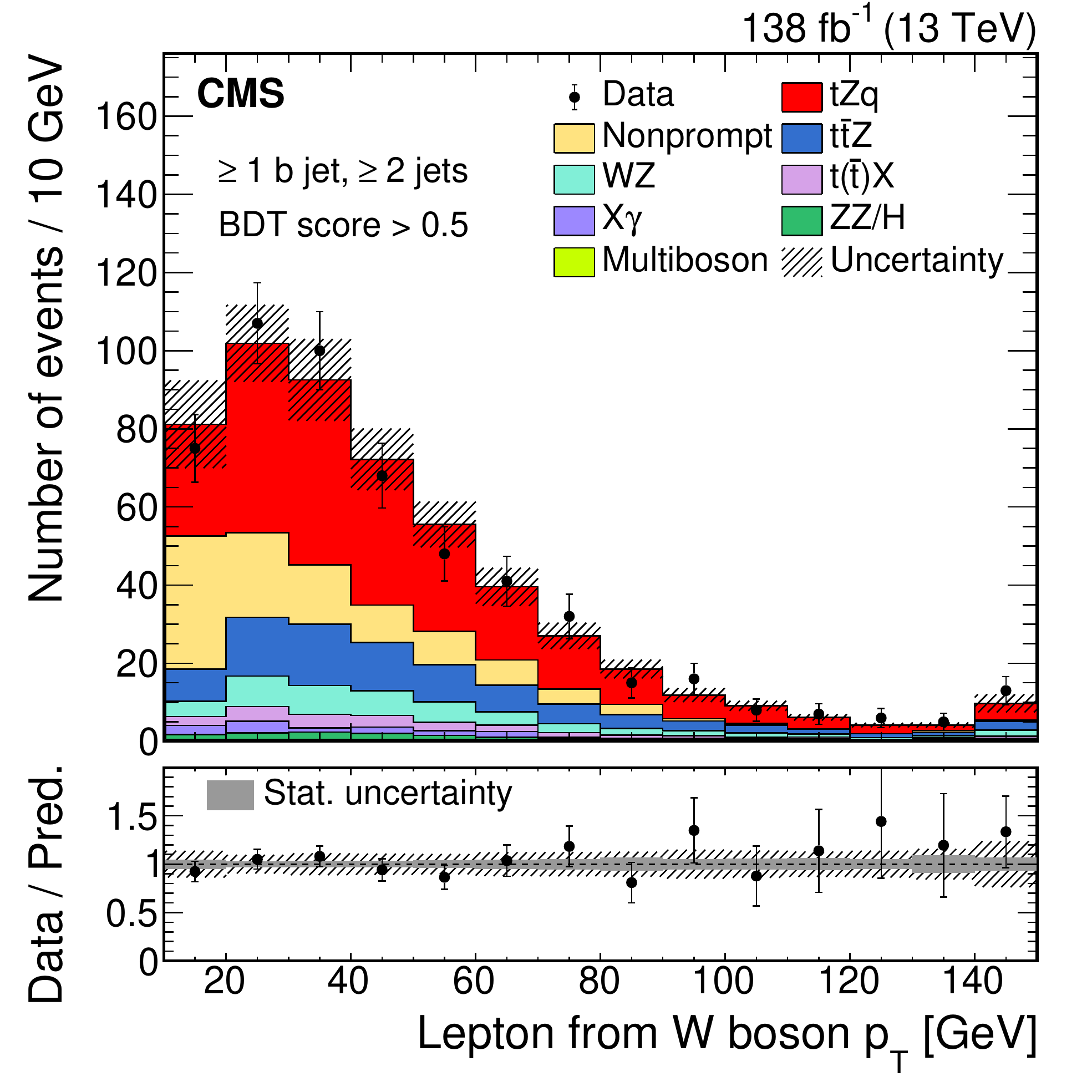}
	\includegraphics[width=0.45\textwidth]{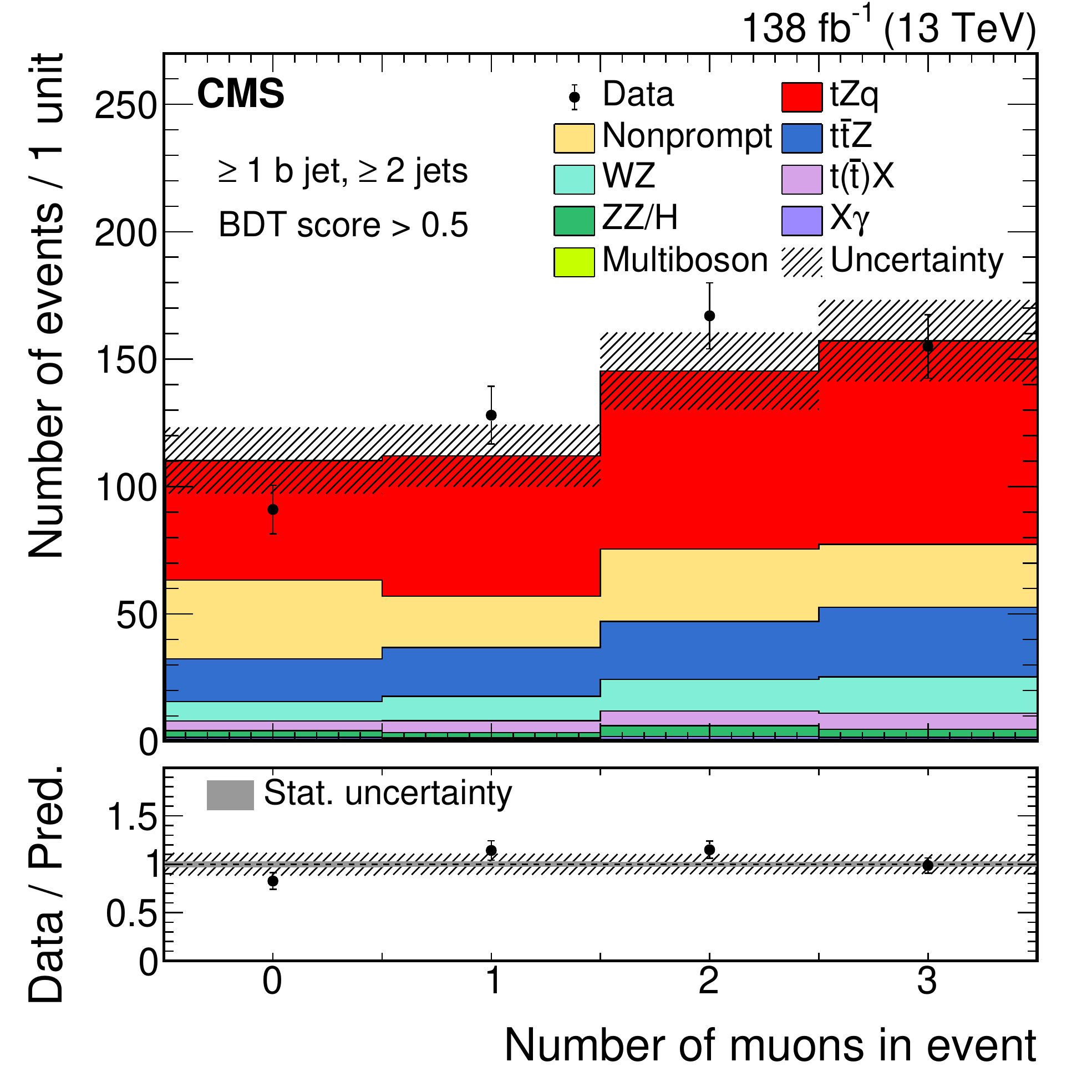} \\
	\includegraphics[width=0.45\textwidth]{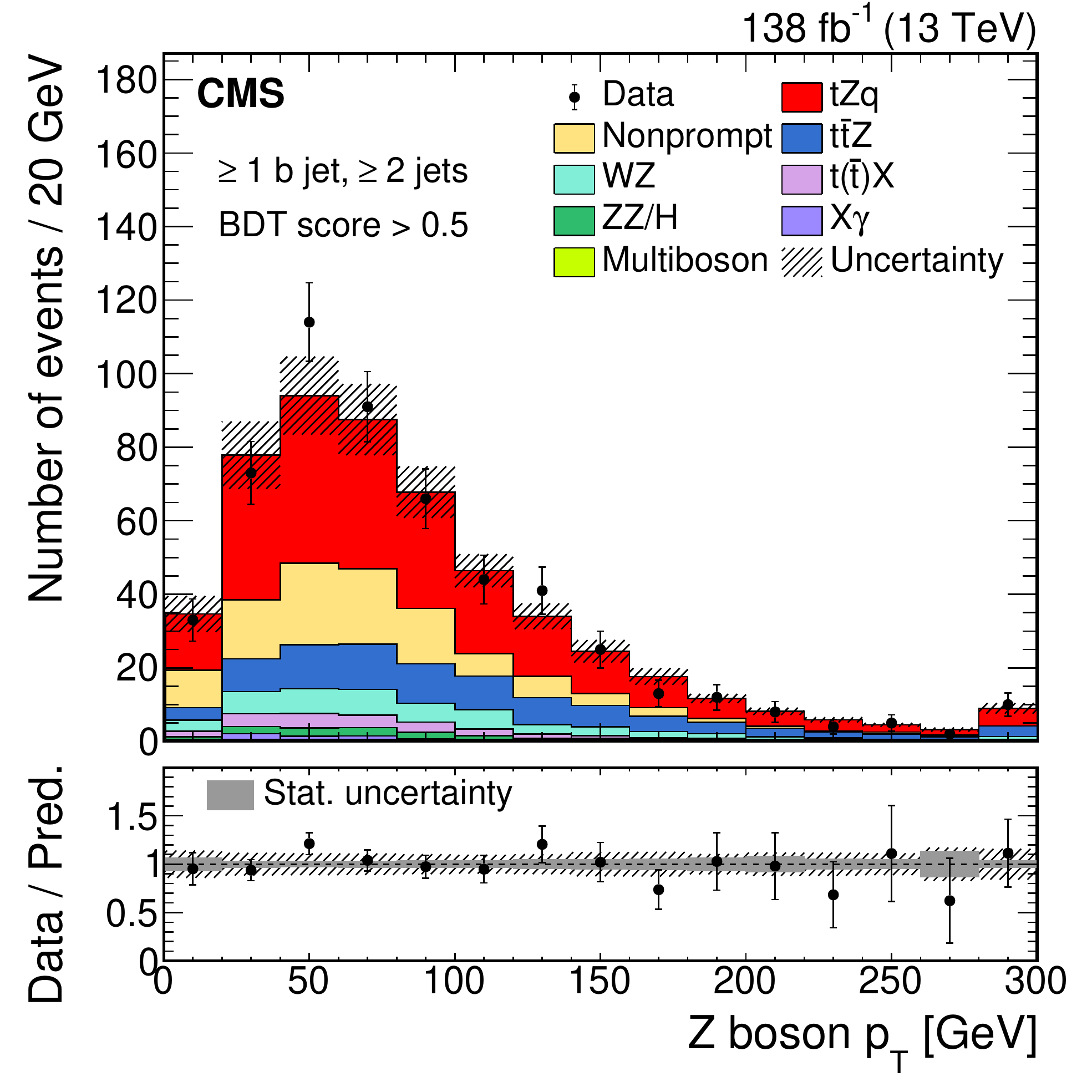}
	\includegraphics[width=0.45\textwidth]{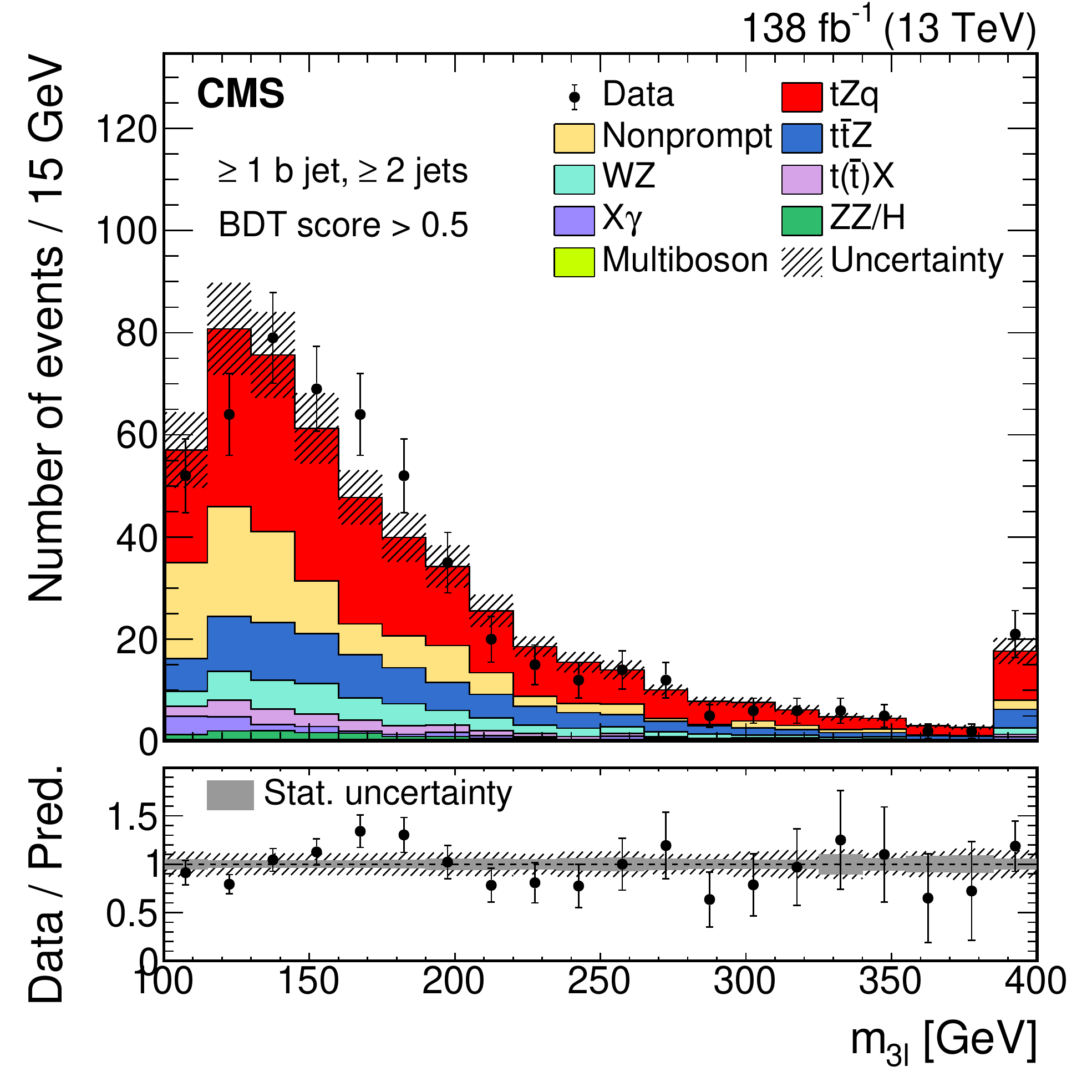} \\
	\caption{Prefit distributions at the detector level of some of the important variables used in the \tZq analysis from a \tZq-enriched region for the data (points) and predictions (colored histograms).
    The selection criteria discussed in Section~\ref{sec:selection} have been used, along with the requirement that the event BDT discriminant be greater than 0.5. The variables shown are as follows: transverse momentum of the lepton associated with the decay of the top quark (upper left), number of muons in the event (upper right), reconstructed transverse momentum of the \PZ boson (lower left) and transverse mass of the \PW boson (lower right).
    The lower panels show the ratio of the data to the sum of the predictions. The vertical lines on the data points represent the statistical uncertainty in the data; the shaded area corresponds to the total uncertainty in the prediction; the \grey area in the ratio indicates the uncertainty related to the limited statistical precision in the prediction.}
	\label{fig:inclusive_signalregion_bdtcut}
\end{figure}

Figure~\ref{fig:inclusive_signalregion_bdtcut} displays some noteworthy kinematic prefit distributions in the \tZq-enriched region, where in addition to the signal selection detailed in Section~\ref{sec:selection}, the BDT score is required to be greater than 0.5. The number of \tZq signal and background events passing this selection are estimated to be 252 and 264, respectively, implying that the contribution of the signal in this \tZq-enriched region is about 49\%. Furthermore, these plots show a good sensitivity to the kinematic properties of the \tZq signal and a good agreement between the data and simulation, with the exception of the \mlll variable, as discussed further in Section~\ref{sec:differential}.
The imperfect simulation of reconstruction inefficiencies and detector acceptance effects (combined and \labelled as ``detector level'' effects) could result in discrepancies between the shape and normalization of the measured distributions and the simulated ones.
The good agreement between the measured and simulated distributions in the figure shows that these possible effects are relatively small in this phase space.
Furthermore, the number of events in the data associated with the \tZq process means that a differential cross section measurement is possible, once the detector-level effects are corrected for,
as described in Section~\ref{sec:differential}.

In addition to the inclusive \tZq cross section, the production cross sections of a top quark (\tplusZq), top antiquark (\tminusZq), and their ratio $R$ are determined.
To this end, the signal regions are split further based on the charge of the lepton associated with the top quark (or antiquark) decay, and the fit procedure is modified to simultaneously extract the signal strengths for the top quark and antiquark processes. For the measurement of the ratio $R$, the fit procedure is modified further to directly obtain the best fit result for the ratio.
The measured signal strengths for the separate top quark and antiquark cross sections are:
\begin{equation*}
\begin{aligned}
\mu_{\tplusZq} &= \EXssTZ, \\
\mu_{\tminusZq} &= \EXssTbarZ. \\
\end{aligned}
\end{equation*}

Using the \MGvATNLO 5FS predictions for the cross sections, these signal strengths translate into:
\begin{equation*}
\begin{aligned}
\sigmatop &= \EXxsecTZ, \\
\sigmaantitop &= \EXxsecTbarZ, \\
\Rtz &= \Rvalue. \\
\end{aligned}
\end{equation*}

The relative systematic uncertainty is reduced in the ratio measurement due to partial correlations. Although currently dominated by the statistical uncertainties, these results show promise for a future precise determination (at the high-luminosity LHC) of the top quark to antiquark production cross section ratio in the rare process \tZq, similarly to what has already been obtained for $t$-channel single top quark production~\cite{Sirunyan_singletop}.

\section{Measurements of the differential cross sections and the spin asymmetry}
\label{sec:differential}
Differential \tZq cross section measurements are performed as functions of several observables at the parton and particle levels, as defined in Section~\ref{sec:genlevel}.
The selected observables are potentially sensitive to beyond-SM effects and can provide information on the modeling of the \tZq process.
Observables based only on lepton kinematic properties are the transverse momenta of the \PZ boson, \zpt, and the lepton from the top quark, \topleppt, the invariant mass of the three leptons, \mlll, and the difference in azimuthal angle between the two leptons from the \PZ boson decay, \delphill.
Other selected variables rely on the top quark reconstruction and include the cosine of the top quark polarization angle, \costheta, and the invariant mass of the \PZSys system, \mtz.
The last two observables are the \jprimpt and \jprimeta at the particle level of the recoiling jet.

A likelihood-based unfolding procedure is performed to separately measure the cross section
$\sigma_{\tZq,\ k}$, in each kinematic region defined by each generator-level bin $k \in \{1,2,... \}$,
where the generator level here corresponds to truth information at either the parton or particle levels,
defined in the next section.
The unfolding procedure accounts for the finite resolution and limited acceptance of the detector and, at the parton level, for hadronization effects.
Compared to the inclusive measurement, not one but multiple signal parameters associated with the different generator-level bins
are extracted in a multidimensional maximum likelihood fit.
The first term in Eq.~(\ref{eq:expected_events}) is therefore replaced by a sum
over signal contributions from the generator-level bins,
\begin{equation}
    s_{i}\left(\stZq,\ \vec{\theta}\right) \to \sum_k s_{i,\ k}\left(\sigma_{\tZq,\ k},\ \vec{\theta}\right).
\end{equation}

To extract the signal in multiple kinematic regions requires not only the separation between the \tZq and background processes,
but also the mutual separation of the \tZq contributions coming from different generator-level bins corresponding to the different kinematic regions.
For this reason, an alternative categorization of events in the signal region
with respect to the inclusive measurement and a more
elaborate classifier have been developed.

\subsection{Parton- and particle-level definitions}
\label{sec:genlevel}

The binning of the various observables at the generator level is optimized
based on a trade-off between the expected number of \tZq events in each
bin at the detector level, the bin width, and
the stability and purity of the response matrix
that relates the generator-level distributions to the detector-level distributions in the simulated \tZq signal.
The stability is based on all reconstructed events and defined as the fraction of events from a generator-level bin that are observed in the corresponding detector-level bin.
The purity is based on all reconstructed events and defined as the fraction of events from a detector-level bin that belong to the corresponding generator-level bin.

\begin{figure}[htb!]
  \centering
  \includegraphics[width=0.45\textwidth]{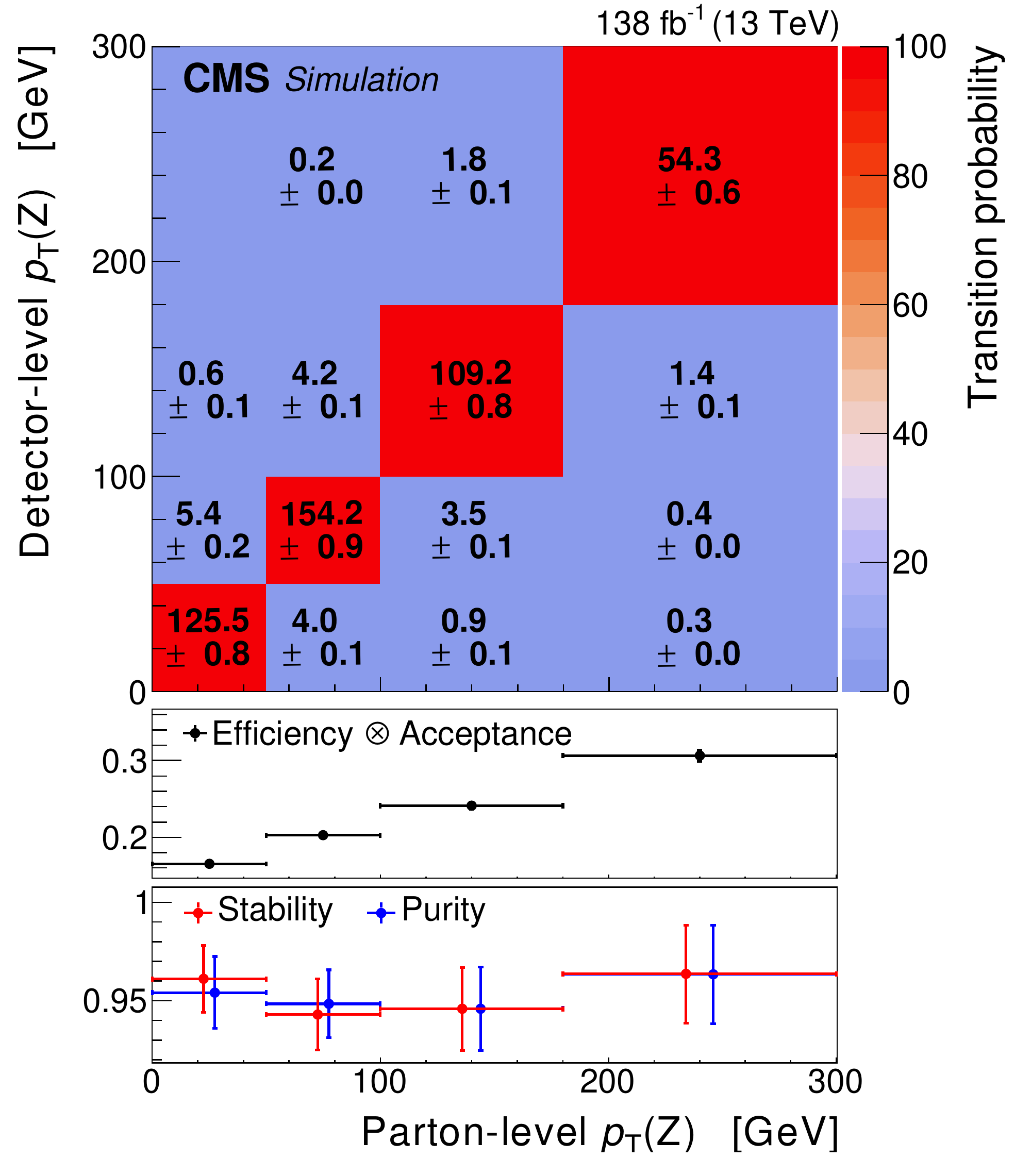}
  \includegraphics[width=0.45\textwidth]{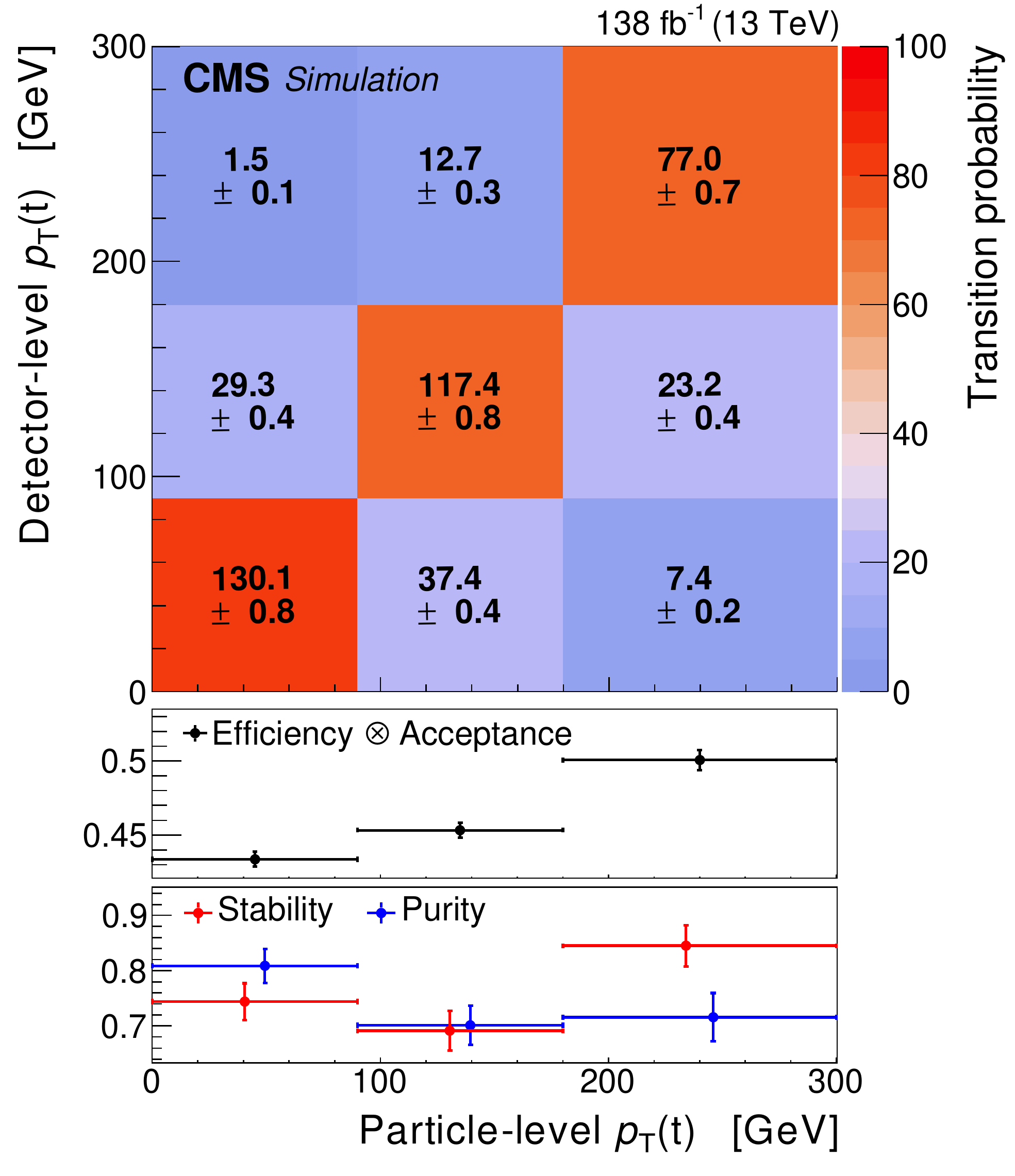}
\caption{Response matrices for \zpt at the parton level (left) and \toppt at the particle level (right) for \tZq events in the full and visible phase space, respectively.
The expected number of reconstructed events is given for each bin.
The \colour indicates the transition probability for an event in a generator-level bin to have a reconstructed value corresponding to a given detector-level bin.
The efficiency times acceptance values of reconstructing events are plotted in the middle panels.
The lower panels show the stability and purity values as defined in the text. The vertical bars on the points give the statistical uncertainty and the horizontal bars show the bin width.}
\label{fig:response}
\end{figure}

Observables associated only with leptons generally have good
measurement resolution, and a total of four bins is chosen.
This is also motivated by the number of events in the data set and the purity and stability values above 95\%.
For observables involving jets, a total of three
bins is chosen to account for the poorer resolution compared to the lepton observables, and to lessen the effects from statistical fluctuations.
The corresponding purity and stability values are above 55\%.
With this choice of binning, the application of a regularization procedure was found to be unnecessary.
Examples of two response matrices, for the \zpt at the parton level and \toppt at the particle level, are shown in Fig.~\ref{fig:response}.

Generator-level definitions at the parton and particle levels are used.
At the parton level, the measurement is performed in the full phase
space of events with three prompt leptons.
Parton-level objects are defined based on event generator particles
after ISR and FSR and before hadronization.
The generated on-shell top quark is selected and the lepton from its
decay is identified. The leptons that are not associated with the decay
products of the top quark are assigned to the OSSF lepton pair.
The quark that recoils against the virtual \PW boson is identified as
the quark with \flavour \PQu, \PQd, \PQs, or \PQc.
In the case of an ambiguity, the one with the highest \pt is chosen.

The particle-level definition aims at minimizing
the dependency on the choice of the generator
and reducing the uncertainty associated with the
extrapolation to the detector level.
A collection of so-called ``dressed'' leptons is defined
through a clustering process involving prompt electrons or muons
and photons that do not arise from hadronic decays using the anti-\kt algorithm with $R = 0.1$~\cite{Cacciari:2008gp, Cacciari:2011ma}.
Jets are clustered from all stable (lifetime $>$30\unit{ps})
particles, excluding prompt leptons but including neutrinos from hadronic decays, with the anti-\kt algorithm and $R = 0.4$.
Using the ghost-clustering method~\cite{Cacciari_2008},
\PQb hadrons are scaled to have an infinitely small momentum and included
in the clustering process.
A jet is \labelled as a \PQb-tagged jet if such a ghost \PQb hadron
is clustered inside it.
The \ptvecmiss is defined as the vector \ptvec sum of all neutrinos from \PW, \PZ, or prompt \PGt decays.
Further selection criteria on \pt, $\eta$, and $\DR(j,\ell)$
are applied to these objects, and the events are selected and reconstructed
using the same requirements and algorithms as in the signal region at the detector level.
The measurement at the particle level is hence performed in a fiducial phase space, leading to the fact that the absolute differential cross sections are reduced by a factor of about two with respect to the parton-level measurements.
Reconstructed \tZq events outside the fiducial phase space make up about 7\% of all reconstructed \tZq events.
They are included in the signal extraction and scaled in the fit with the integrated differential cross section.

\subsection{Signal extraction and fit strategy}
\label{sec:differentialfit}

In the measurement of the differential \tZq cross sections, a multiclass neural network,
implemented via the \Tensorflow~\cite{tensorflow2015-whitepaper}
package, is used.
The neural network contains 22 input and five output
nodes to distinguish \tZq events from \ttZ, \WZ, \ttX,
and all other backgrounds. 
The most important input variables are shown in Fig.~\ref{fig:SRFeatures}.
The separation into multiple output nodes is based on the distinct nature of
the different backgrounds and provides an improved isolation
of \tZq events compared to a binary classifier.

\begin{figure}[htb!]
  \centering
  \includegraphics[width=0.8\textwidth]{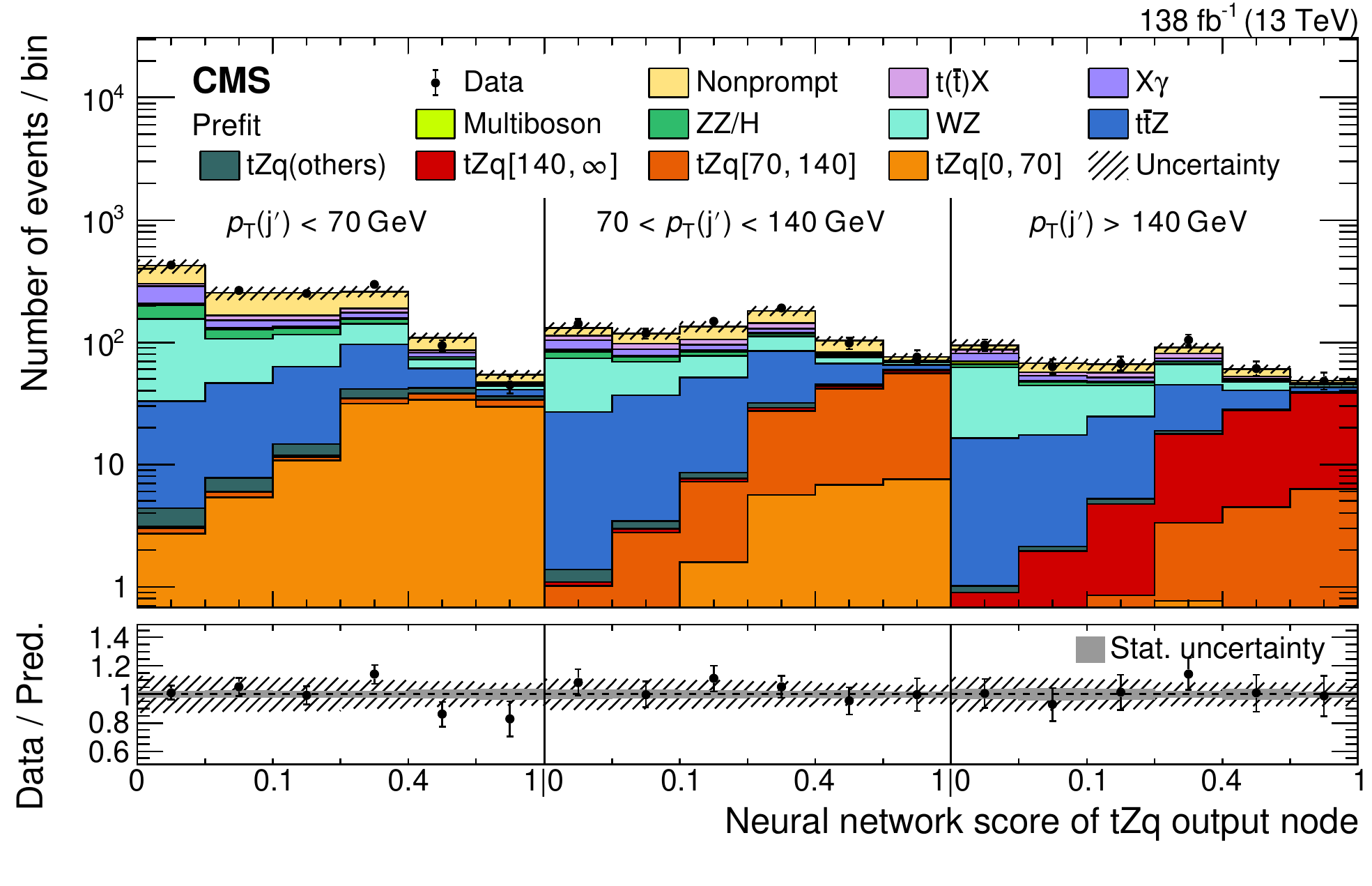}\\
  \includegraphics[width=0.8\textwidth]{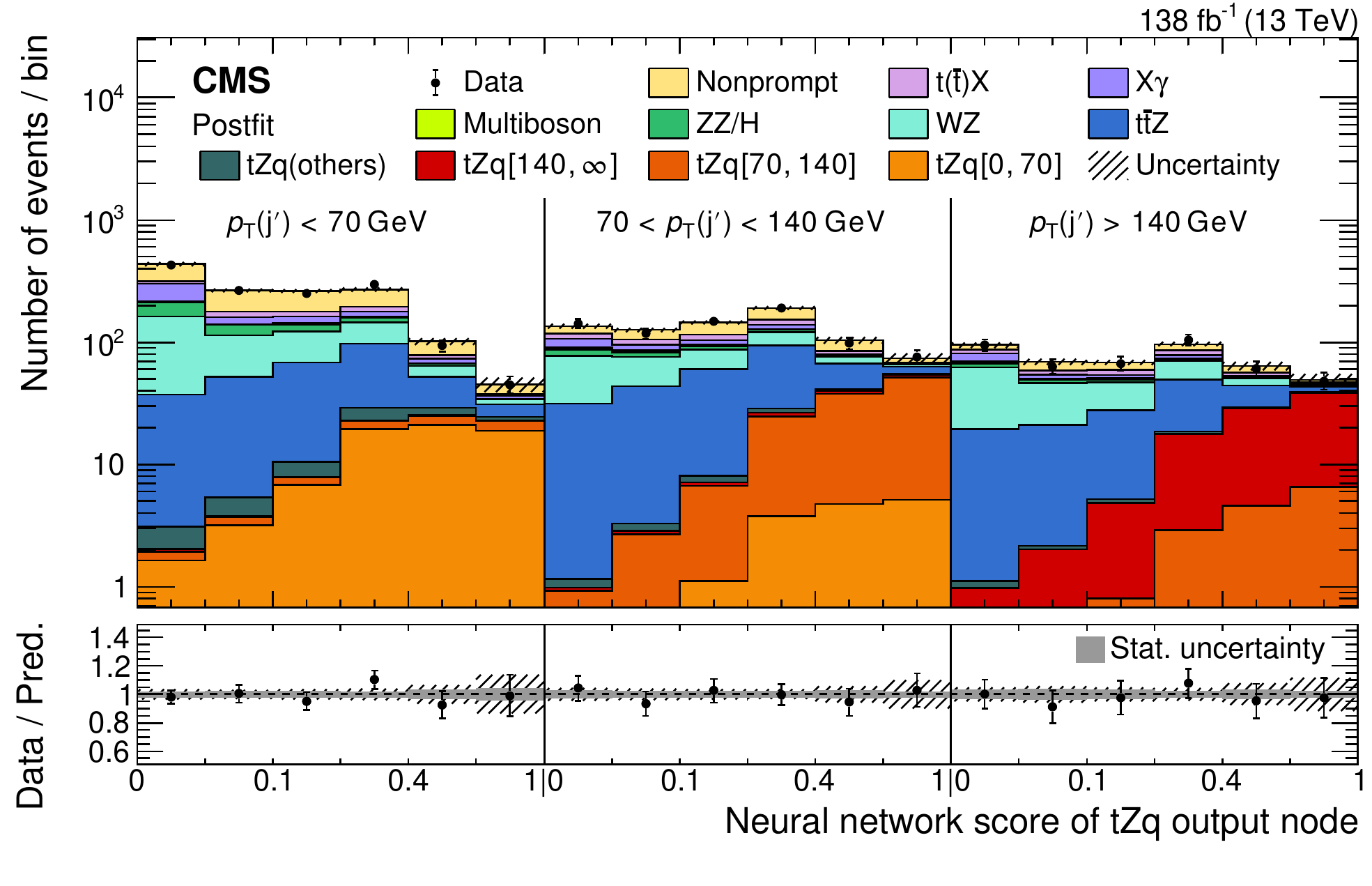}
\caption{Prefit (upper) and postfit (lower) distributions of the neural network score from the \tZq output node for events in the signal region with fewer than four jets, used for the \jprimpt differential cross section measurement at the particle level.
The data are shown by the points and the predictions by the \coloured histograms.
The vertical lines on the points represent the statistical uncertainty in the data, and the hatched region the total uncertainty in the prediction.
The events are split into three subregions based on the value of \jprimpt measured at the detector level.
Three different \tZq templates, defined by the same intervals of
\jprimpt at the particle level and shown in different shades of orange and red,
are used to model the contribution of each particle-level bin.
Reconstructed \tZq events that are outside the fiducial phase space
are \labelled as ``\tZq(others)'' and represent a minor contribution.
The lower panels show the ratio of the data to the sum of the predictions,
with the \grey band indicating the uncertainty from the finite number of MC events.}
\label{fig:SRFit0}
\end{figure}

\begin{figure}[htb!]
  \centering
  \includegraphics[width=0.8\textwidth]{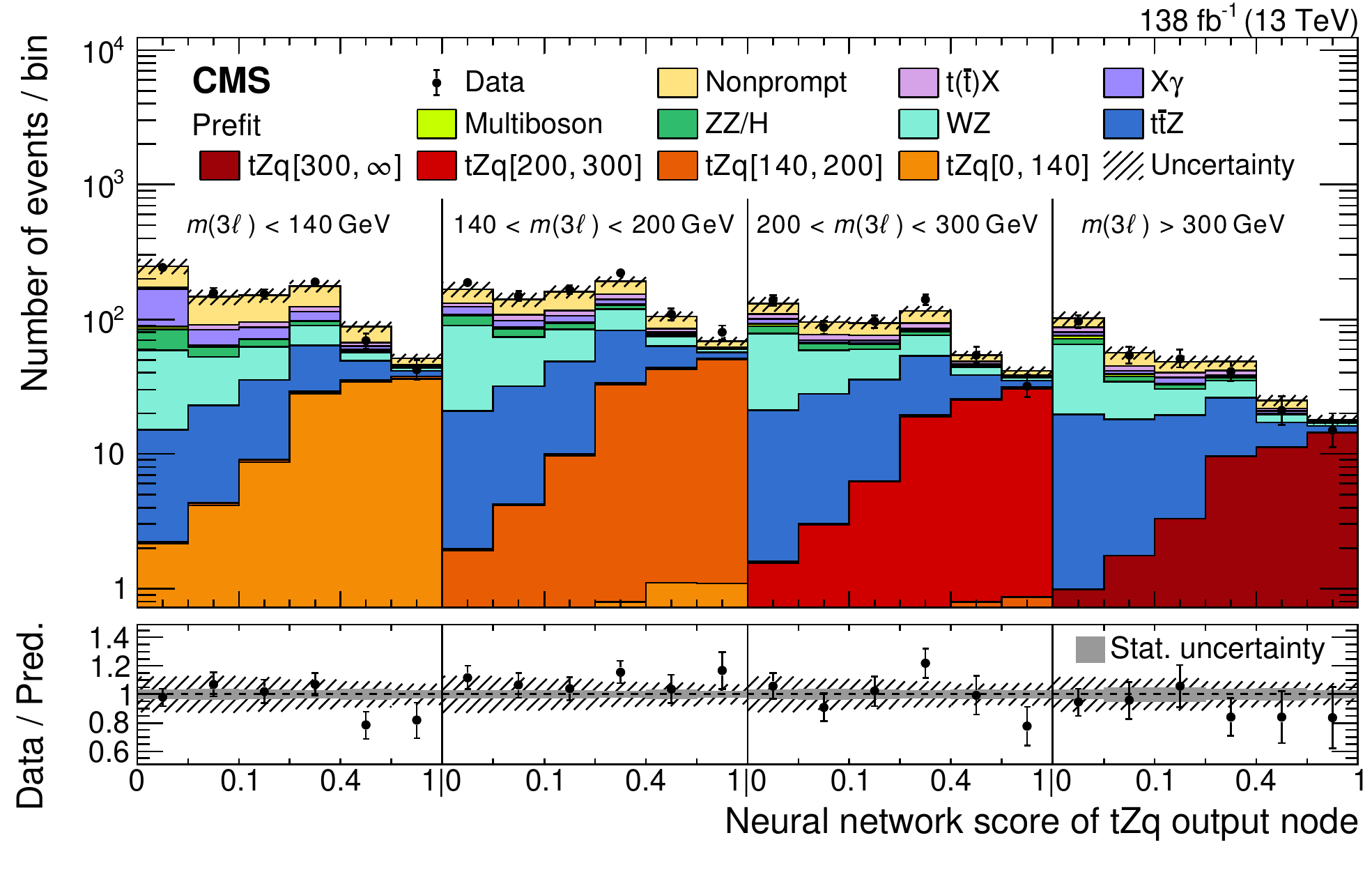}\\
  \includegraphics[width=0.8\textwidth]{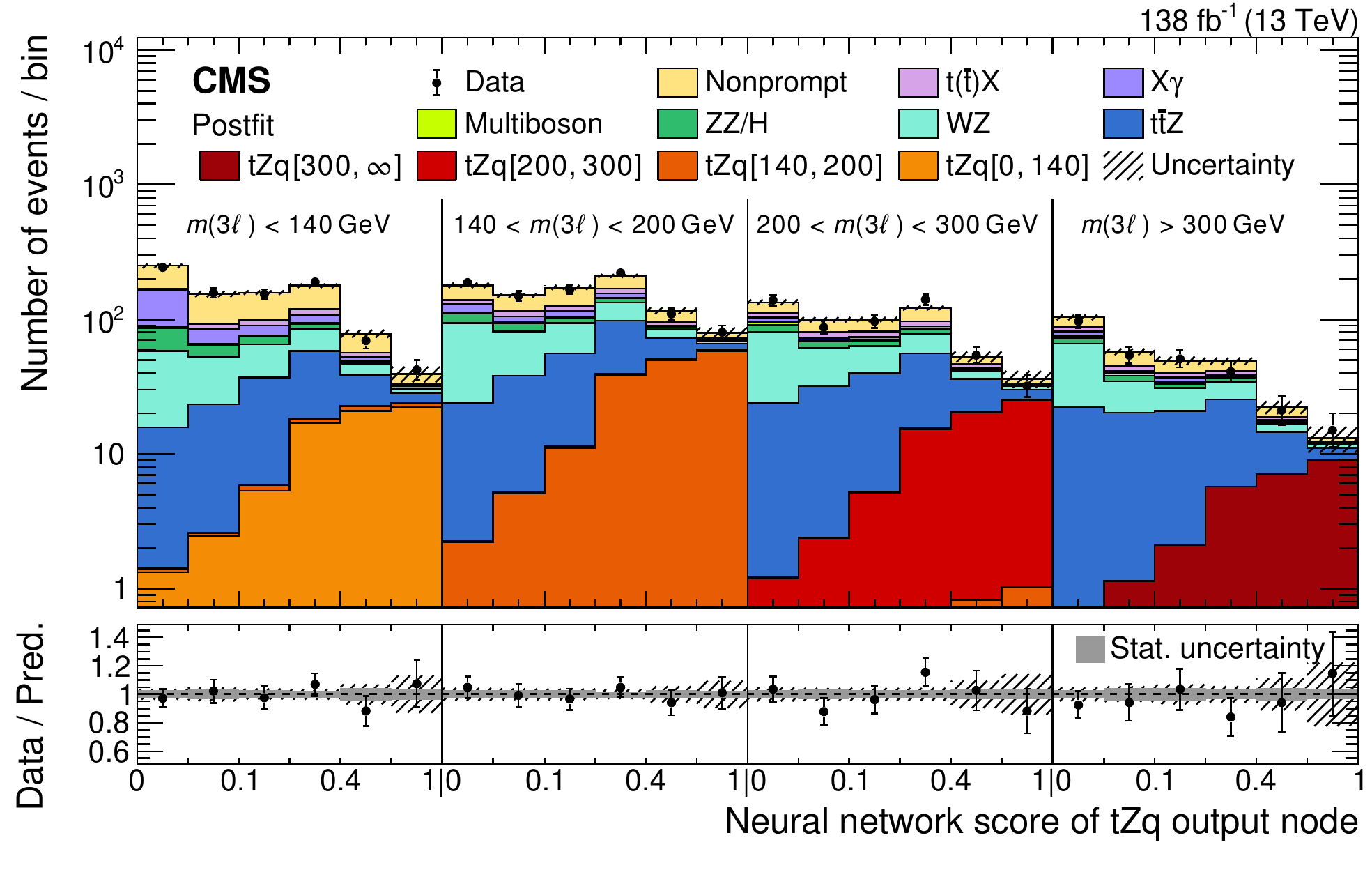}
\caption{
Prefit (upper) and postfit (lower) distributions of the neural network score from the \tZq output node for events in the signal region with fewer than four jets, used for the \mlll differential cross section measurement at the parton level.
The data are shown by the points and the predictions by the \coloured histograms.
The vertical lines on the points represent the statistical uncertainty in the data, and the hatched region the total uncertainty in the prediction.
The events are split into four subregions based on the value of \mlll measured at the detector level.
Four different \tZq templates, defined by the same intervals of
\mlll at the parton level and shown in different shades of orange and red,
are used to model the contribution of each parton-level bin.
The lower panels show the ratio of the data to the sum of the predictions,
with the \grey band indicating the uncertainty from the finite number of MC events.}
\label{fig:SRFit1}
\end{figure}

For the signal extraction, events in the signal region
(defined in Section~\ref{sec:selection}) are additionally
required to have fewer than four central jets.
Events with four or more central jets are dominated by \ttZ events
and used as a control region ($\ttZ \to 3\ell$).
The remaining events in the signal region are split into three subregions
(four for observables only involving leptons)
based on the value of the observable at the detector level,
using the bin ranges from the definition of the parton- and particle-level bins.
As a result, the \tZq contribution from each parton- and particle-level bin of the considered observable
is enriched in the corresponding subregion at the detector level.
To isolate the \tZq events from various background events, each subregion is binned in the
neural network score of the \tZq output node.
As an example, the prefit and postfit \jprimpt and \mlll distributions in the signal region are shown in
Figs.~\ref{fig:SRFit0} and~\ref{fig:SRFit1}, respectively.

To further constrain the normalization of various backgrounds,
additional complementary control regions with three or four leptons are defined,
as described in Section~\ref{sec:backgrounds},
and included in the signal extraction.
Events in these control regions are additionally required to have at least
two jets to minimize the uncertainty associated with the extrapolation of the various backgrounds
to the signal region.
Events in the $\ttZ\to 3\ell$ control region are fit as a function
of the neural network score of the \ttZ output node to separate the \ttZ
process, while events in the \WZ control regions are fit as a function
of \mtw. The \Zg control region consists of two
bins that are included in the fit, which are determined from the number of selected electrons.
Events in the $\ttZ\to 4\ell$ control region are separated into three bins in the fit,
as a function of the number of \PQb-tagged jets in the event.
Events in the \ZZ and nonprompt-lepton control regions are fit in one bin.

\subsection{Results}

\begin{figure}[htb!]
  \centering
  \includegraphics[width=0.45\textwidth]{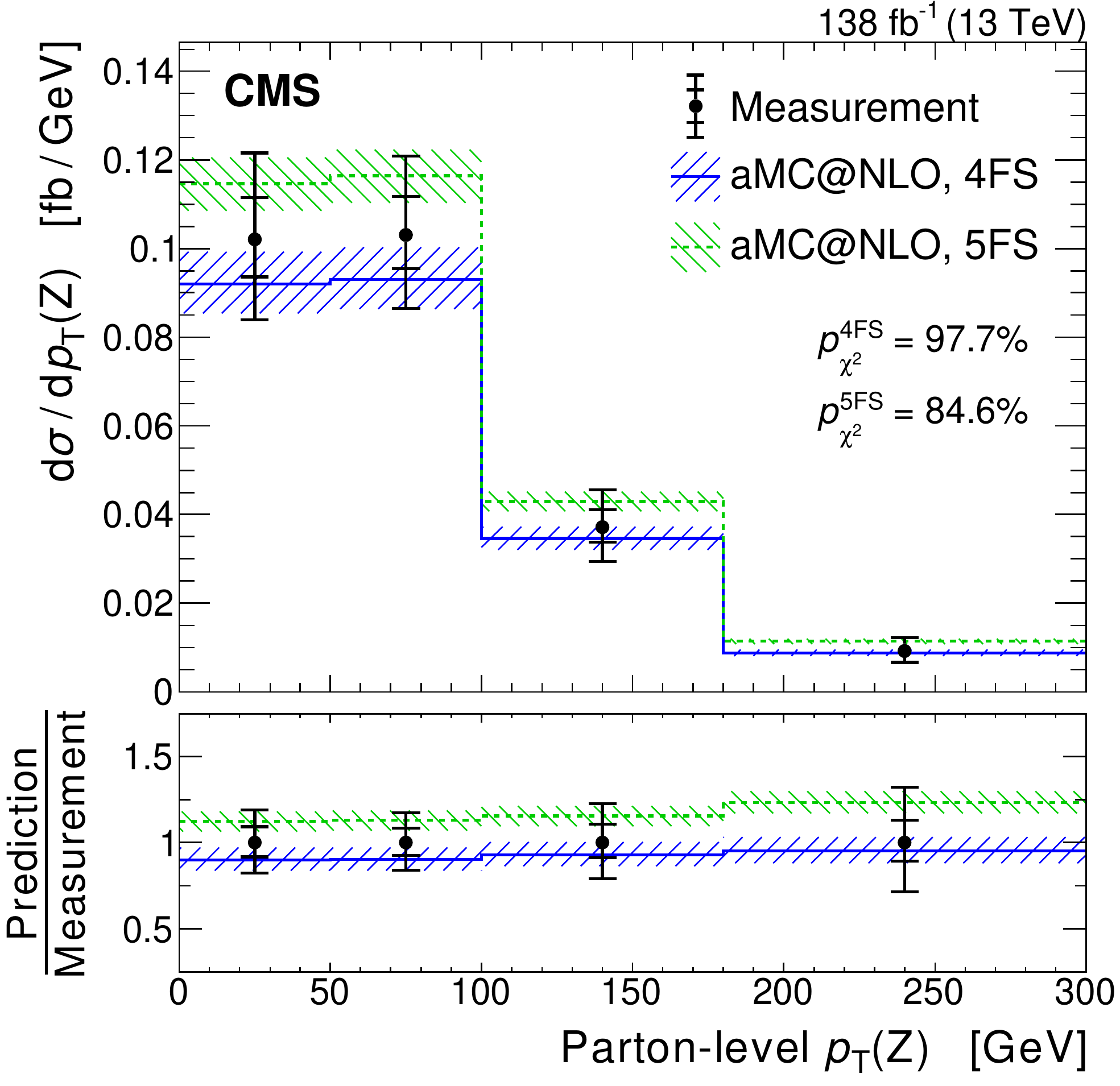}
  \includegraphics[width=0.45\textwidth]{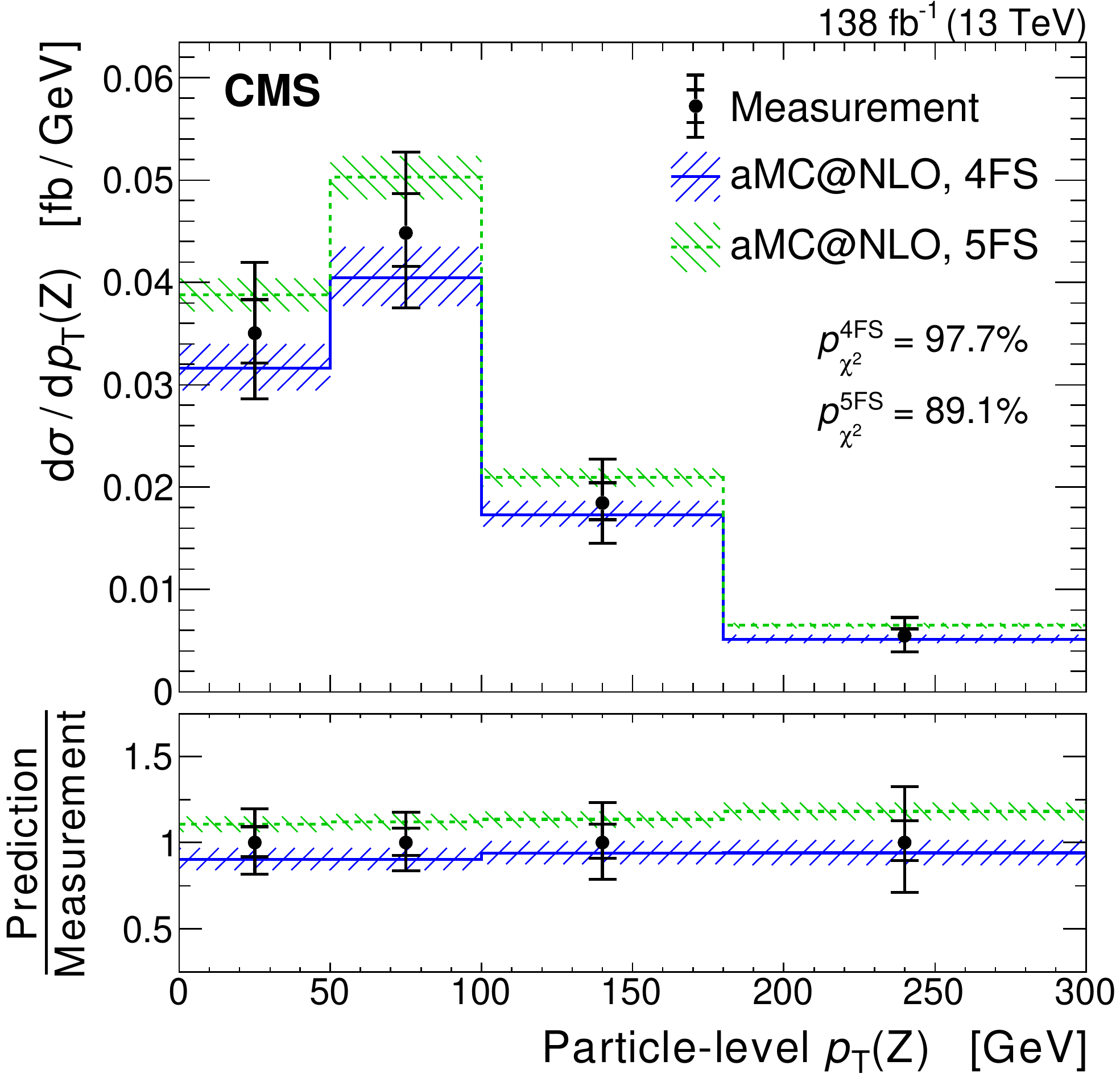} \\
  \includegraphics[width=0.45\textwidth]{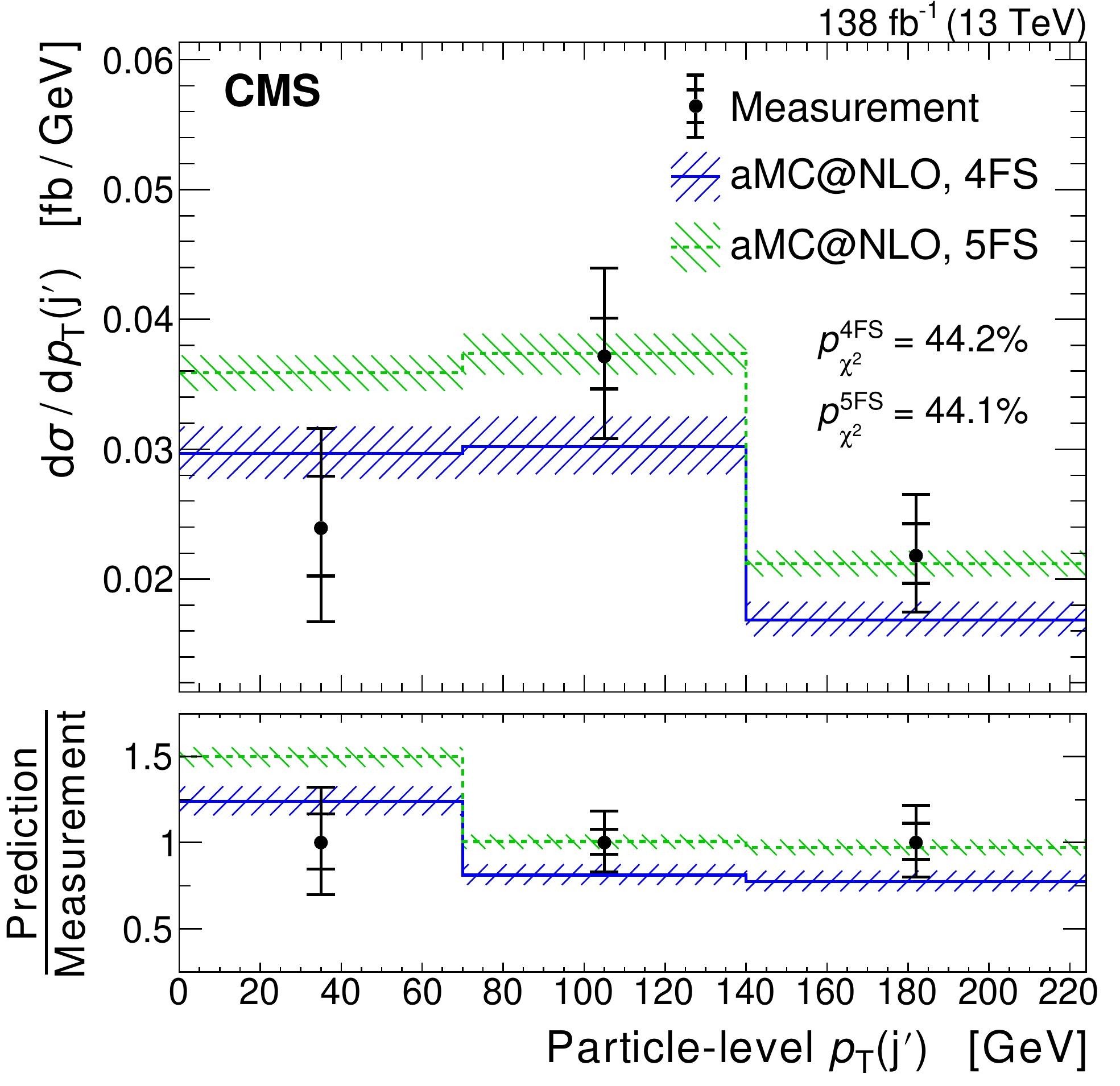}
  \includegraphics[width=0.45\textwidth]{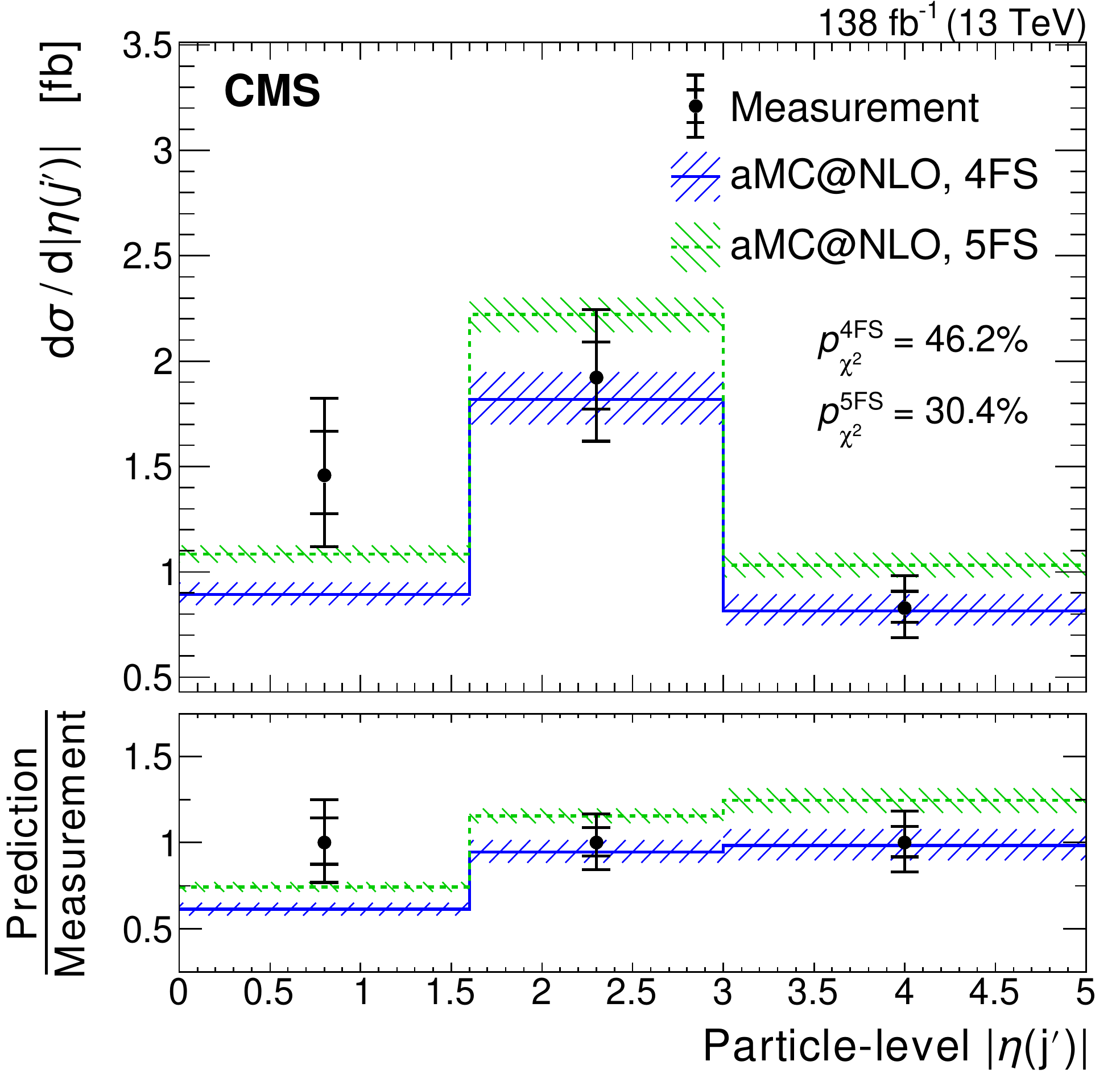}
\caption{Absolute differential cross sections as a function of
\zpt measured at the parton (upper left) and particle levels (upper right), as
well as a function of \jprimpt (lower left) and \jprimeta (lower right) at the particle level.
The observed values are shown as black points, with the inner and outer vertical bars giving the systematic and total uncertainties, respectively.
The SM predictions for the \tZq process are based on
events simulated in the 5FS (green) and 4FS (blue). The $p$-values of the \chit tests
are given to quantify their compatibility with the measurement.
The lower panels show the ratio of the simulation to the measurement.}
\label{fig:result0}
\end{figure}

\begin{figure}[p!]
  \centering
  \includegraphics[width=0.45\textwidth]{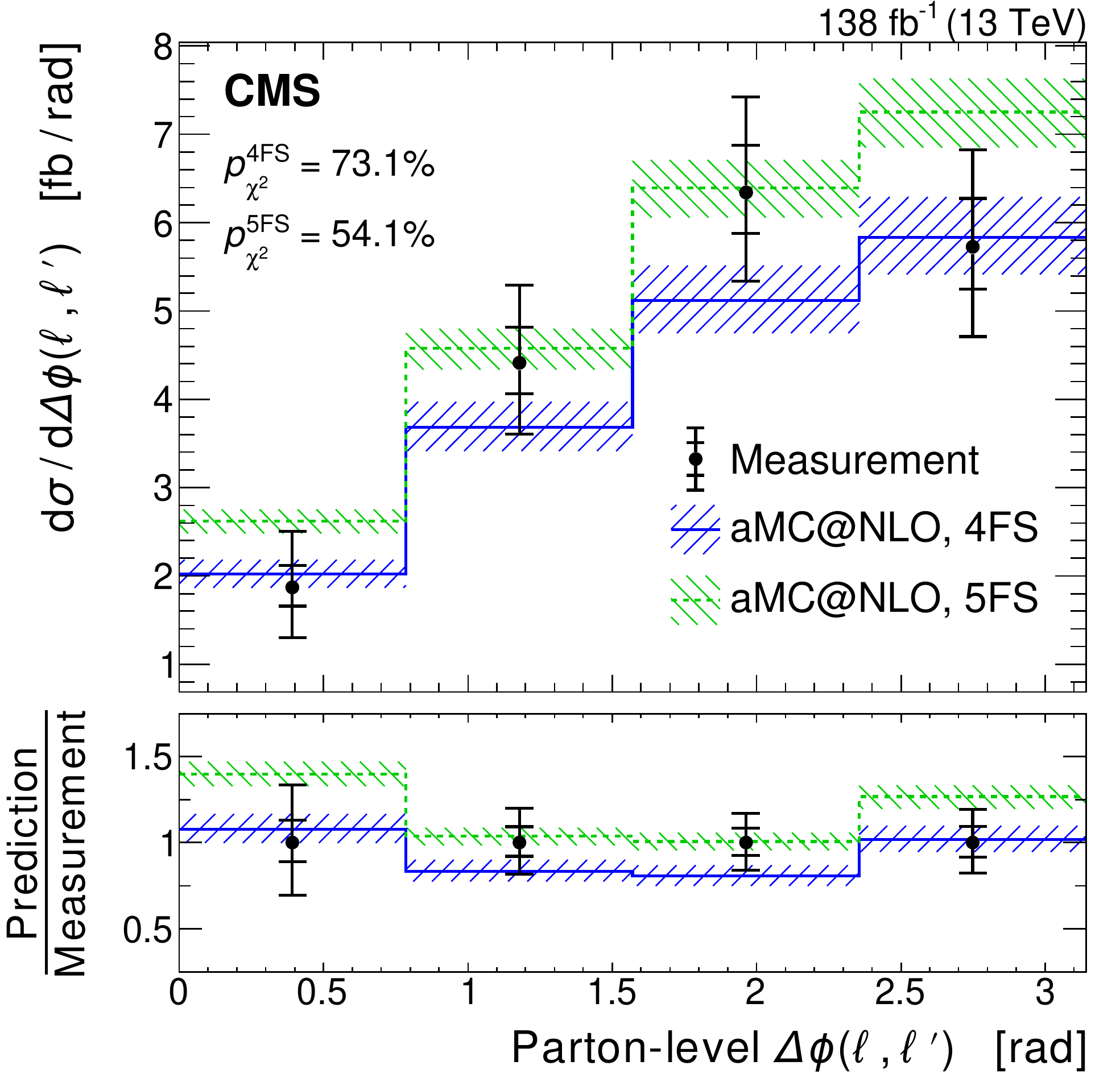}
  \includegraphics[width=0.45\textwidth]{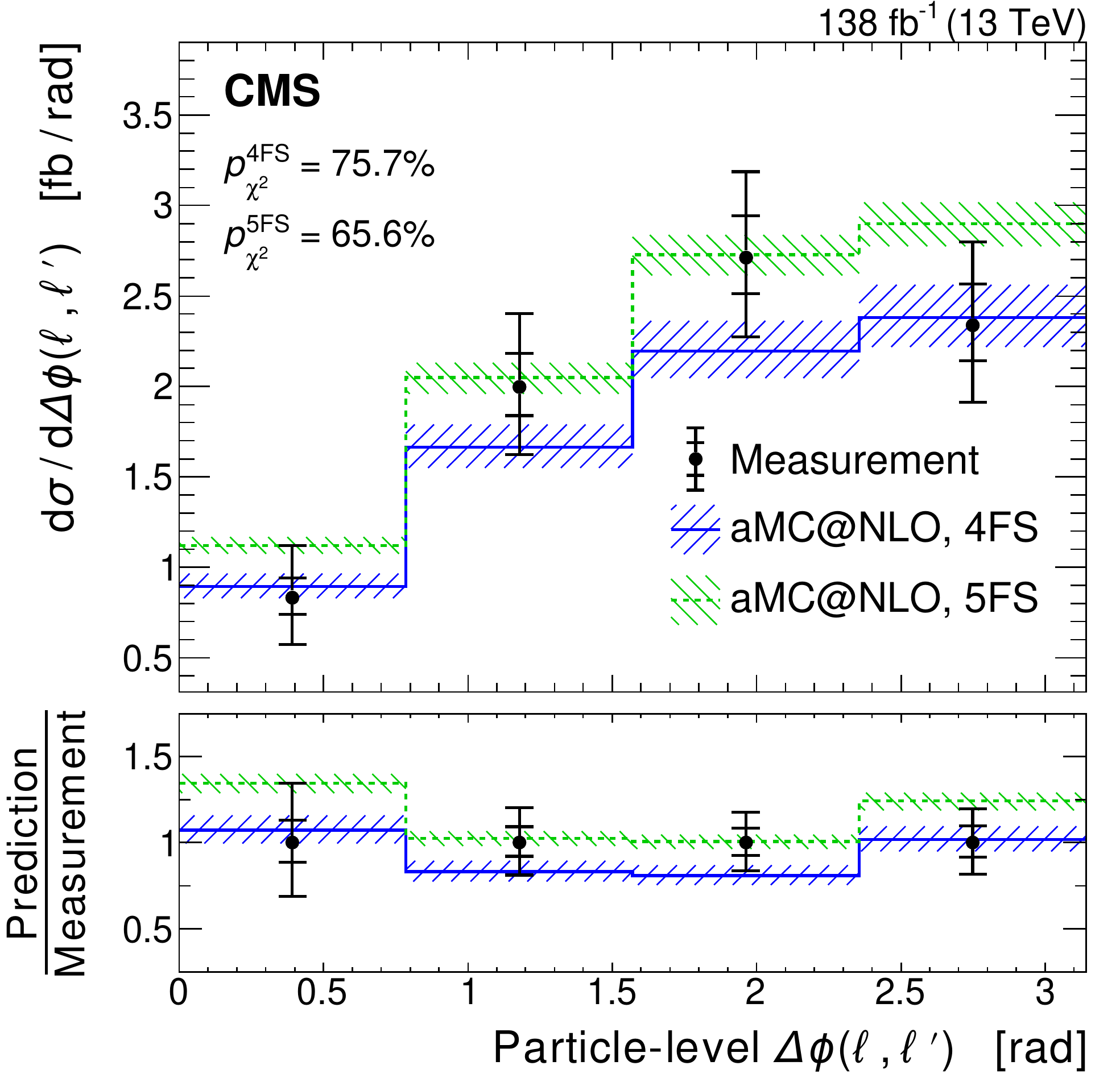}\\
  \includegraphics[width=0.45\textwidth]{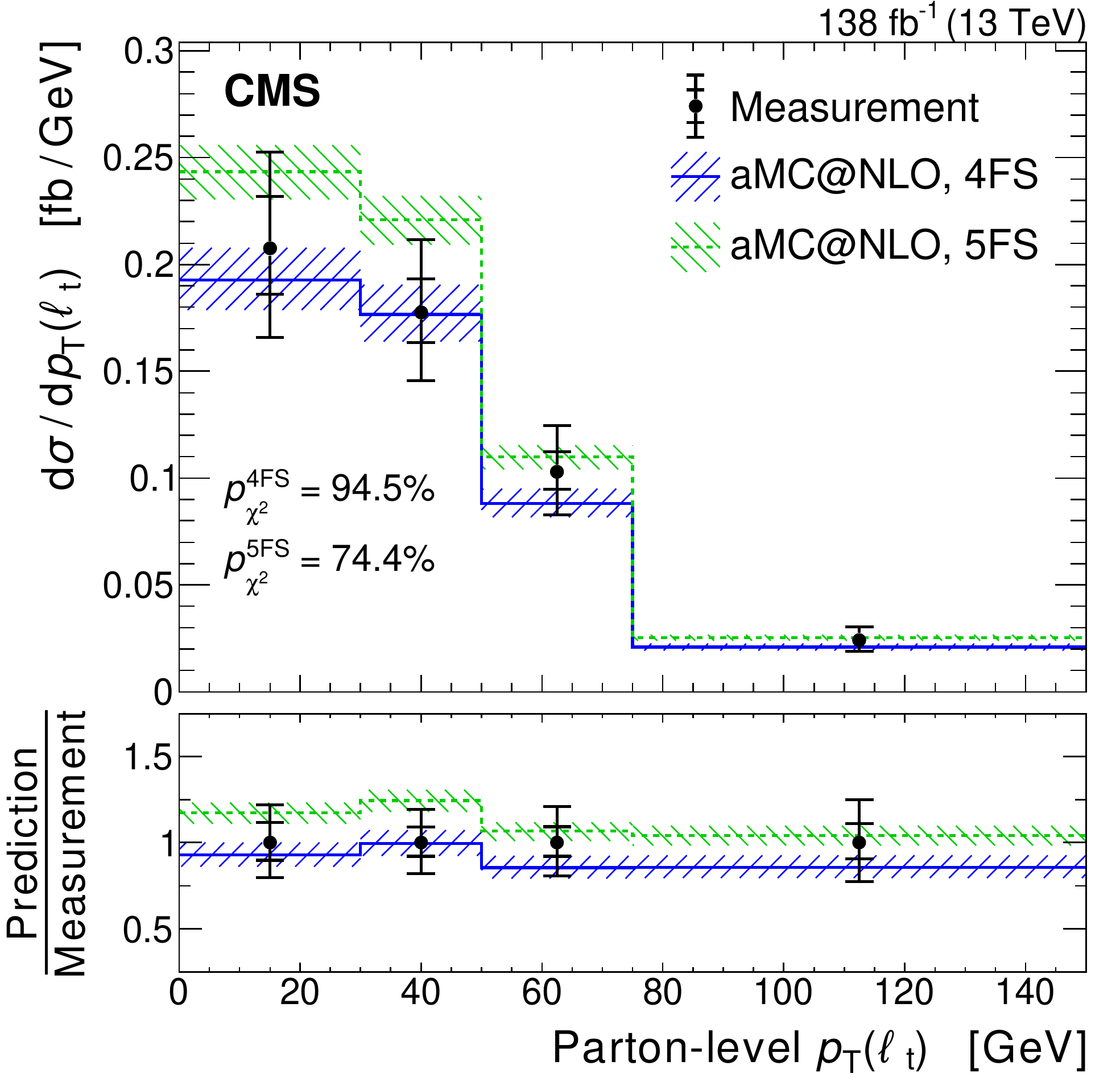}
  \includegraphics[width=0.45\textwidth]{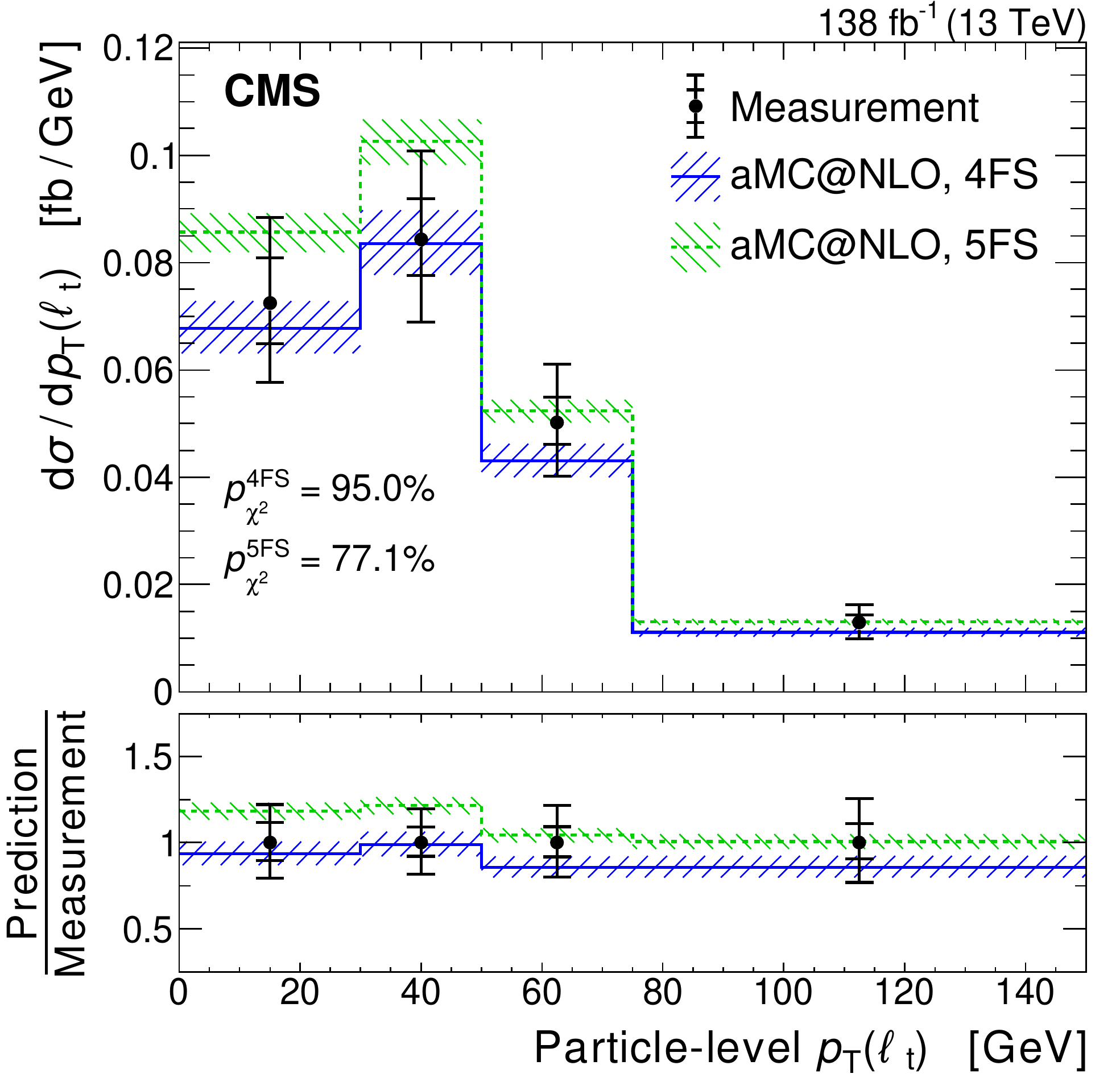}\\
  \includegraphics[width=0.45\textwidth]{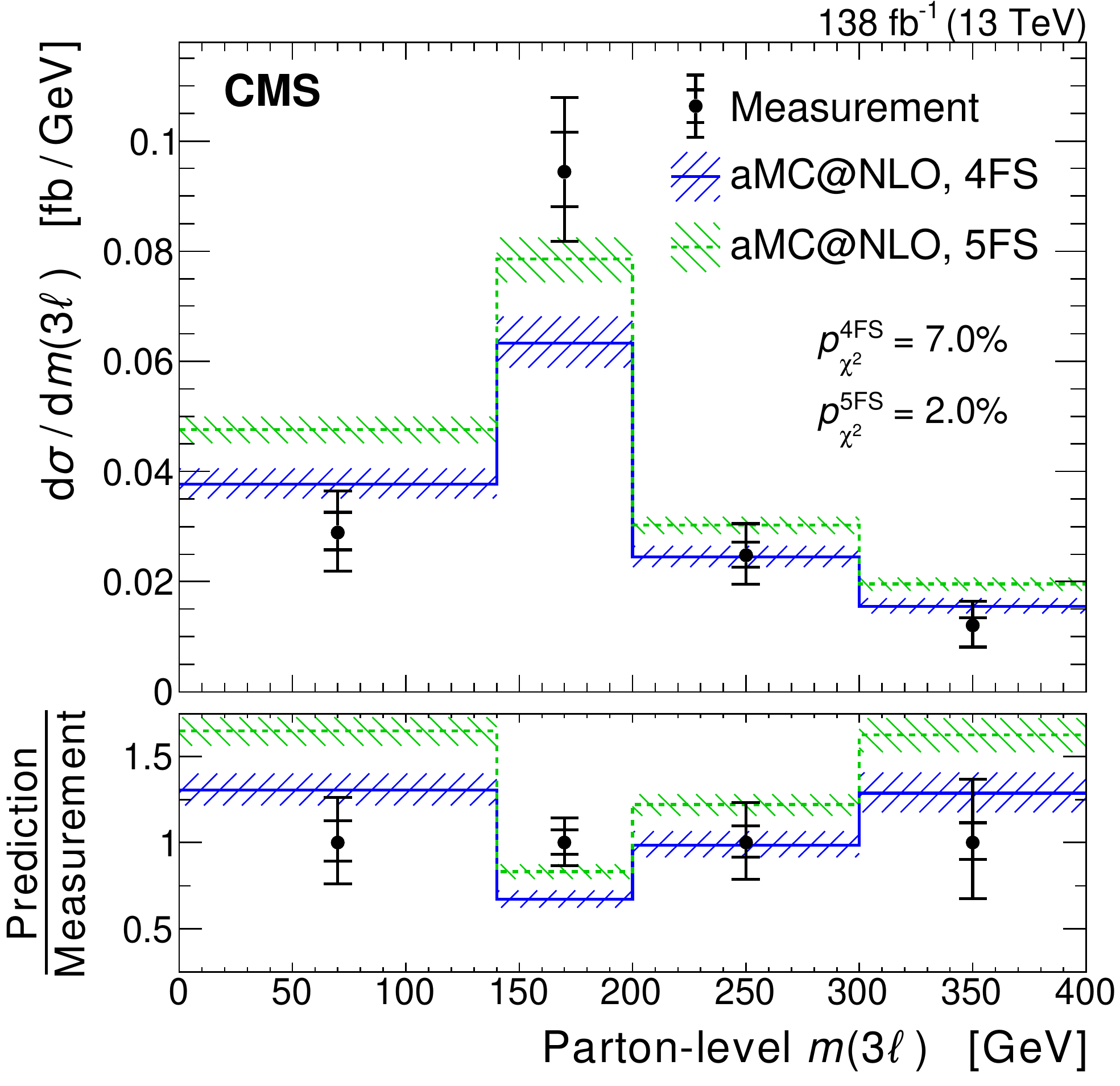}
  \includegraphics[width=0.45\textwidth]{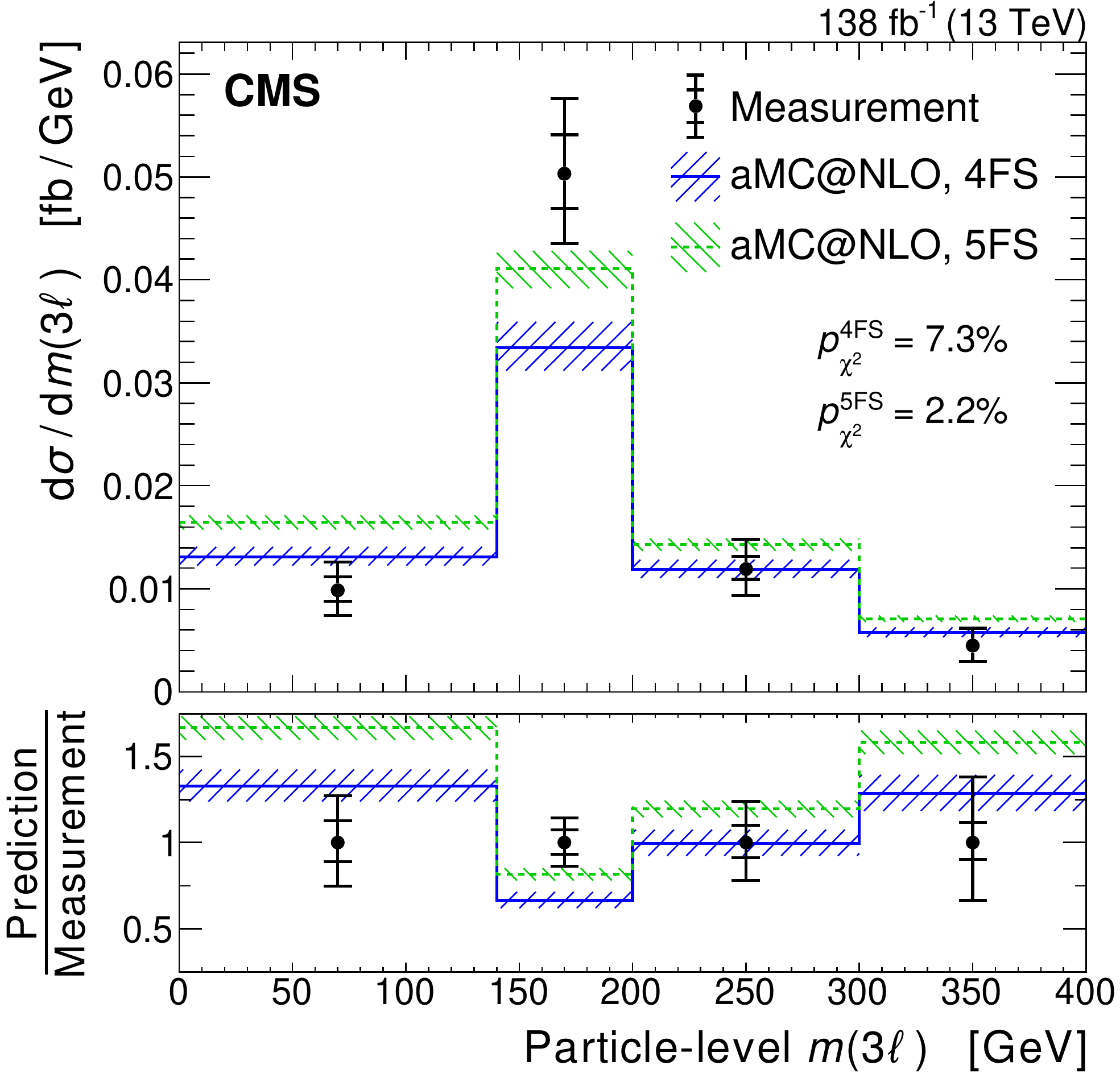}
\caption{Absolute differential cross sections at the parton (left)
and particle level (right) measured as a function of \delphill (upper), \topleppt
(middle) and \mlll (lower).
The observed values are shown as black points, with the inner and outer vertical bars giving the systematic and total uncertainties, respectively.
The SM predictions for the \tZq process are based on
events simulated in the 5FS (green) and 4FS (blue). The $p$-values of the \chit tests
are given to quantify their compatibility with the measurement.
The lower panels show the ratio of the simulation to the measurement.}
\label{fig:result1}
\end{figure}

\begin{figure}[p!]
  \centering
  \includegraphics[width=0.45\textwidth]{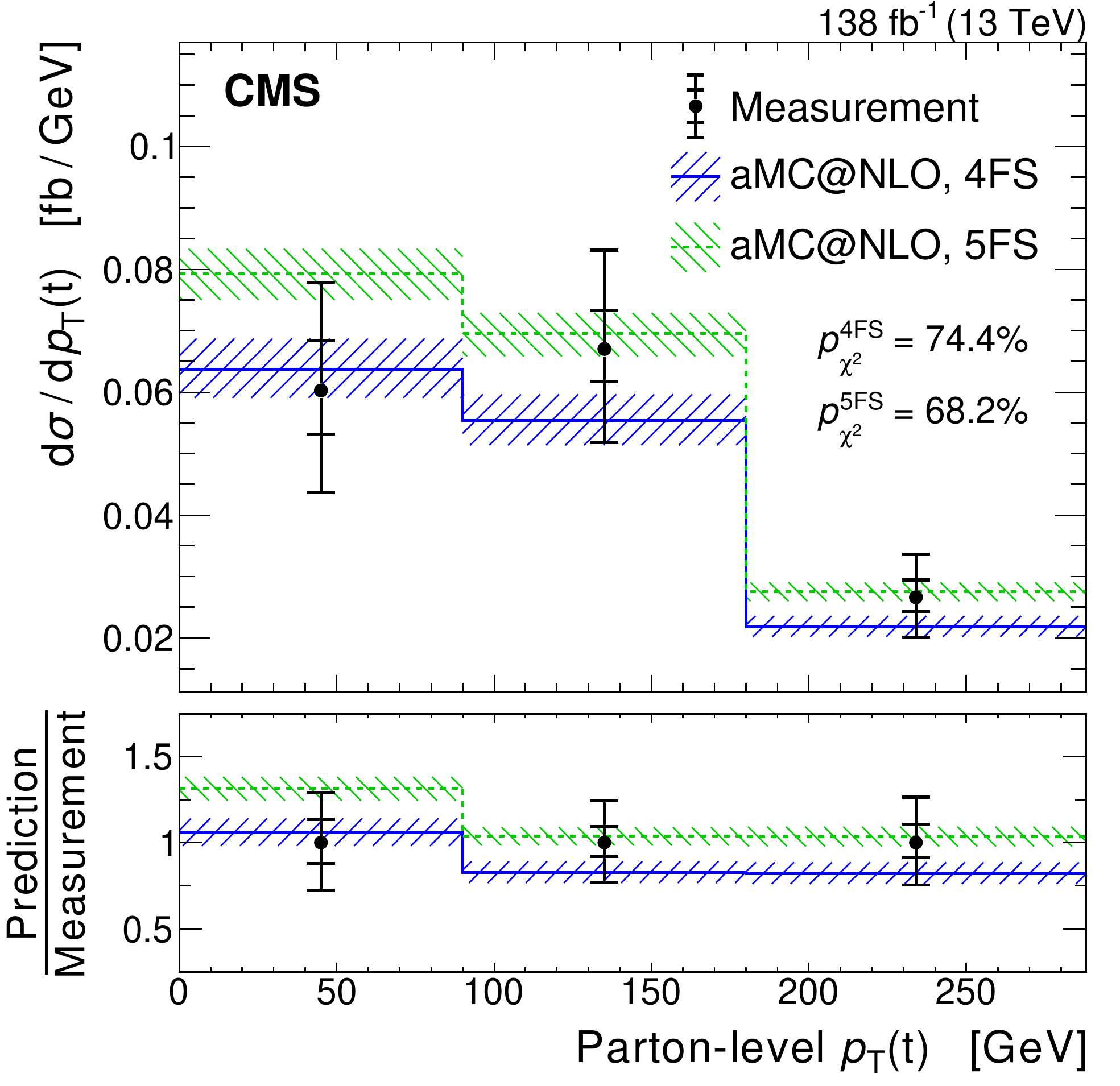}
  \includegraphics[width=0.45\textwidth]{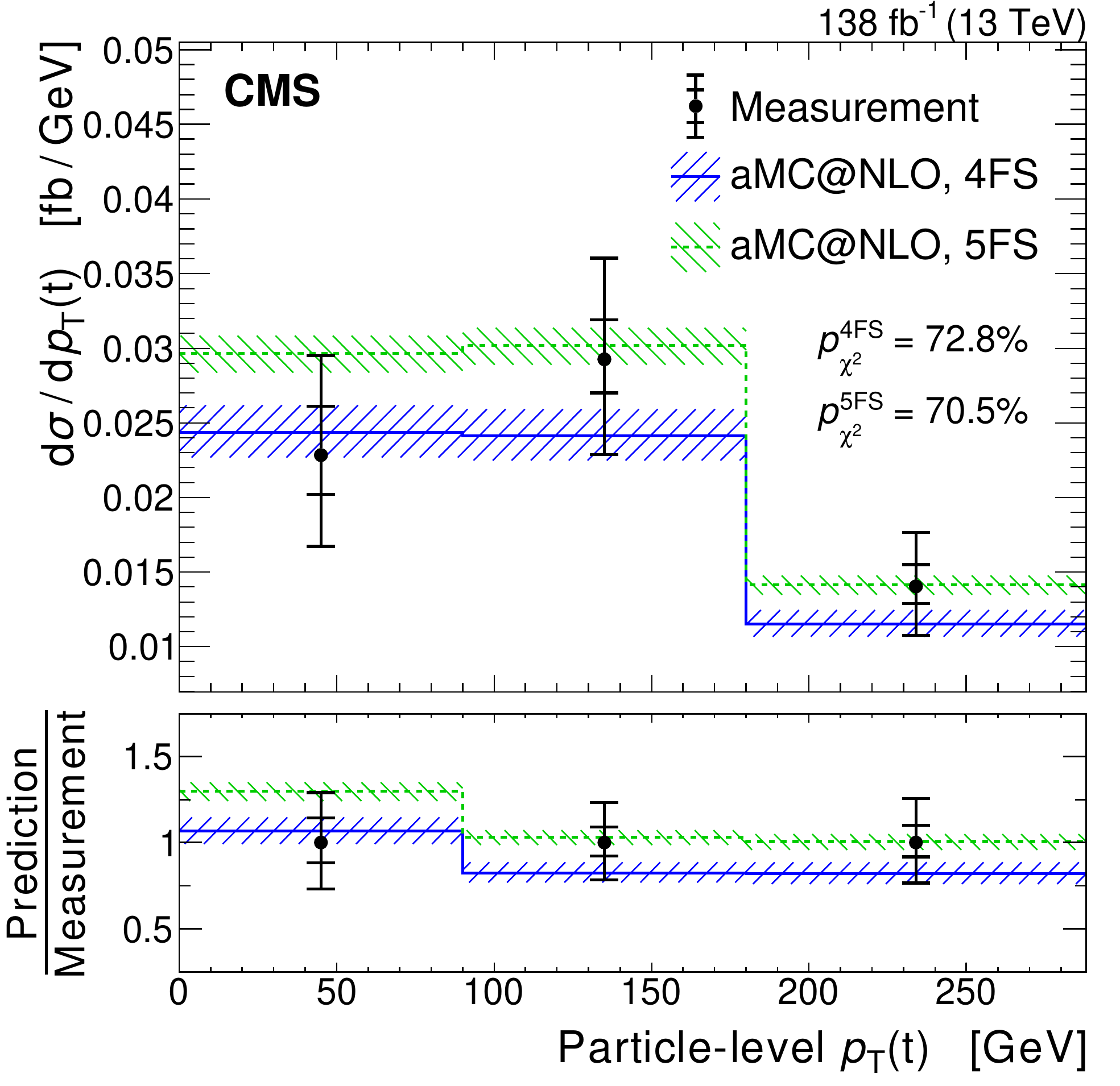}\\
  \includegraphics[width=0.45\textwidth]{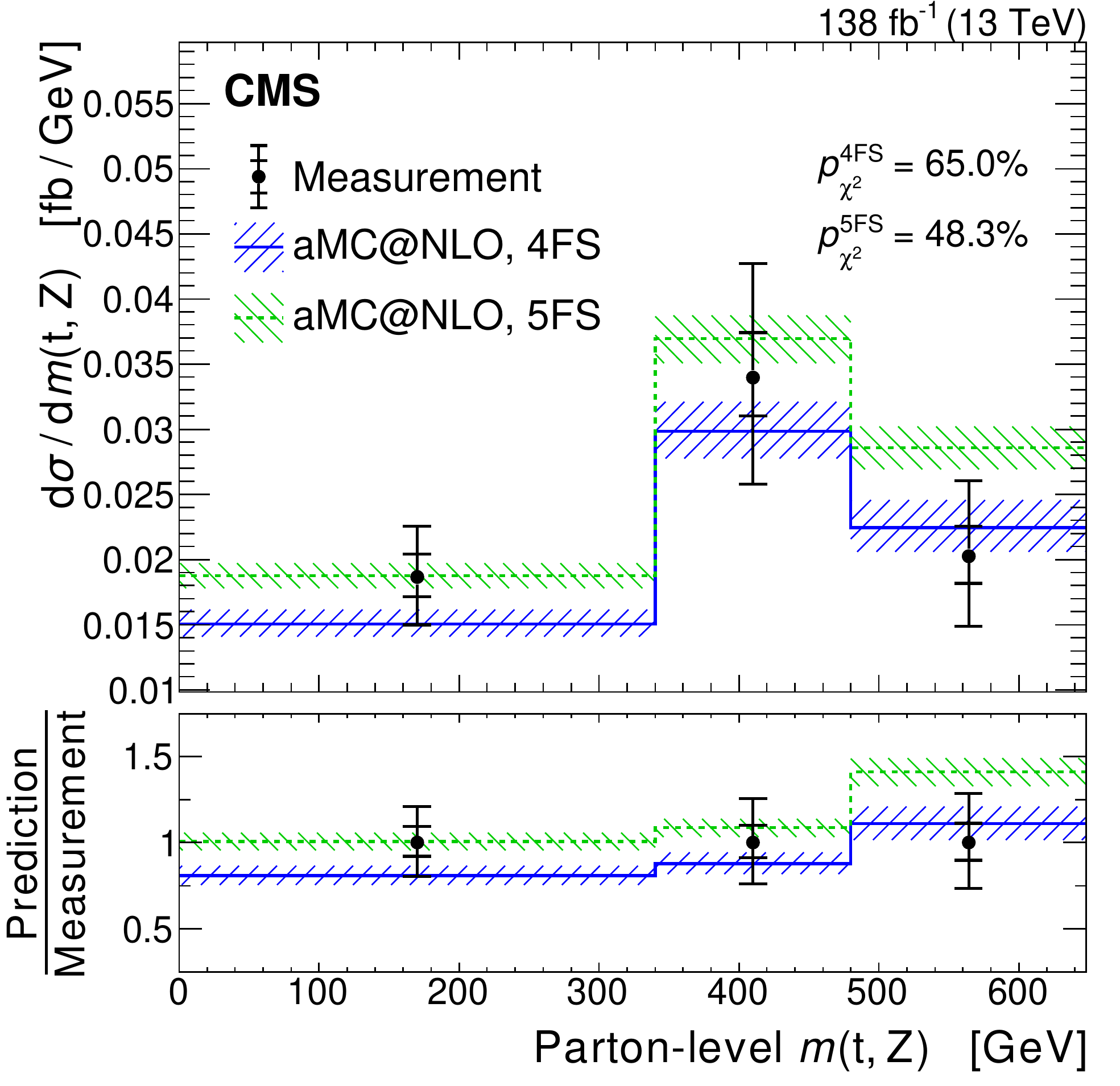}
  \includegraphics[width=0.45\textwidth]{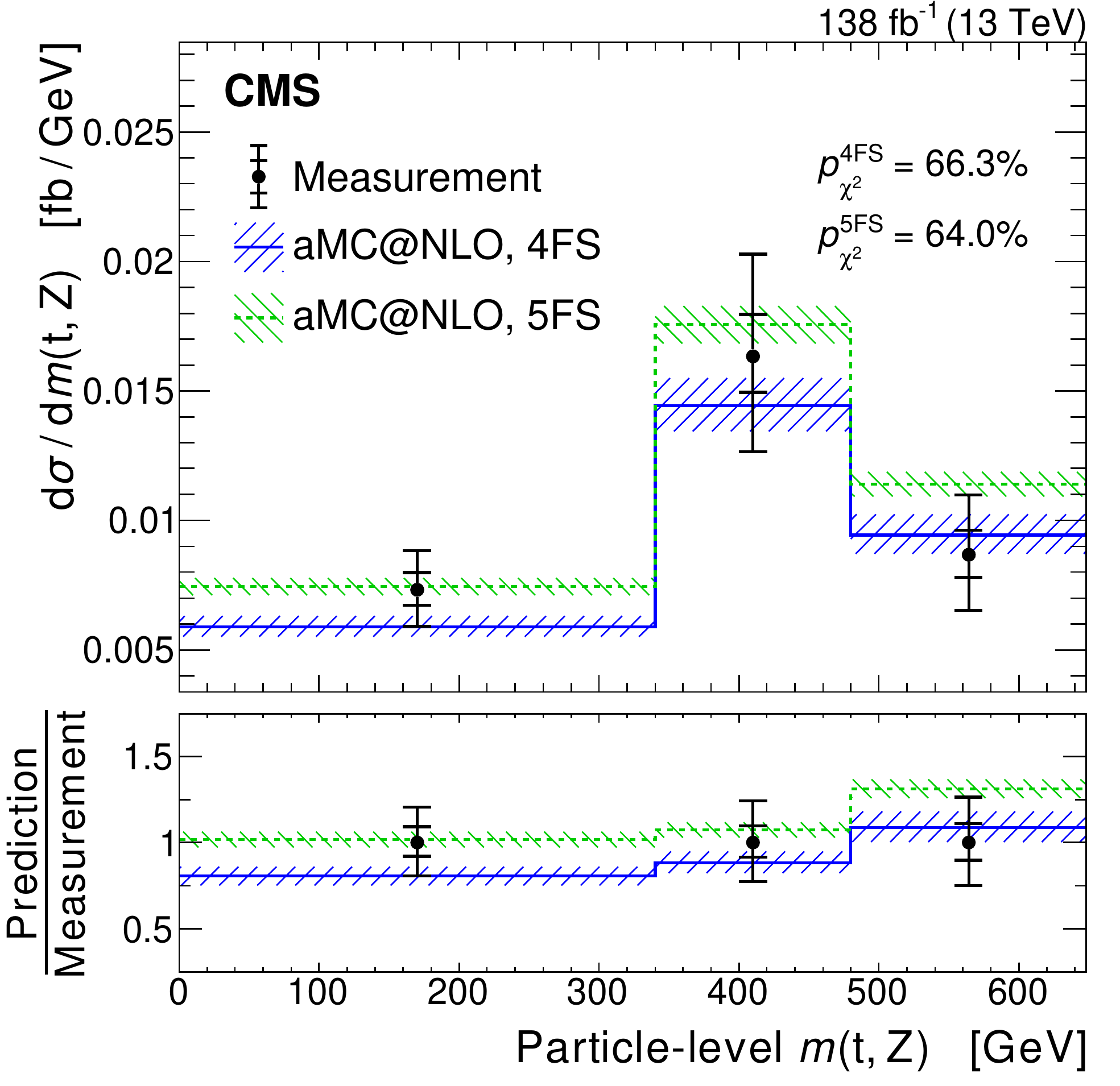}\\
  \includegraphics[width=0.45\textwidth]{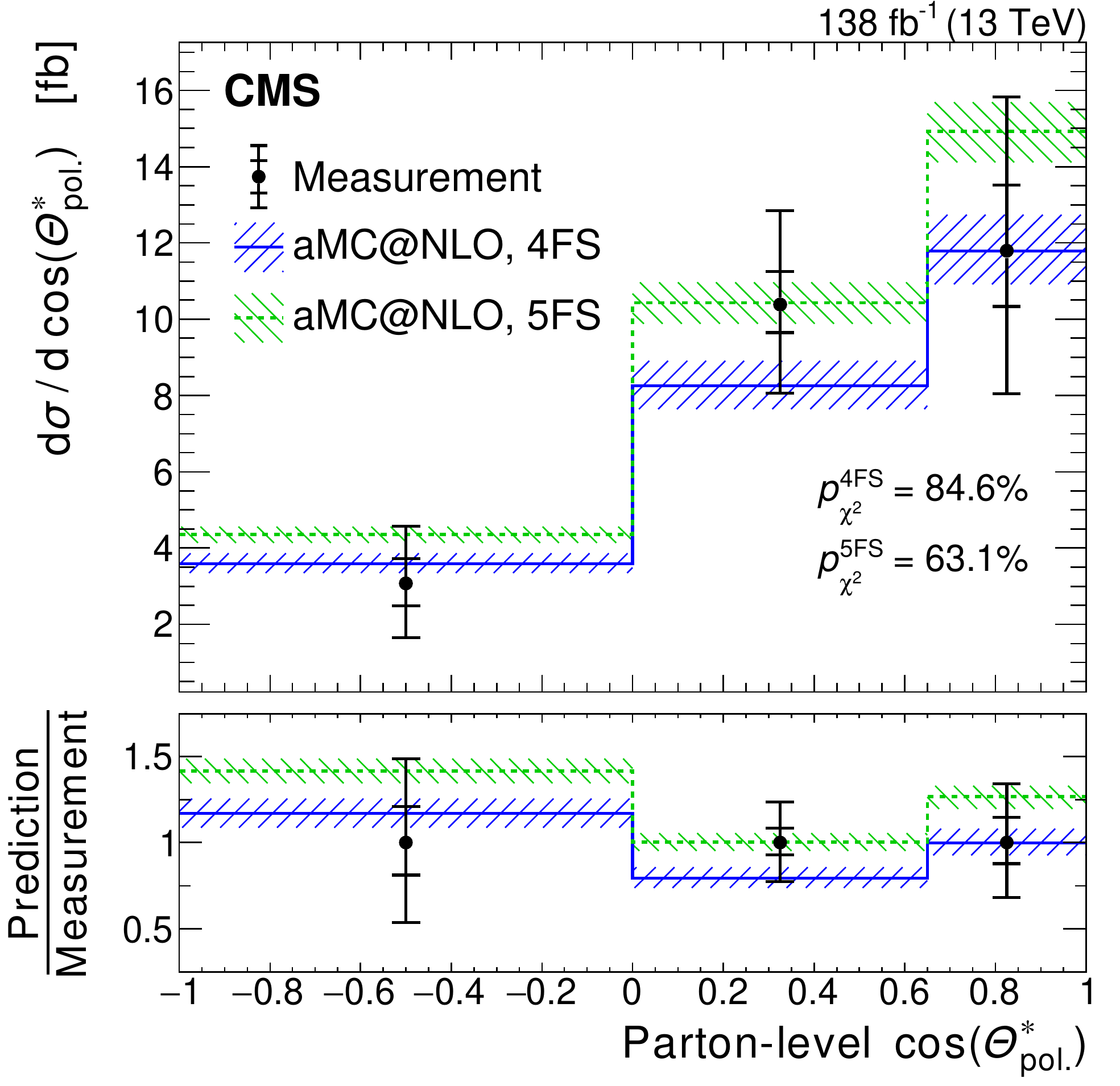}
  \includegraphics[width=0.45\textwidth]{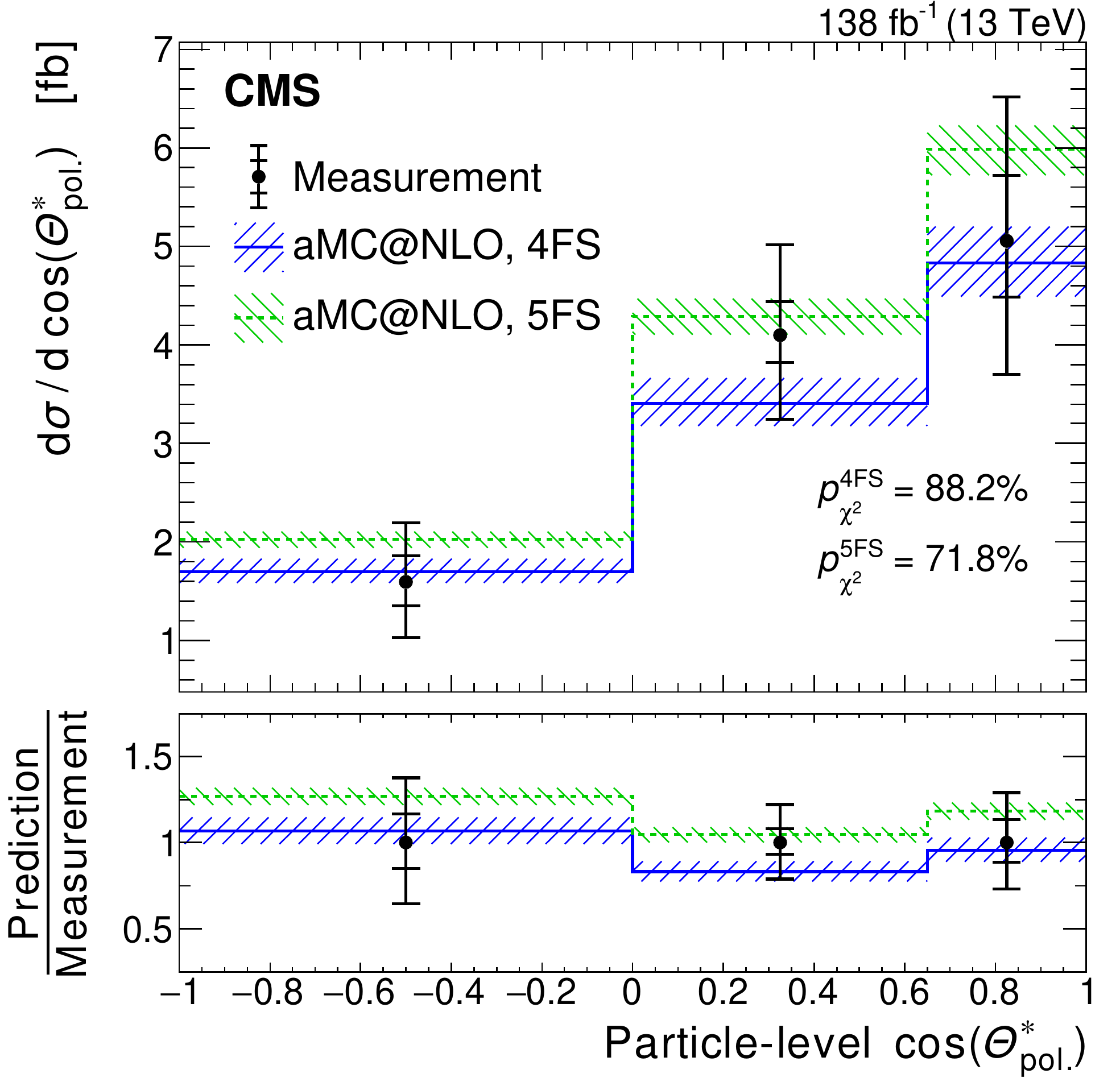}
\caption{Absolute differential cross sections at the parton (left)
and particle level (right) measured as a function of \toppt (upper), \mtz (middle)
and \costheta (lower).
The observed values are shown as black points, with the inner and outer vertical bars giving the systematic and total uncertainties, respectively.
The SM predictions for the \tZq process are based on
events simulated in the 5FS (green) and 4FS (blue). The $p$-values of the \chit tests
are given to quantify their compatibility with the measurement.
The lower panels show the ratio of the simulation to the measurement.}
\label{fig:result2}
\end{figure}

The measured absolute differential cross sections are shown in
Figs.~\ref{fig:result0}--\ref{fig:result2}.
The last bin of each distribution also includes the overflow contributions.

\begin{figure}[htbp!]
  \centering
  \includegraphics[width=0.45\textwidth]{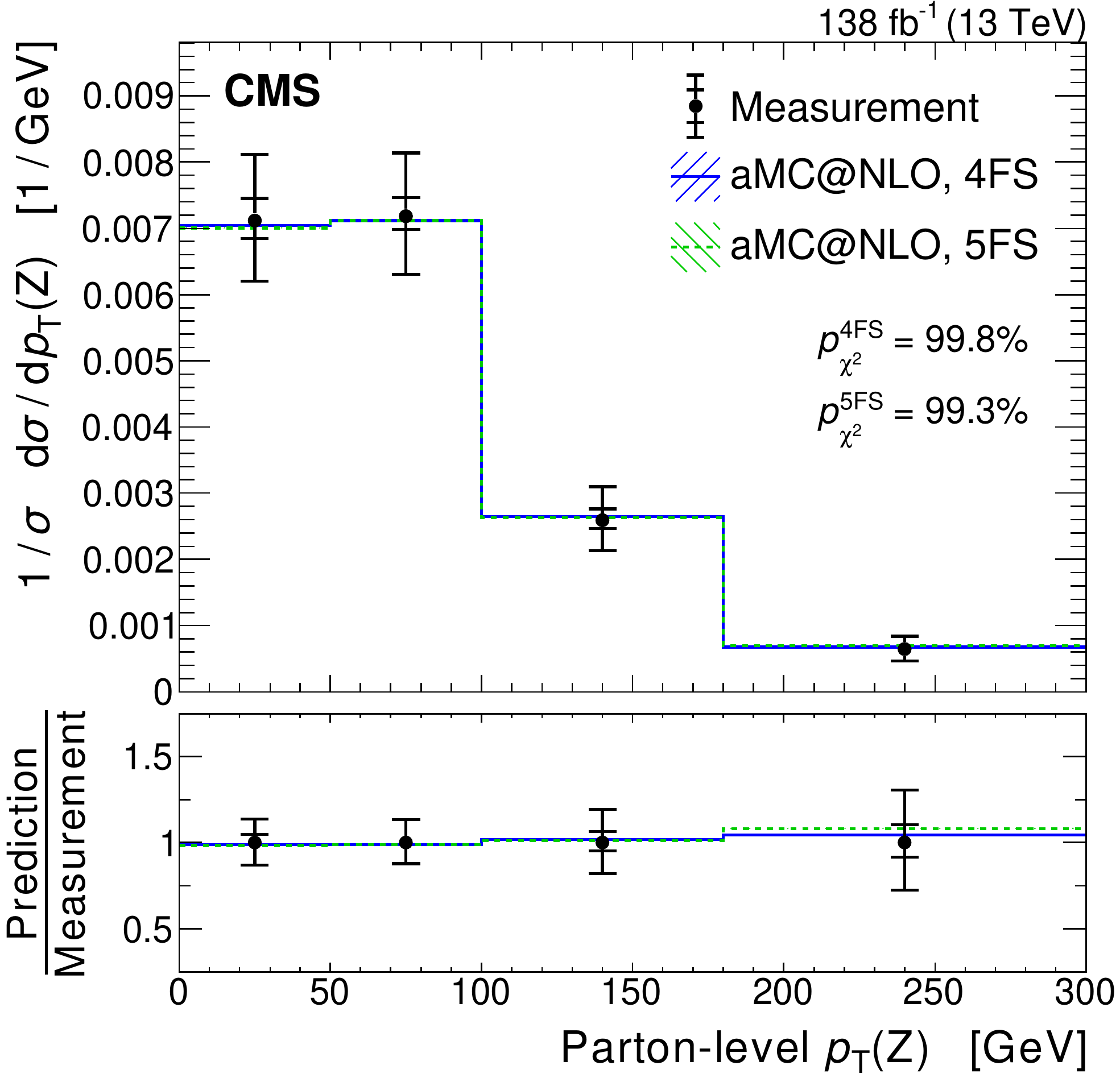}
  \includegraphics[width=0.45\textwidth]{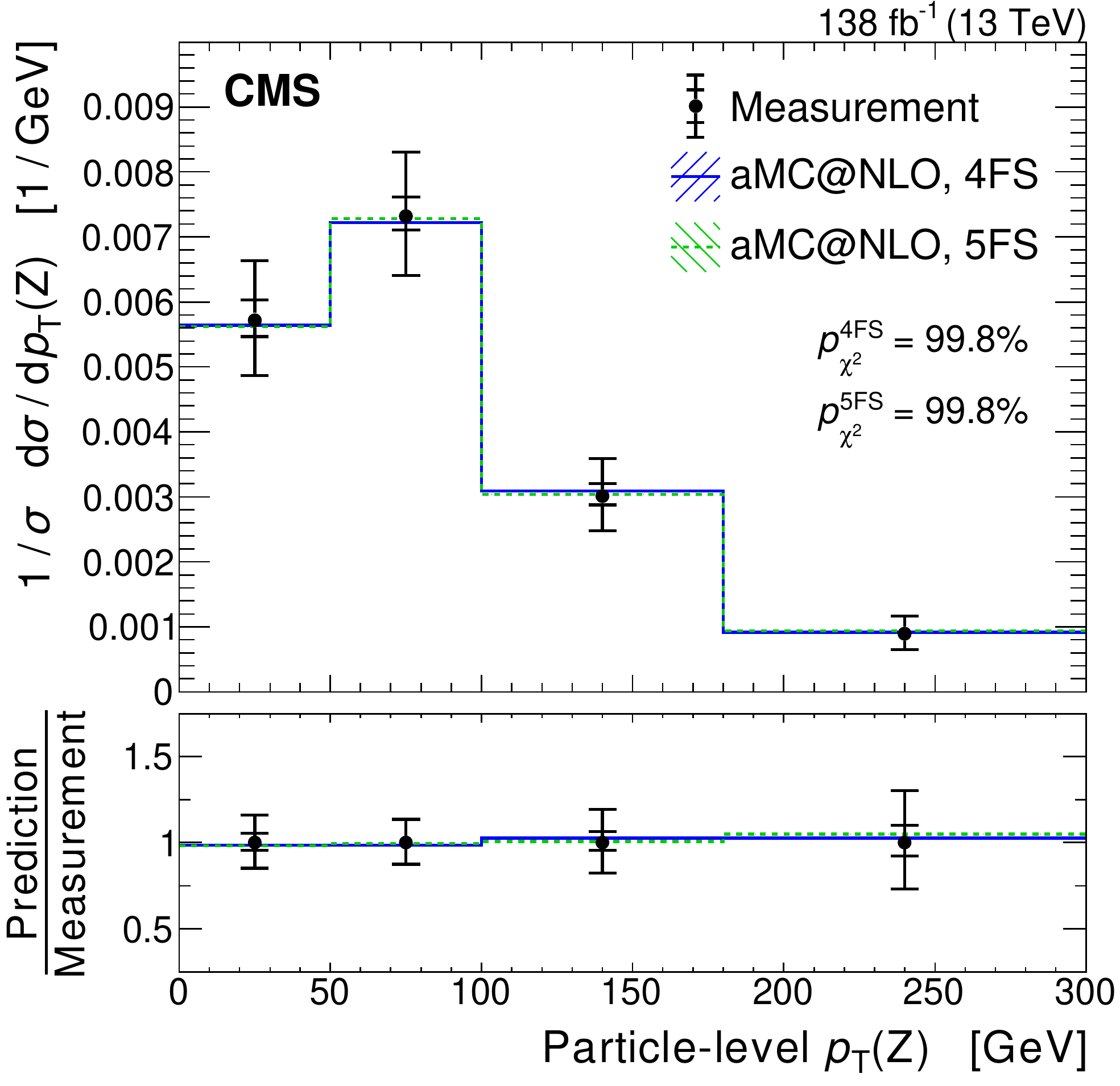} \\
  \includegraphics[width=0.45\textwidth]{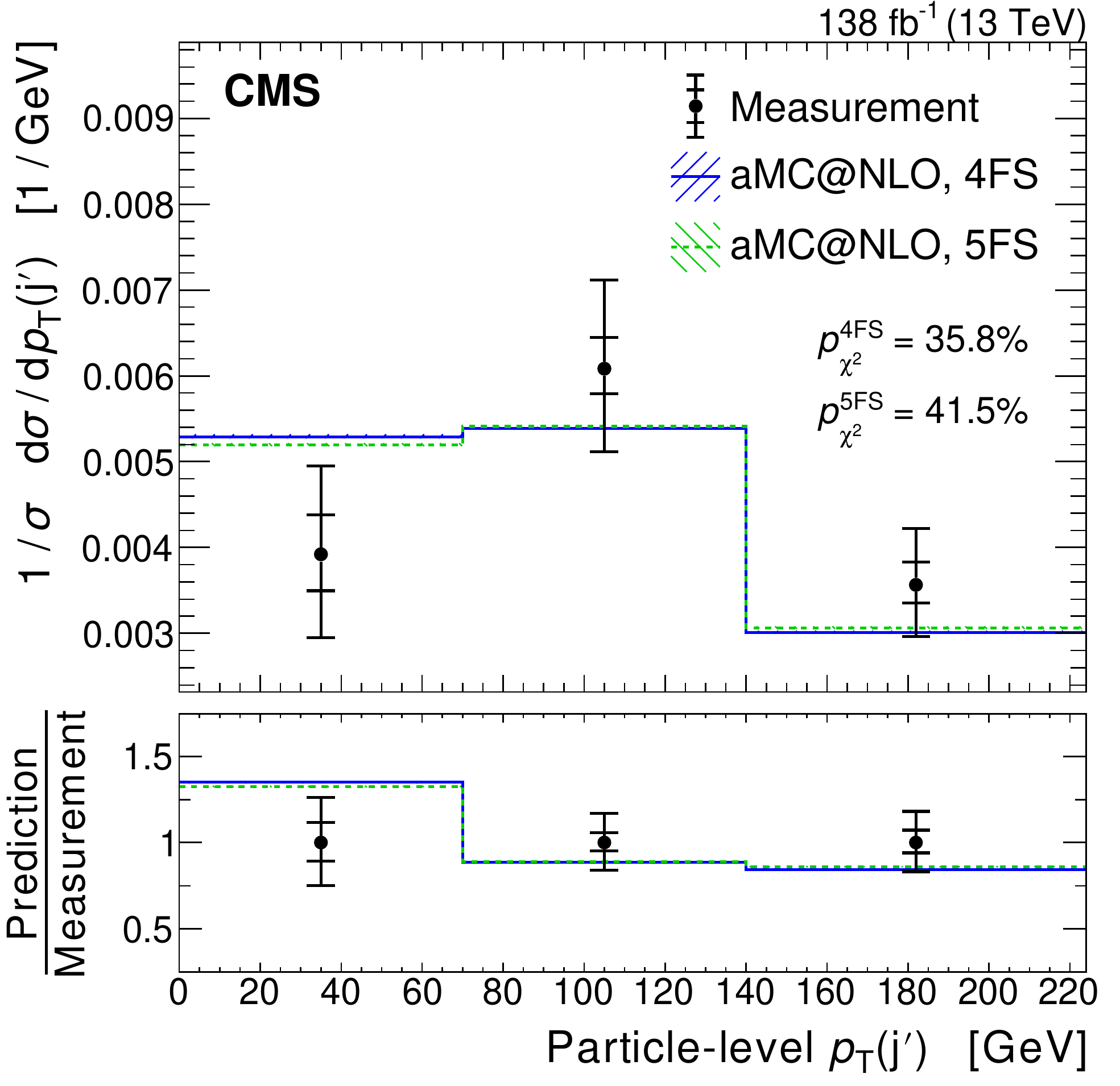}
  \includegraphics[width=0.45\textwidth]{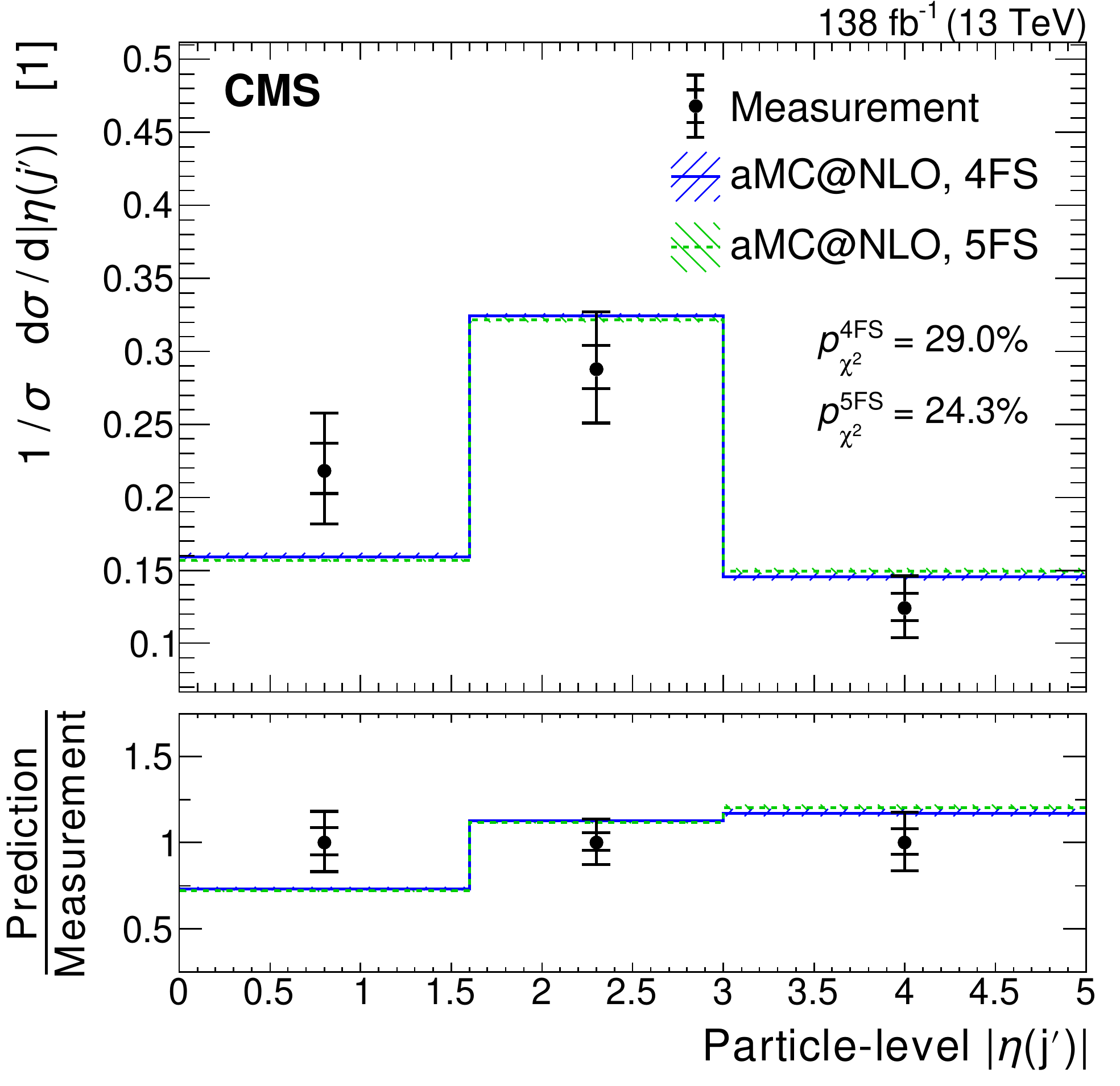}
\caption{Normalized differential cross sections measured as a function of \zpt
at the parton (upper left) and particle level (upper right), as well as a function of \jprimpt (lower left)
and \jprimeta (lower right) at the particle level.
The observed values are shown as black points, with the inner and outer vertical bars giving the systematic and total uncertainties, respectively.
The SM predictions for the \tZq process are based on
events simulated in the 5FS (green) and 4FS (blue). The $p$-values of the \chit tests
are given to quantify their compatibility with the measurement.
The lower panels show the ratio of the simulation to the measurement.}
\label{fig:result_norm0}
\end{figure}

\begin{figure}[p!]
  \centering
  \includegraphics[width=0.45\textwidth]{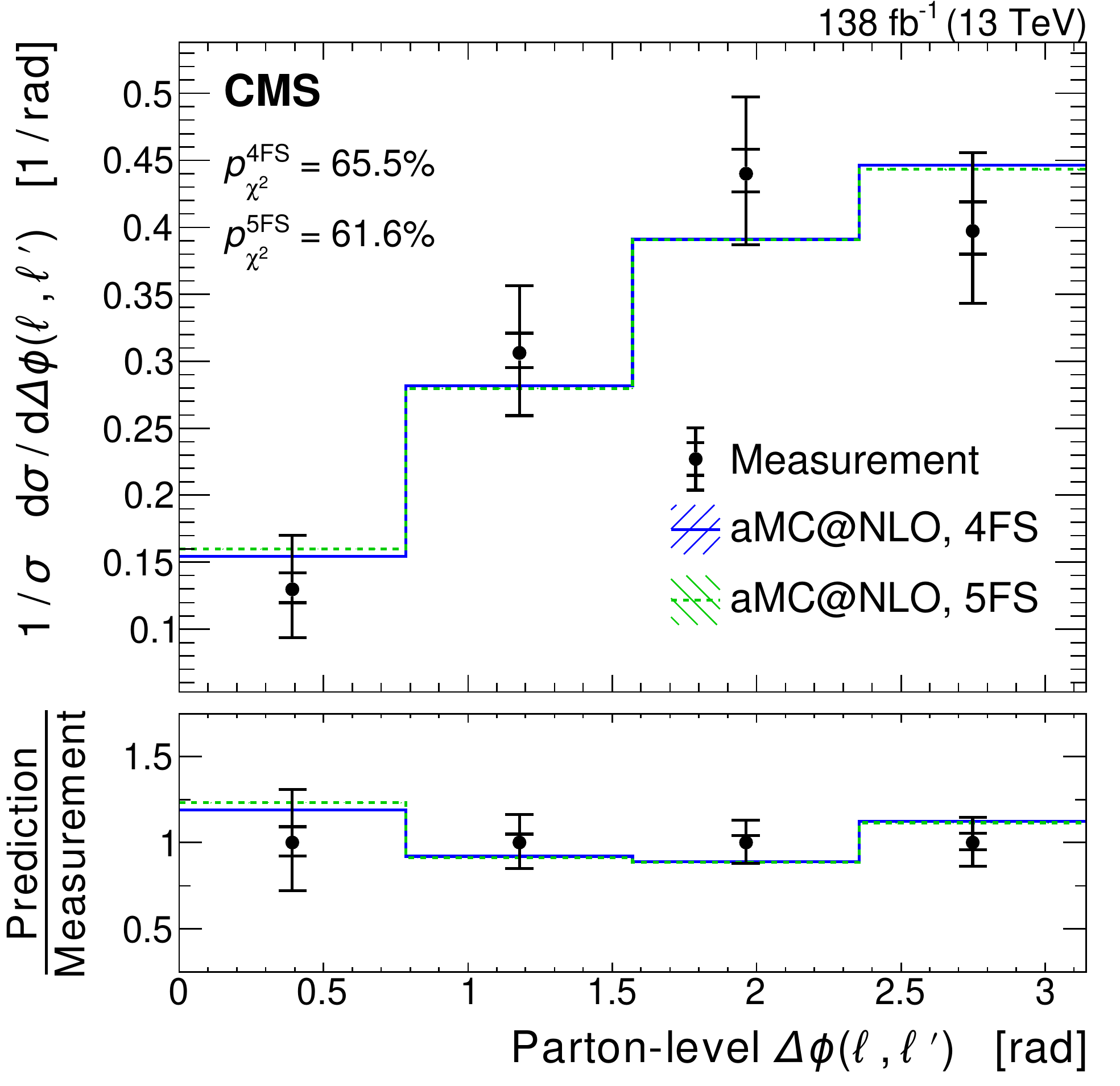}
  \includegraphics[width=0.45\textwidth]{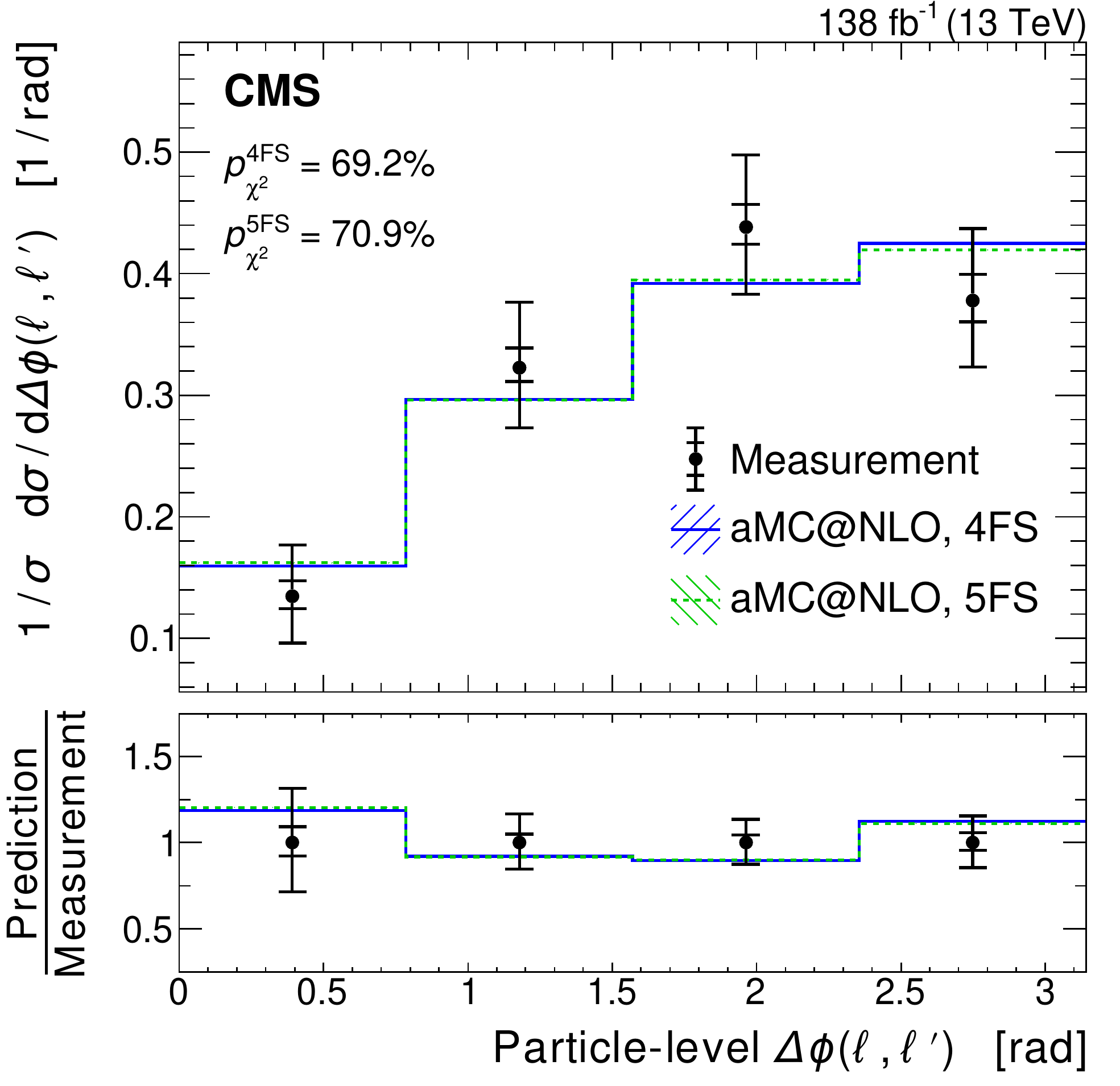}\\
  \includegraphics[width=0.45\textwidth]{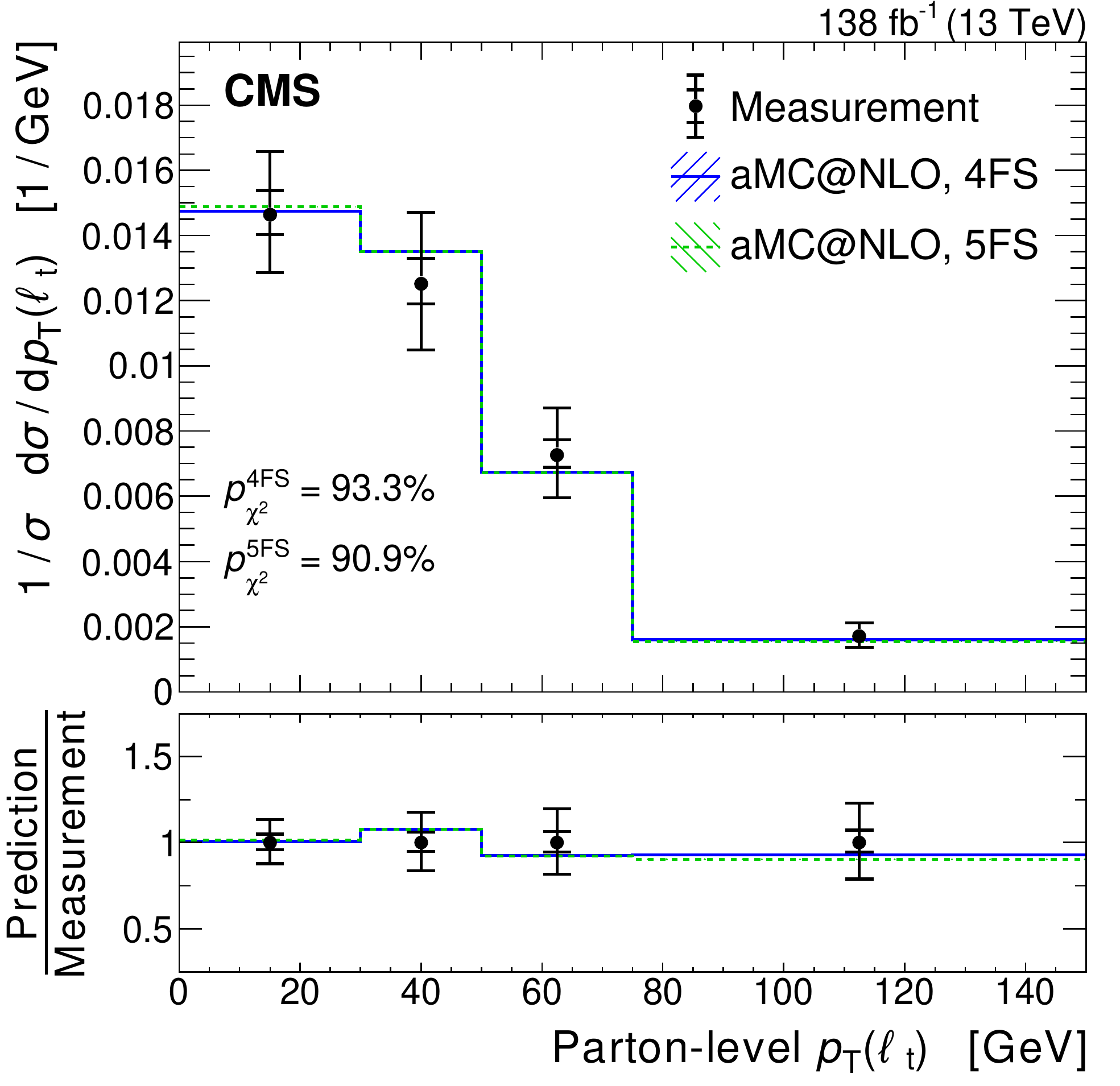}
  \includegraphics[width=0.45\textwidth]{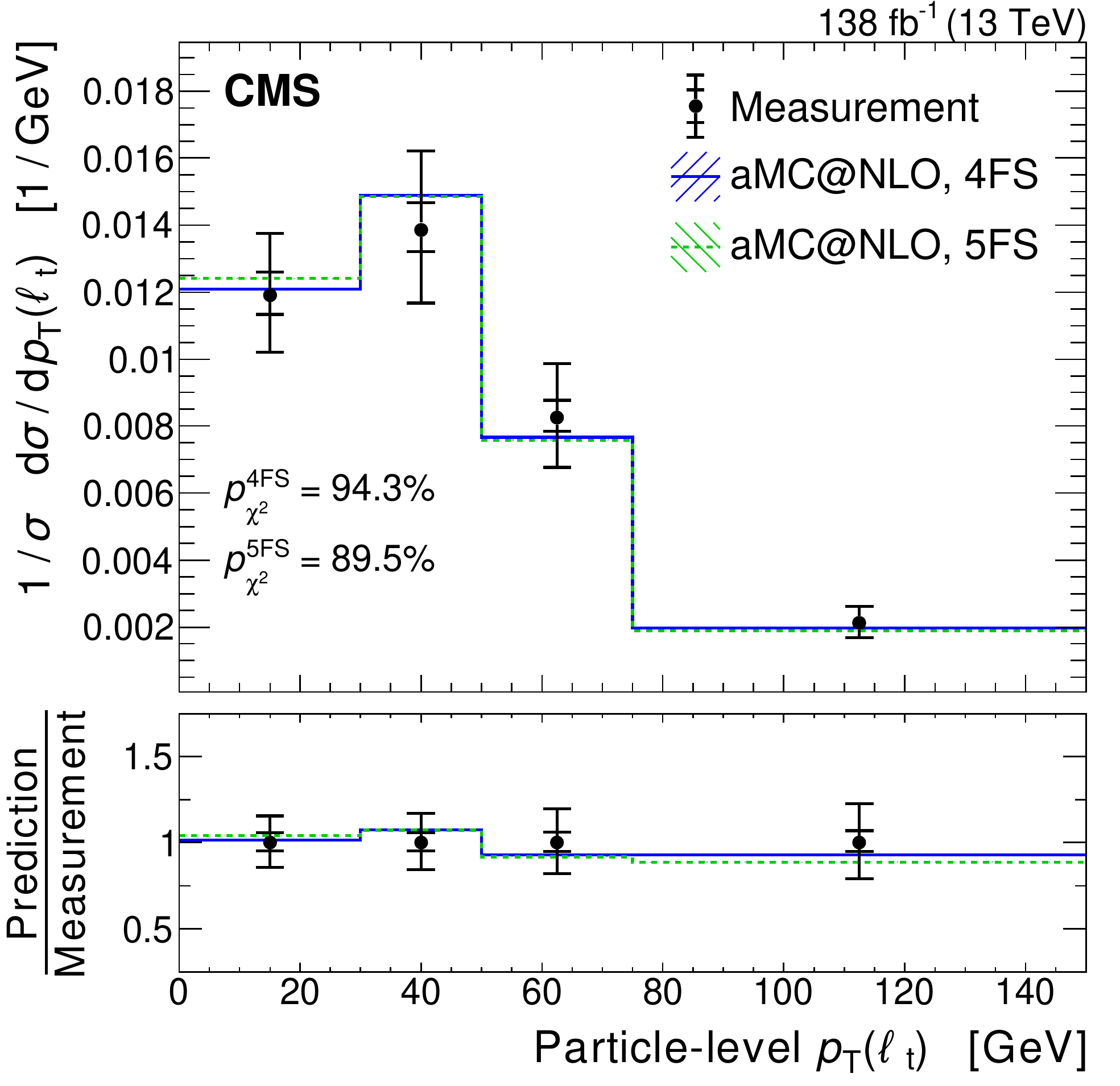}\\
  \includegraphics[width=0.45\textwidth]{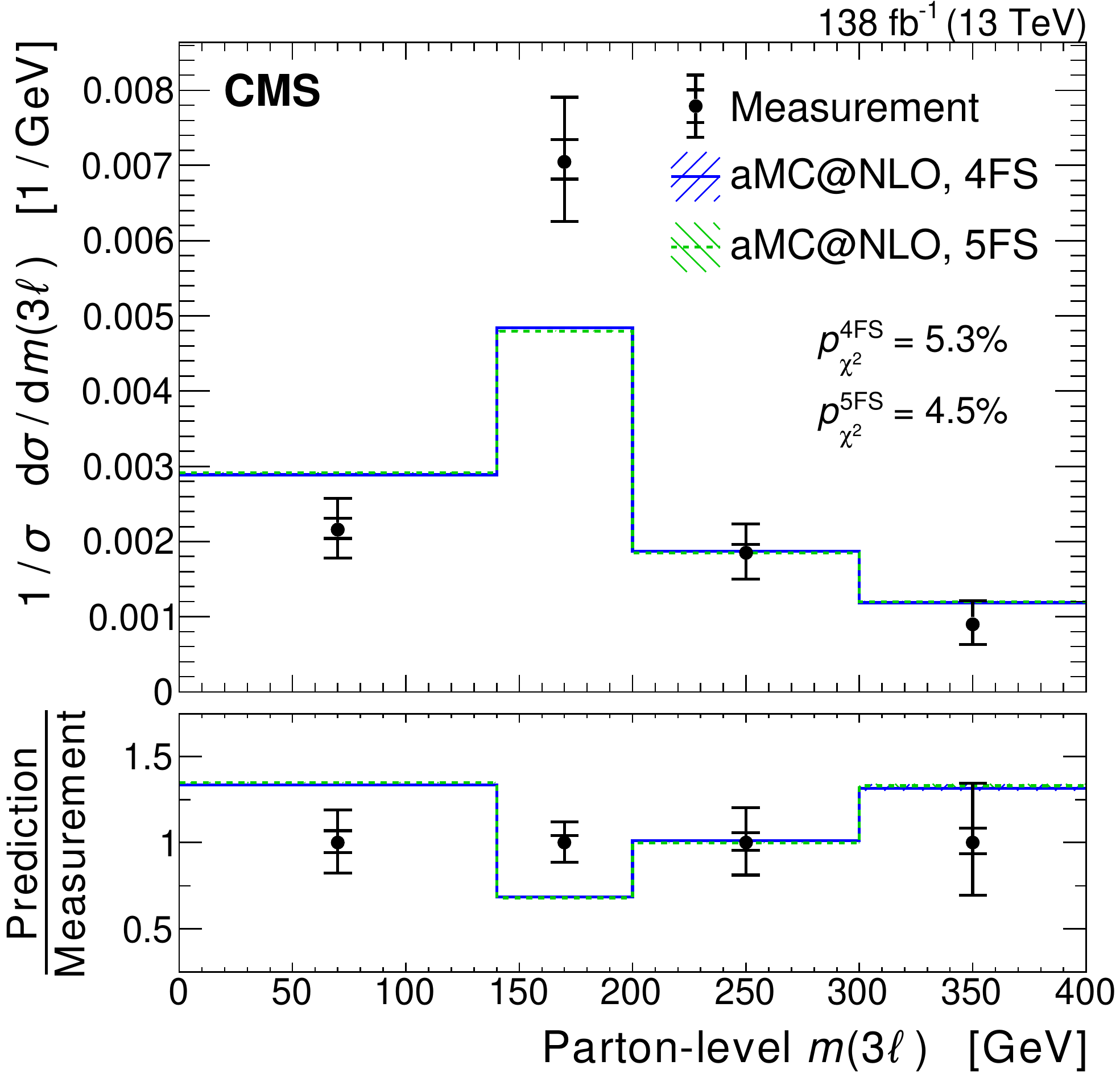}
  \includegraphics[width=0.45\textwidth]{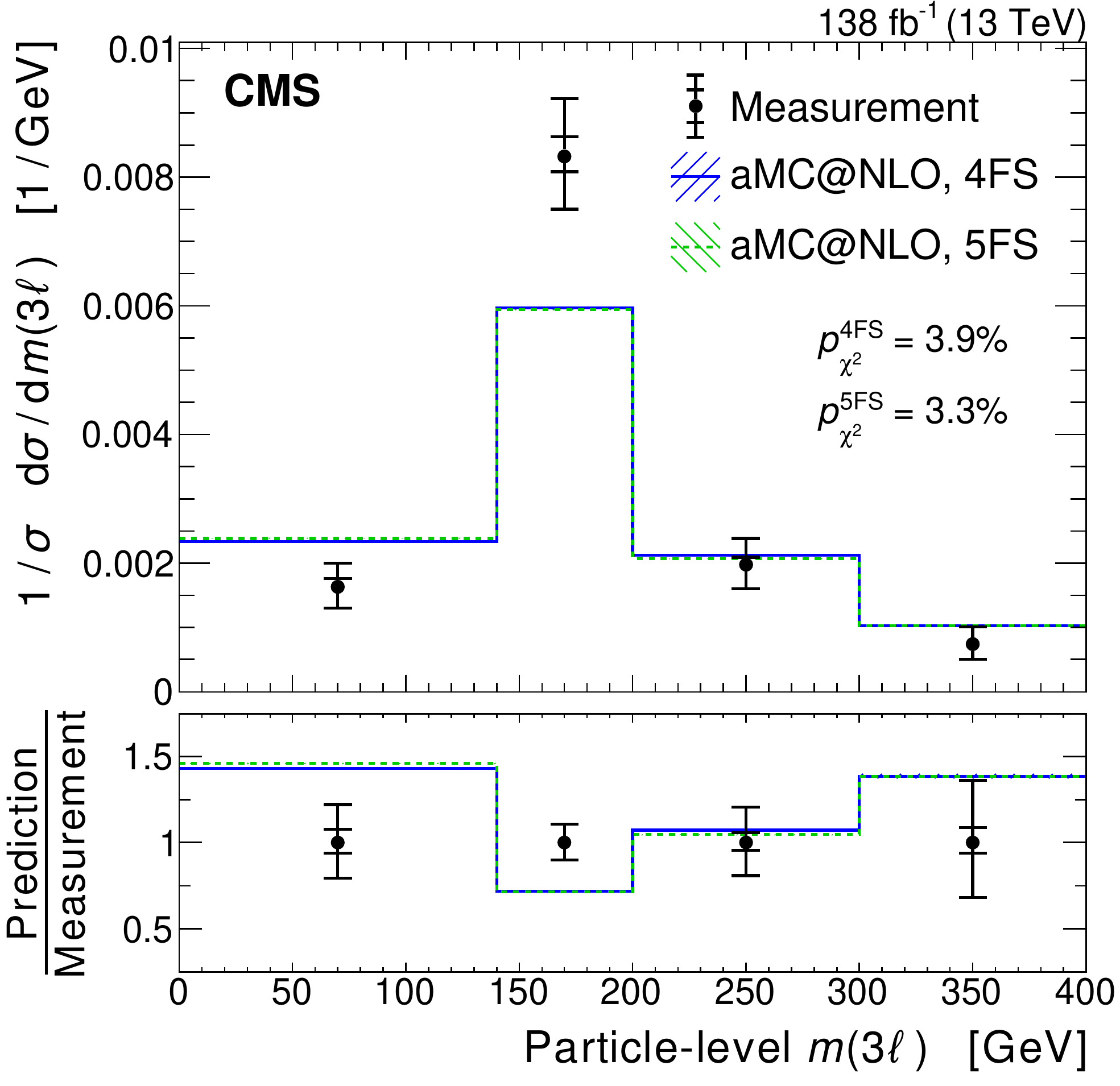}
\caption{Normalized differential cross sections measured at the parton (left)
and particle level (right) as a function of \delphill (upper), \topleppt (middle)
and \mlll (lower).
The observed values are shown as black points, with the inner and outer vertical bars giving the systematic and total uncertainties, respectively.
The SM predictions for the \tZq process are based on
events simulated in the 5FS (green) and 4FS (blue). The $p$-values of the \chit tests
are given to quantify their compatibility with the measurement.
The lower panels show the ratio of the simulation to the measurement.}
\label{fig:result_norm1}
\end{figure}

\begin{figure}[p!]
  \centering
  \includegraphics[width=0.45\textwidth]{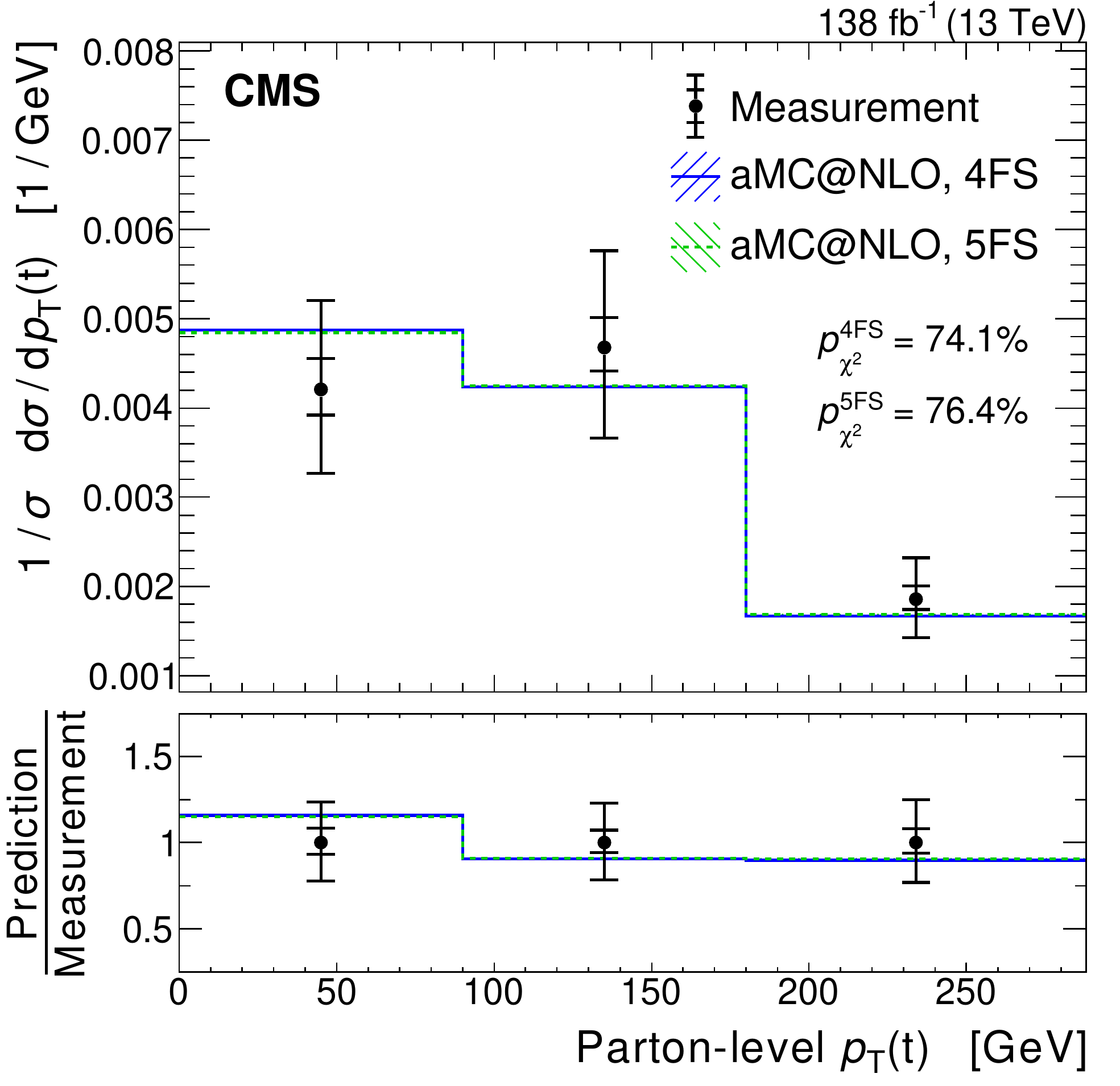}
  \includegraphics[width=0.45\textwidth]{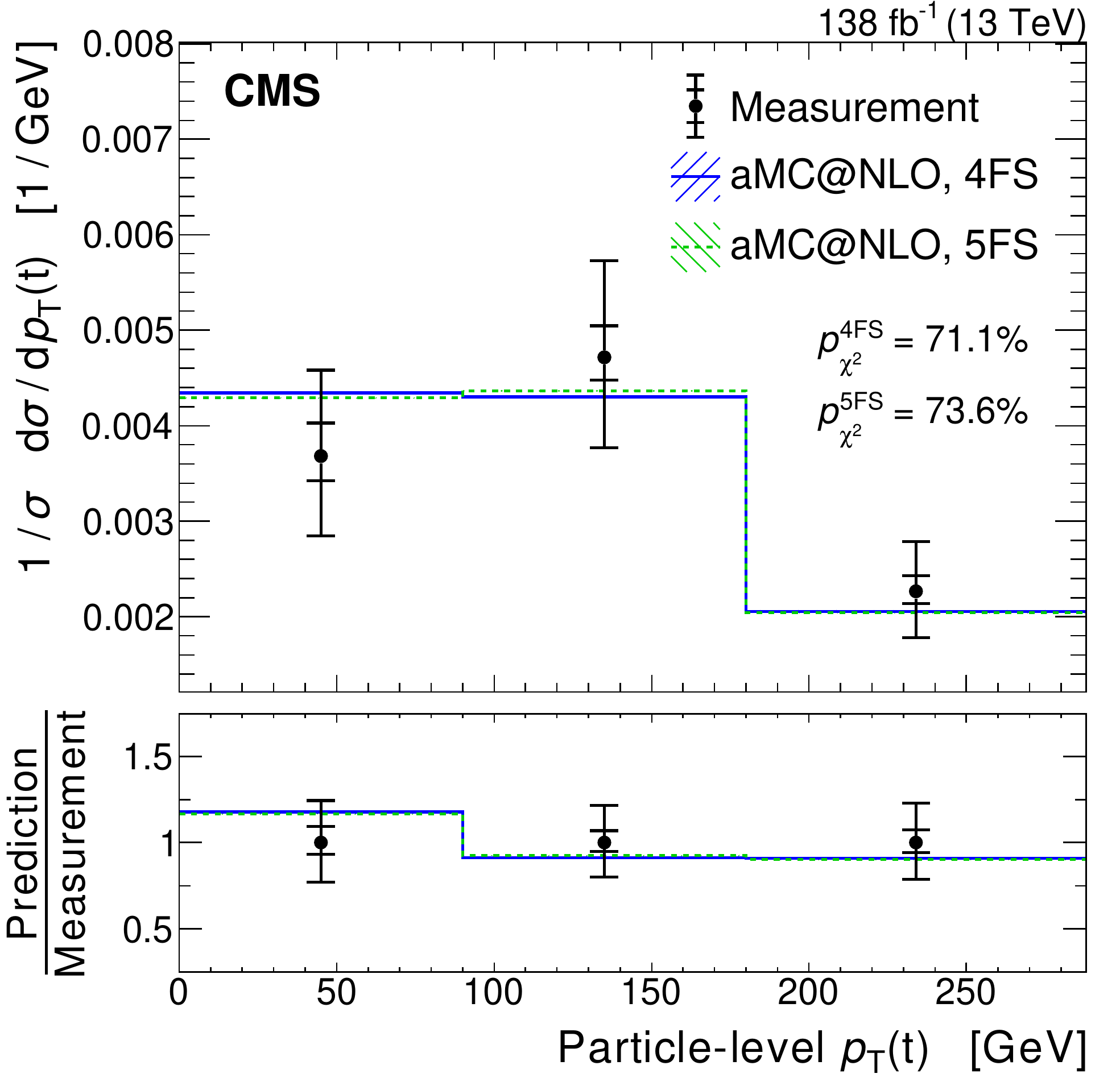}\\
  \includegraphics[width=0.45\textwidth]{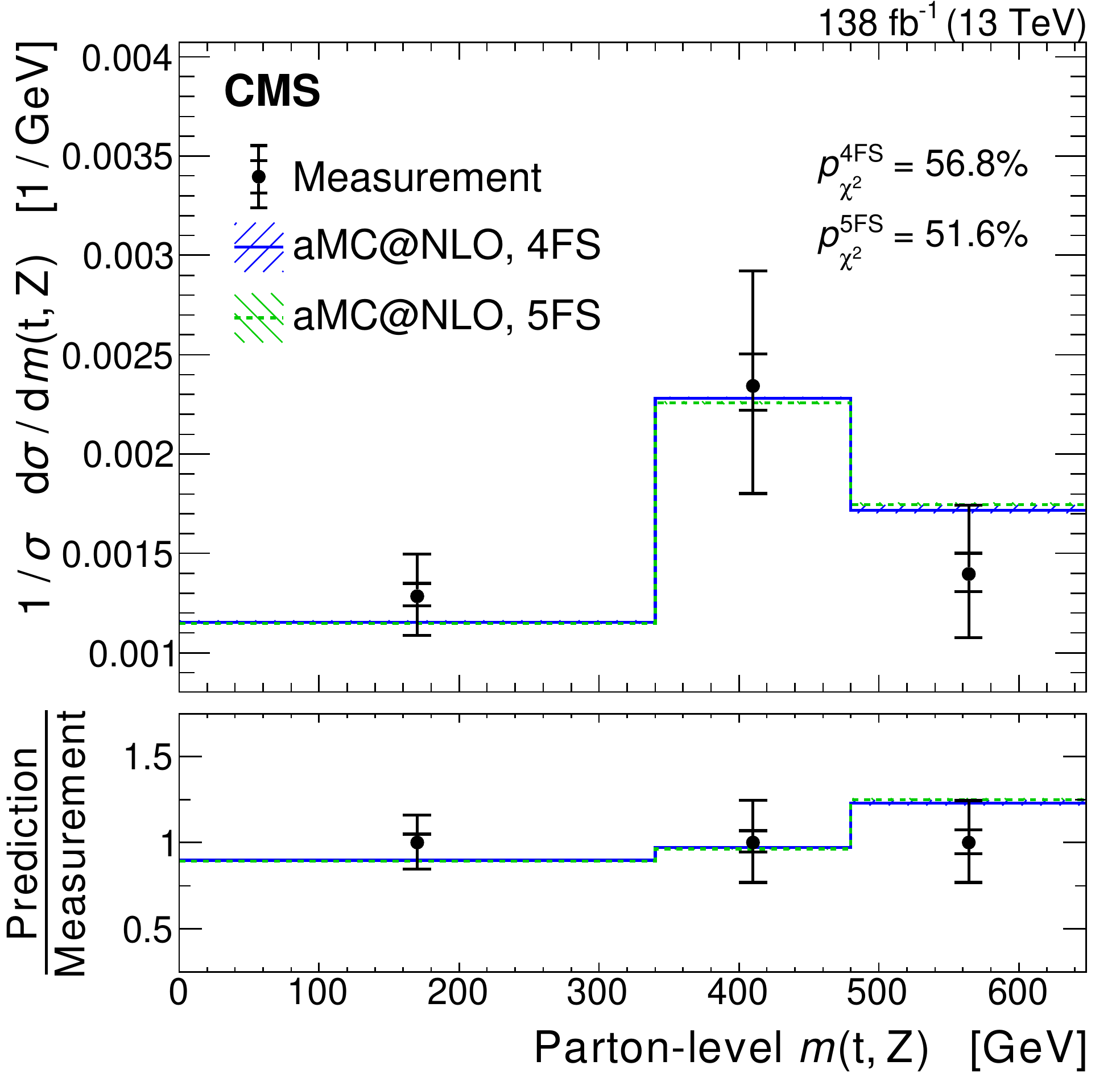}
  \includegraphics[width=0.45\textwidth]{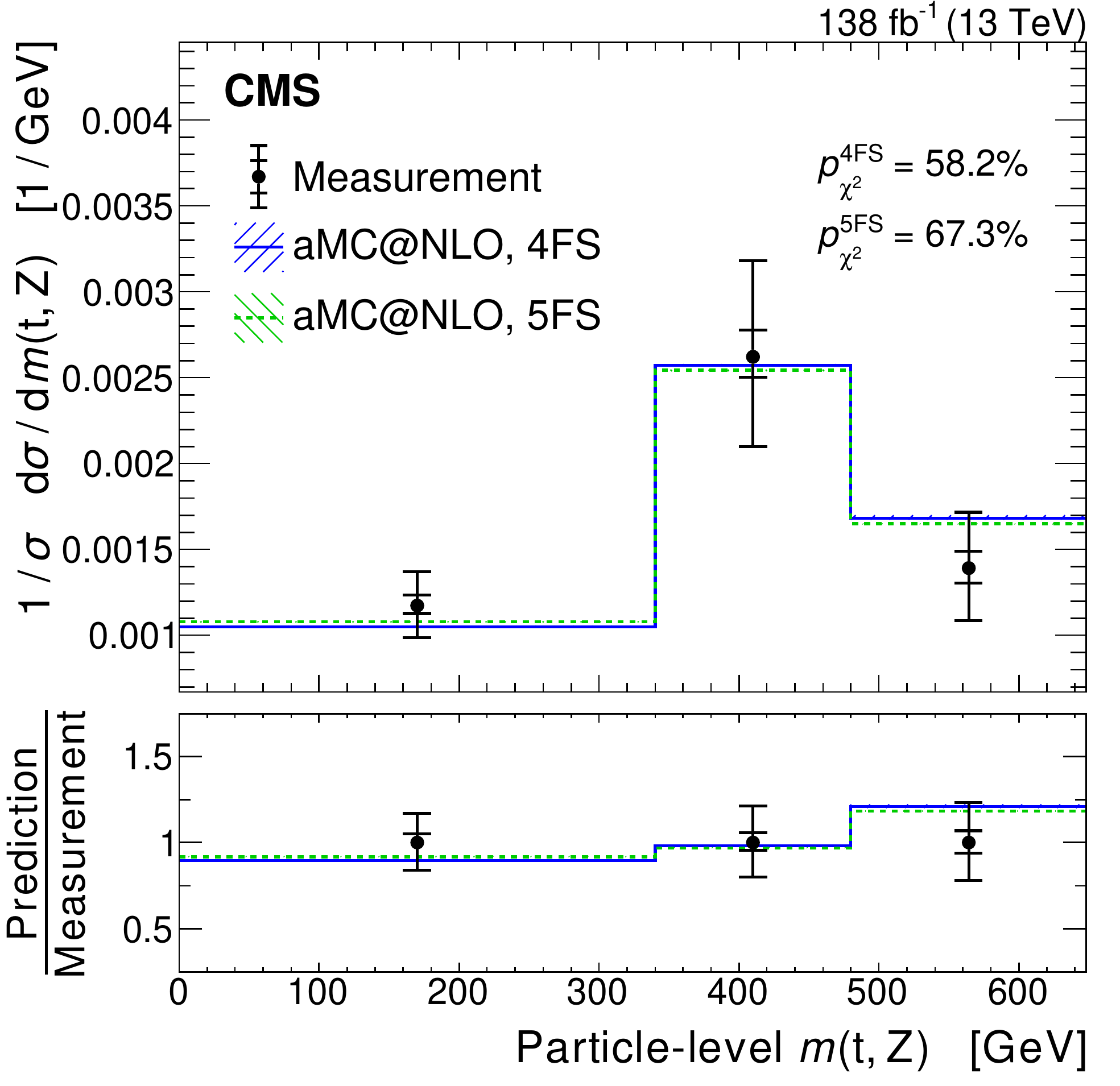}\\
  \includegraphics[width=0.45\textwidth]{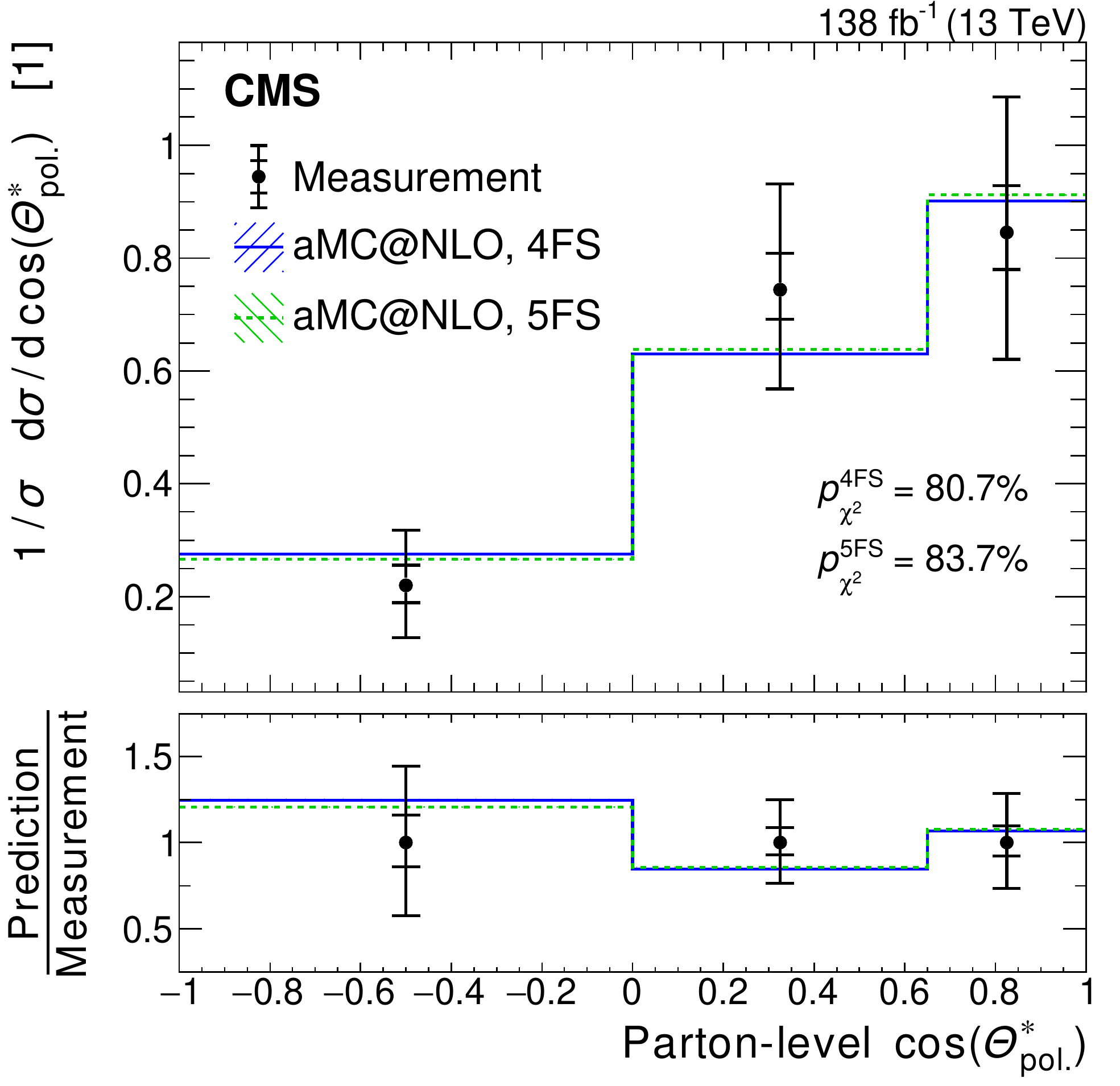}
  \includegraphics[width=0.45\textwidth]{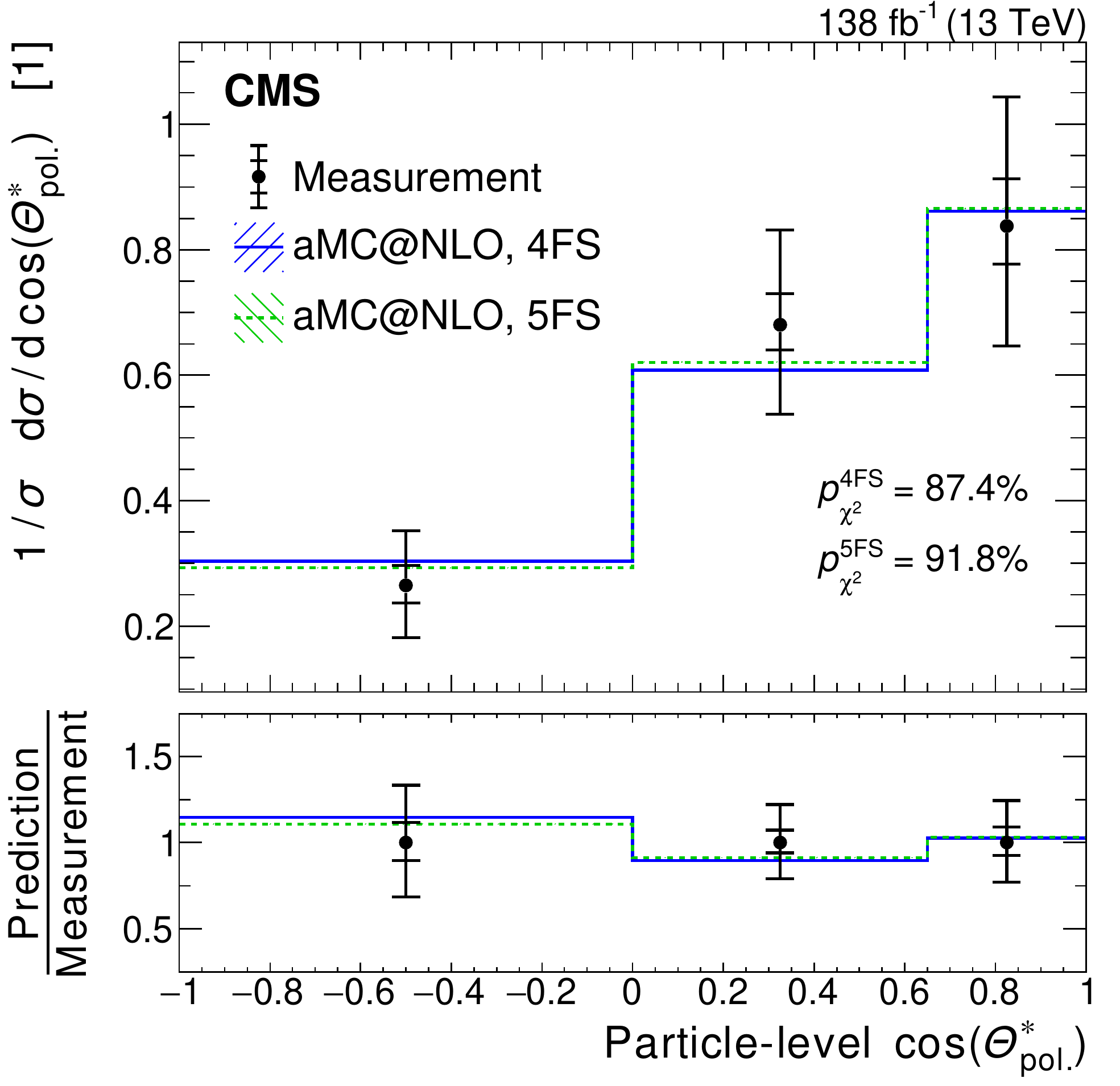}
\caption{Normalized differential cross sections measured at the parton (left) and
particle level (right) as a function of \toppt (upper), \mtz (middle) and \costheta
(lower).
The observed values are shown as black points, with the inner and outer vertical bars giving the systematic and total uncertainties, respectively.
The SM predictions for the \tZq process are based on
events simulated in the 5FS (green) and 4FS (blue). The $p$-values of the \chit tests
are given to quantify their compatibility with the measurement.
The lower panels show the ratio of the simulation to the measurement.}
\label{fig:result_norm2}
\end{figure}

The full covariance matrix is obtained from the fit and the normalized differential cross sections, shown in Figs.~\ref{fig:result_norm0}--\ref{fig:result_norm2}, are calculated by dividing each absolute differential cross section value by the sum of the values from all the bins.
In general, observables that include quarks in their definition are measured to a precision of around 20--30\% in each bin. For observables that are associated with leptons only the uncertainty goes down to 15\% in some bins.
As a cross-check, the differential cross section of each distribution
is integrated, extrapolated, and compared to the result of the inclusive measurement.
In all cases, the results are in good agreement within
the associated uncertainties.

The measured distributions are compared with theoretical predictions of \tZq
at NLO in QCD for the 4FS and 5FS. 
Uncertainties in these predictions include the effects from
the ME factorization and renormalization scales, PDF, ISR, and FSR, as discussed before. 
For the normalization of the 4FS a cross section of \THxsecFourF is used, as obtained from the \MGvATNLO generator.

Although the 5FS predicts larger cross sections~\cite{Maltoni_2012, Lim_2016} as compared with the 4FS, the absolute differential cross sections measurements are compatible with both calculations within the uncertainties.
The two methods yield similar results for the normalized differential cross sections, both of which are compatible with the measurement.
The only exception is the normalized  \mlll differential cross section, shown in the lower plots of Fig.~\ref{fig:result_norm1}, where neither scheme is able to describe the measurement around 175\GeV.
The level of agreement between the unfolded measurements and the theoretical predictions is quantified using $p$-values from a \chit test, summarized in Table~\ref{tab:chi2}, where the full covariance matrix for the measurement and each prediction is considered.

In the measurement of the spin asymmetry, the fit of \costheta at the parton level is reparameterized according to Eq.~(\ref{eq:spin_asymmetry}) such that
the spin asymmetry is directly used as a free parameter in the fit.
Apart from the spin asymmetry, the total cross section is left freely floating in the fit as well, to account for the overall normalization. The spin asymmetry is measured as
\begin{equation*}
    \asym = \Avalue,
\end{equation*}

compatible with SM predictions of $\asym^\mathrm{4FS} = 0.44$ (in the 4FS)
and $\asym^\mathrm{5FS} = 0.45$ (in the 5FS) from \MGvATNLO simulations at NLO. 
Uncertainties in the SM predictions from ISR, FSR, PDFs, and renormalization and factorization scales at ME level were found to be negligible with respect to those of the measured value.
Additional prefit and postfit results, the extracted distribution, and the likelihood as a function of the spin asymmetry are given in Appendix~\ref{sec:appendix3}.

The uncertainties in both the differential cross section and spin asymmetry
measurements are dominated by the statistical component. The
leading systematic uncertainties come from the experimental uncertainties,
including the background modeling, \PQb tagging efficiency,
and lepton identification.
For measurements using observables that include quarks in their definition,
the jet energy scale
also represents an important source of systematic uncertainty.
The leading theoretical uncertainties are associated with the renormalization and factorization scales at ME level, and FSR effects.

\section{Summary}
\label{sec:summary}
\begin{figure}[htb!]
	\centering
	\includegraphics[width=0.55\textwidth]{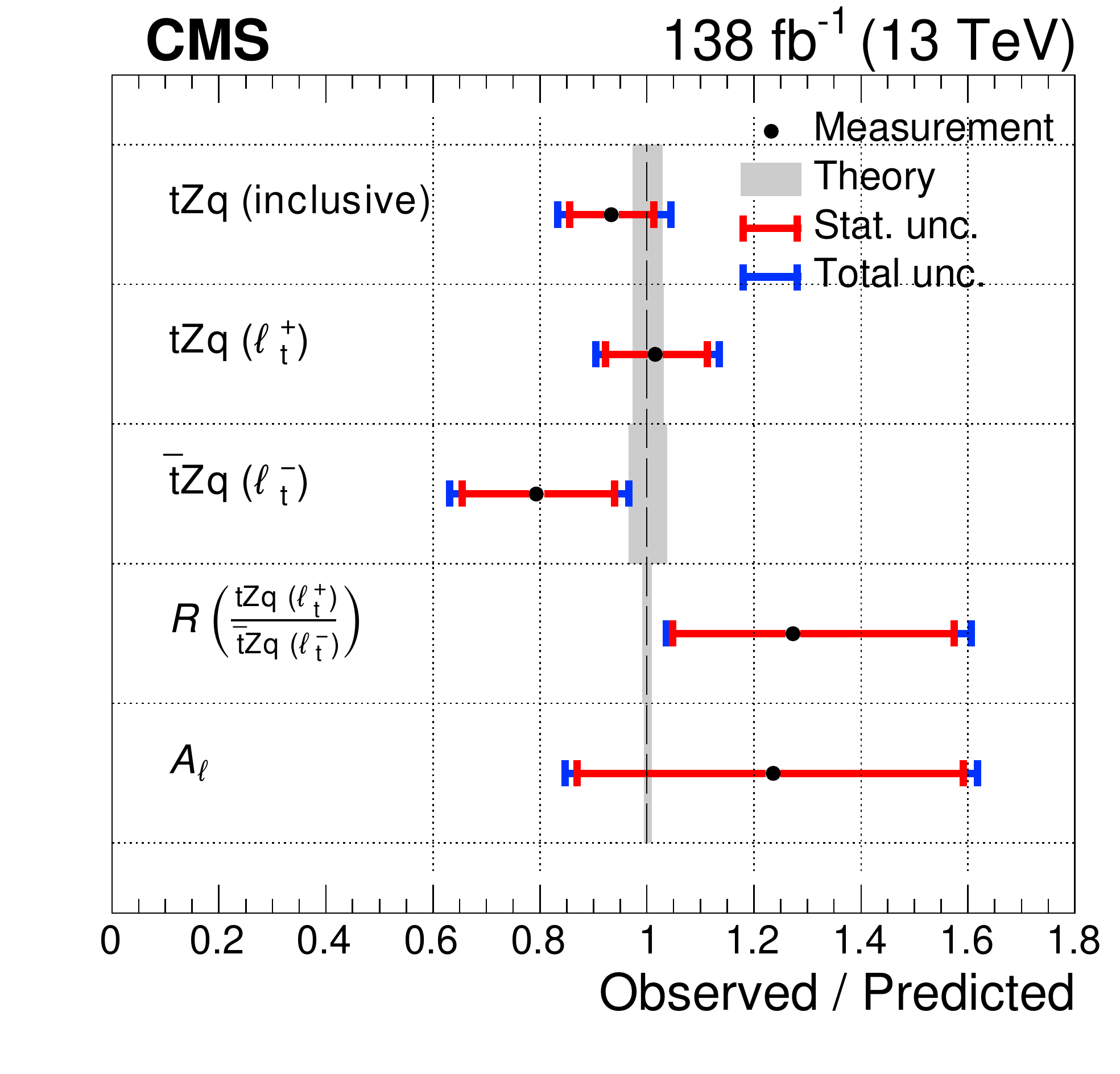}
	\caption{Measured values of the inclusive \tZq cross section (first row),
    top quark and antiquark cross sections (second and third rows)
    and their ratio (fourth row), together with
    the top quark spin asymmetry in the \tZq process (last row).
    Each row shows the ratio between the observed and the SM predicted values.
    The black points show the central values, while the red and blue bars refer to the statistical and total uncertainties in the measurements, respectively.
	Uncertainties in the predictions are indicated by the \grey bands.}
	\label{fig:channels}
\end{figure}

Inclusive and differential cross section measurements of single top quark production in
association with a \PZ boson (\tZq) are presented using events with
three leptons (electrons or muons).
The data sample for this measurement was collected by the CMS experiment at the LHC in proton-proton collisions at a \centre-of-mass energy of 13\TeV and corresponds to an integrated luminosity of \il.
Including nonresonant lepton pairs,
an inclusive cross section of $ \stZq = \EXxsec $ is obtained for dilepton invariant masses greater than 30\GeV.
This result is the most precise inclusive \tZq cross section
measurement to date, with a relative precision about 25\%
better than previously published results.
For the first time, the inclusive \tZq cross sections are also measured separately
for top quark and antiquark production, obtaining $ \sigmatop = \EXxsecTZ $ and
$ \sigmaantitop = \EXxsecTbarZ $, respectively, with the ratio of \Rvalue.
The measured values compared to the theoretical predictions are summarized in Fig.~\ref{fig:channels}.

The differential \tZq cross sections are measured for the first time at the parton
and particle levels using a binned maximum-likelihood-based unfolding.
The studied observables are the transverse momenta of the top quark, the \PZ boson, and the lepton
associated with the top quark decay, as well as the invariant masses of the three leptons
and the \PZSys system.
Also used as observables are the difference in azimuthal angle between the two
leptons from the \PZ boson decay, the cosine of the top quark polarization angle, and,
at the particle level, the transverse momentum and absolute pseudorapidity of the recoiling jet.
The results are mostly compatible with the standard model predictions using both the
four- and five-flavor schemes, while the sensitivity is not sufficient
to show a preference for one or the other.
From the differential distribution of the top quark polarization angle,
the top quark spin asymmetry is measured to be $\asym = \Avalue$,
in agreement with the standard model prediction.

\begin{acknowledgments}
  We congratulate our colleagues in the CERN accelerator departments for the excellent performance of the LHC and thank the technical and administrative staffs at CERN and at other CMS institutes for their contributions to the success of the CMS effort. In addition, we gratefully acknowledge the computing centers and personnel of the Worldwide LHC Computing Grid and other centers for delivering so effectively the computing infrastructure essential to our analyses. Finally, we acknowledge the enduring support for the construction and operation of the LHC, the CMS detector, and the supporting computing infrastructure provided by the following funding agencies: BMBWF and FWF (Austria); FNRS and FWO (Belgium); CNPq, CAPES, FAPERJ, FAPERGS, and FAPESP (Brazil); MES and BNSF (Bulgaria); CERN; CAS, MoST, and NSFC (China); MINCIENCIAS (Colombia); MSES and CSF (Croatia); RIF (Cyprus); SENESCYT (Ecuador); MoER, ERC PUT and ERDF (Estonia); Academy of Finland, MEC, and HIP (Finland); CEA and CNRS/IN2P3 (France); BMBF, DFG, and HGF (Germany); GSRI (Greece); NKFIA (Hungary); DAE and DST (India); IPM (Iran); SFI (Ireland); INFN (Italy); MSIP and NRF (Republic of Korea); MES (Latvia); LAS (Lithuania); MOE and UM (Malaysia); BUAP, CINVESTAV, CONACYT, LNS, SEP, and UASLP-FAI (Mexico); MOS (Montenegro); MBIE (New Zealand); PAEC (Pakistan); MSHE and NSC (Poland); FCT (Portugal); JINR (Dubna); MON, RosAtom, RAS, RFBR, and NRC KI (Russia); MESTD (Serbia); SEIDI, CPAN, PCTI, and FEDER (Spain); MOSTR (Sri Lanka); Swiss Funding Agencies (Switzerland); MST (Taipei); ThEPCenter, IPST, STAR, and NSTDA (Thailand); TUBITAK and TAEK (Turkey); NASU (Ukraine); STFC (United Kingdom); DOE and NSF (USA).
  
  \hyphenation{Rachada-pisek} Individuals have received support from the Marie-Curie program and the European Research Council and Horizon 2020 Grant, contract Nos.\ 675440, 724704, 752730, 758316, 765710, 824093, 884104, and COST Action CA16108 (European Union); the Leventis Foundation; the Alfred P.\ Sloan Foundation; the Alexander von Humboldt Foundation; the Belgian Federal Science Policy Office; the Fonds pour la Formation \`a la Recherche dans l'Industrie et dans l'Agriculture (FRIA-Belgium); the Agentschap voor Innovatie door Wetenschap en Technologie (IWT-Belgium); the F.R.S.-FNRS and FWO (Belgium) under the ``Excellence of Science -- EOS" -- be.h project n.\ 30820817; the Beijing Municipal Science \& Technology Commission, No. Z191100007219010; the Ministry of Education, Youth and Sports (MEYS) of the Czech Republic; the Deutsche Forschungsgemeinschaft (DFG), under Germany's Excellence Strategy -- EXC 2121 ``Quantum Universe" -- 390833306, and under project number 400140256 - GRK2497; the Lend\"ulet (``Momentum") Program and the J\'anos Bolyai Research Scholarship of the Hungarian Academy of Sciences, the New National Excellence Program \'UNKP, the NKFIA research grants 123842, 123959, 124845, 124850, 125105, 128713, 128786, and 129058 (Hungary); the Council of Science and Industrial Research, India; the Latvian Council of Science; the Ministry of Science and Higher Education and the National Science Center, contracts Opus 2014/15/B/ST2/03998 and 2015/19/B/ST2/02861 (Poland); the Funda\c{c}\~ao para a Ci\^encia e a Tecnologia, grant CEECIND/01334/2018 (Portugal); the National Priorities Research Program by Qatar National Research Fund; the Ministry of Science and Higher Education, projects no. 14.W03.31.0026 and no. FSWW-2020-0008, and the Russian Foundation for Basic Research, project No.19-42-703014 (Russia); the Programa Estatal de Fomento de la Investigaci{\'o}n Cient{\'i}fica y T{\'e}cnica de Excelencia Mar\'{\i}a de Maeztu, grant MDM-2015-0509 and the Programa Severo Ochoa del Principado de Asturias; the Stavros Niarchos Foundation (Greece); the Rachadapisek Sompot Fund for Postdoctoral Fellowship, Chulalongkorn University and the Chulalongkorn Academic into Its 2nd Century Project Advancement Project (Thailand); the Kavli Foundation; the Nvidia Corporation; the SuperMicro Corporation; the Welch Foundation, contract C-1845; and the Weston Havens Foundation (USA).\end{acknowledgments}

\bibliography{auto_generated}

\numberwithin{figure}{section}
\numberwithin{table}{section}
\appendix

\section{Validation of the misidentification-rate method in simulation}
\label{sec:appendix}

\begin{figure}[htb!]
	\centering
	\includegraphics[width=0.45\textwidth]{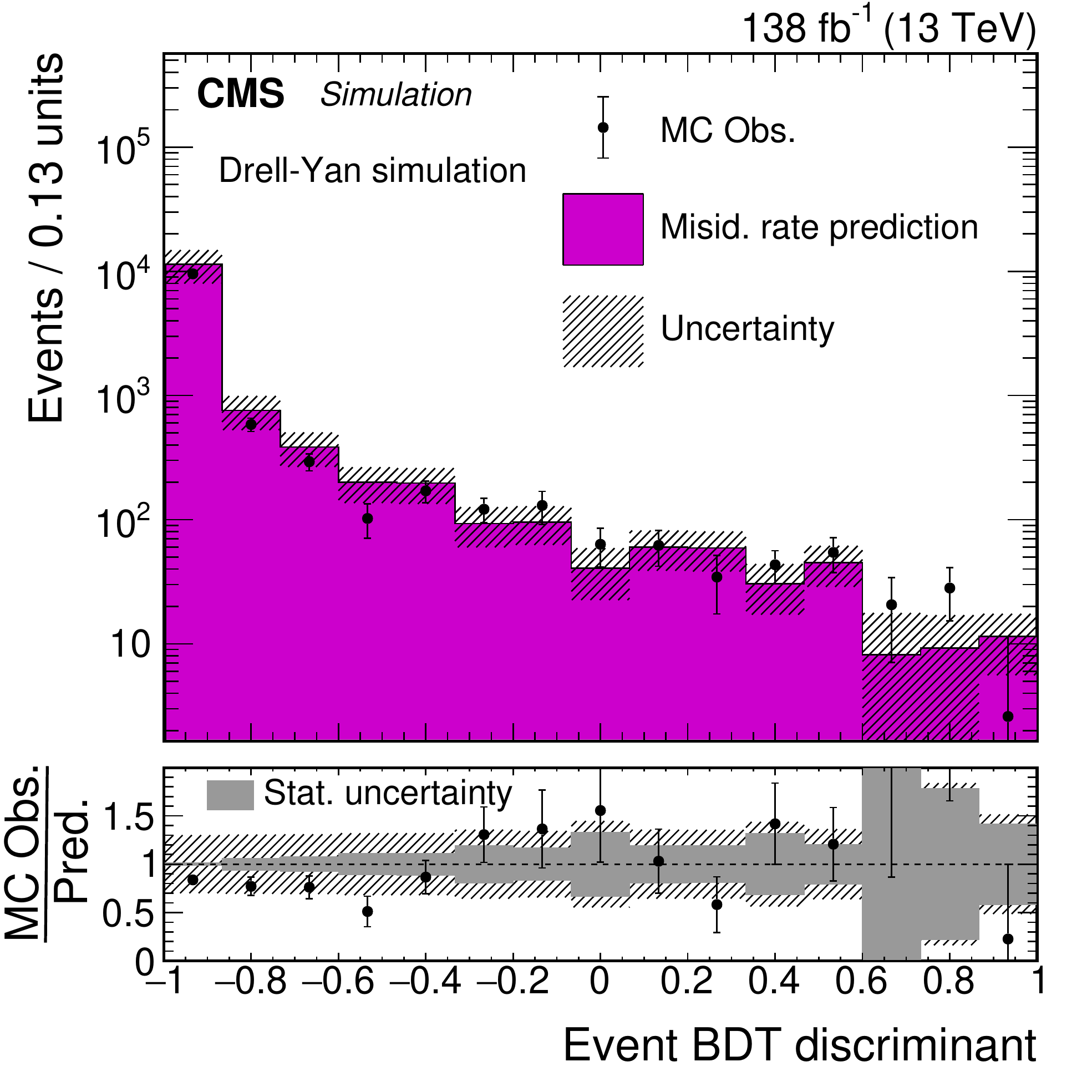}
	\includegraphics[width=0.45\textwidth]{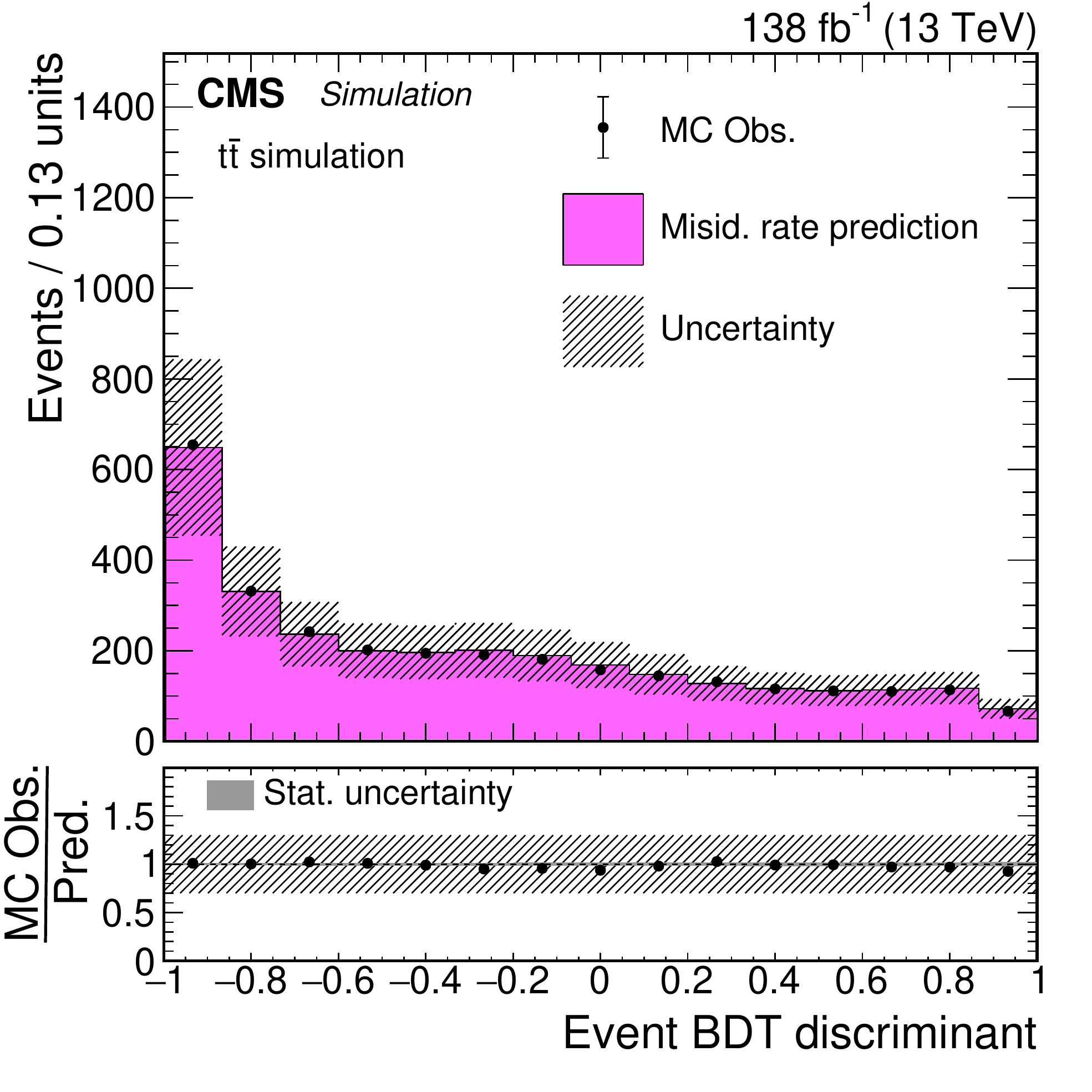}
	\caption{
    The event BDT discriminant distributions for simulated (left) trilepton DY events and (right) \ttbar events.
    The black points give the nonprompt-lepton predictions from the simulation, while the \coloured histograms represent a similar prediction estimated with the misidentification rate method applied to the same events.
    The lower panels plot the ratio of the observed MC prediction to the prediction from the misidentification rate method.
    The vertical bars on the points show the statistical uncertainty in the observed MC distribution.
    The hatched bands represent the total uncertainty in the misidentification rate predictions and the shaded band its statistical component in the ratio.
    }
	\label{fig:closuretests}
\end{figure}

Figure~\ref{fig:closuretests} shows the results of the misidentification-rate optimization and validation using
the tight-to-loose ratio method, as described in Section~\ref{sec:backgrounds}.
This cross-check is fully simulation-based: the misidentification rate is measured in simulated QCD multijet samples and evaluated separately in simulated DY and \ttbar samples.
The event BDT discriminant distributions are shown for (left) simulated trilepton events from the DY process, and (right) simulated \ttbar events.
The black points are the nonprompt-lepton predictions from the simulation, and the colored histograms the predictions using the tight-to-loose ratio method applied to the same simulated events. The lower panels give the ratio of the two predictions.
The good agreement between the two predictions for the two different types of simulated events validates the method for determining the nonprompt-lepton background.

\section{Hypothesis tests of differential cross sections}
\label{sec:appendix2}
The level of agreement of the measured differential cross sections with the theory predictions is quantified with $p$-values of \chit tests as summarized in Table~\ref{tab:chi2}. The full covariance matrix for the measurement and each prediction is considered.

\begin{table}[!hbt]
  \centering
  \topcaption{
  Summary of the $p$-values from the \chit test between the unfolded measurements and theoretical predictions from the 4FS and 5FS.
  The test is performed on the measurements of the absolute and normalized differential cross sections at the parton and particle levels
  for the observables given in the first column. All numbers are given in percent.
  }
  \label{tab:chi2}
  \begin{tabular}{lcccccccc}
    \multirow{3}{*}{Observable} & \multicolumn{4}{c}{Parton level}  & \multicolumn{4}{c}{Particle level} \\
                                & \multicolumn{2}{c}{Absolute}  & \multicolumn{2}{c}{Normalized} & \multicolumn{2}{c}{Absolute}  & \multicolumn{2}{c}{Normalized}\\
                                & 4FS   & 5FS   & 4FS   & 5FS   & 4FS   & 5FS   & 4FS   & 5FS \\    [\cmsTabSkip]
    \zpt                        & 97.7  & 84.6  & 99.8  & 99.3  & 97.7  & 89.1  & 99.8  & 99.8  \\
    \delphill                   & 73.1  & 54.1  & 65.5  & 61.6  & 75.7  & 65.6  & 69.2  & 70.9  \\
    \topleppt                   & 94.5  & 74.4  & 93.3  & 90.9  & 95.0  & 77.1  & 94.3  & 89.5  \\
    \mlll                       & 7.0   & 2.0   & 5.4   & 4.5   & 7.3   & 2.2   & 3.9   & 3.3   \\
    \toppt                      & 74.4  & 68.2  & 74.1  & 76.4  & 72.8  & 70.5  & 71.1  & 73.6  \\
    \mtz                        & 65.0  & 48.3  & 56.8  & 51.6  & 66.3  & 64.0  & 58.2  & 67.3  \\
    \costheta                   & 84.6  & 63.1  & 80.7  & 83.7  & 88.2  & 71.8  & 87.4  & 91.8  \\
    \jprimpt                    & \NA     & \NA     & \NA     & \NA     & 44.2  & 44.1  & 35.8  & 41.5  \\
    \jprimeta                   & \NA     & \NA     & \NA     & \NA     & 46.2  & 30.4  & 29.0  & 24.3  \\
  \end{tabular}
\end{table}

\section{Extraction of the top quark spin asymmetry}
\label{sec:appendix3}

\begin{figure}[htbp!]
  \centering
  \includegraphics[width=0.8\textwidth]{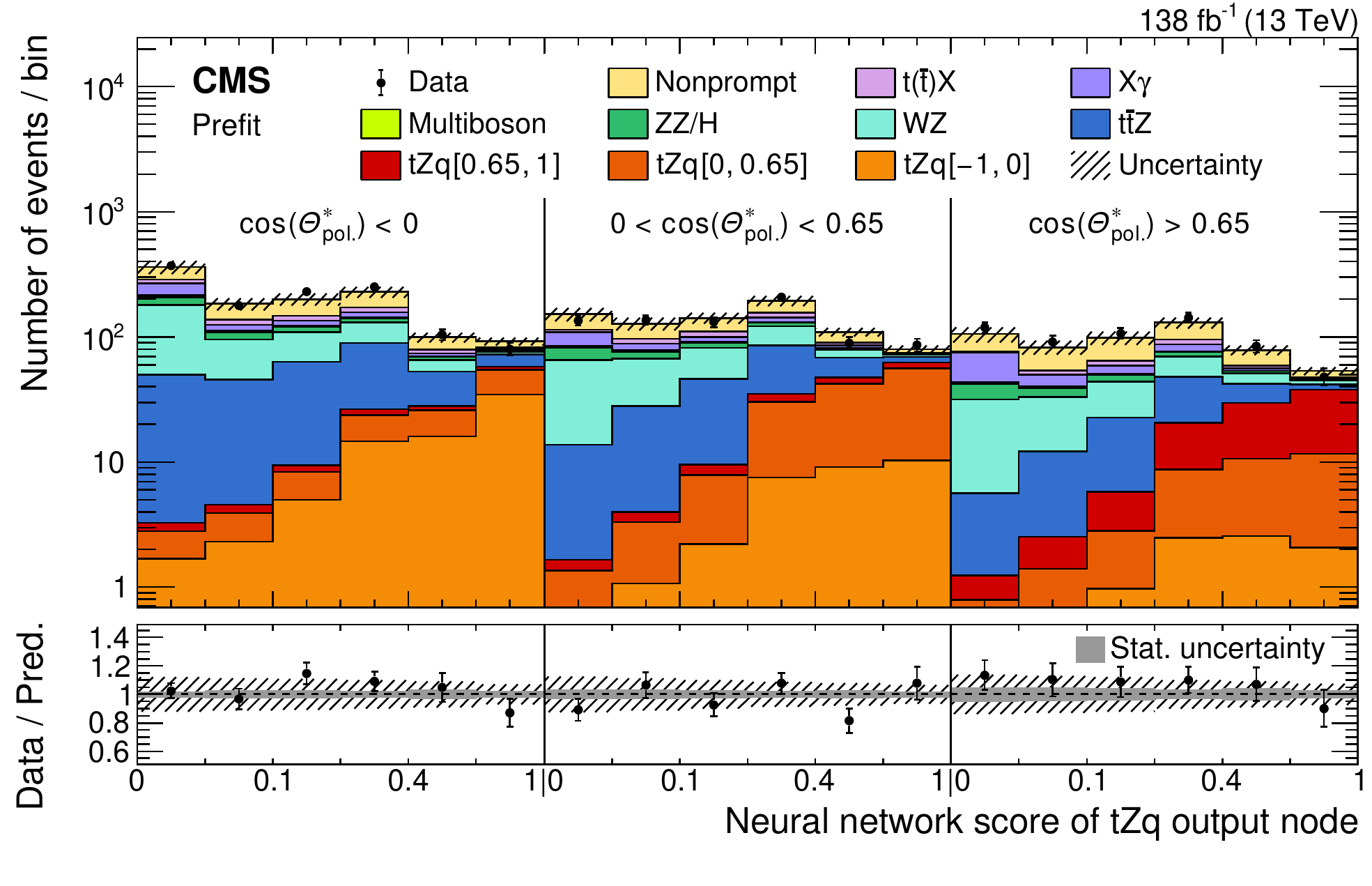}\\
  \includegraphics[width=0.8\textwidth]{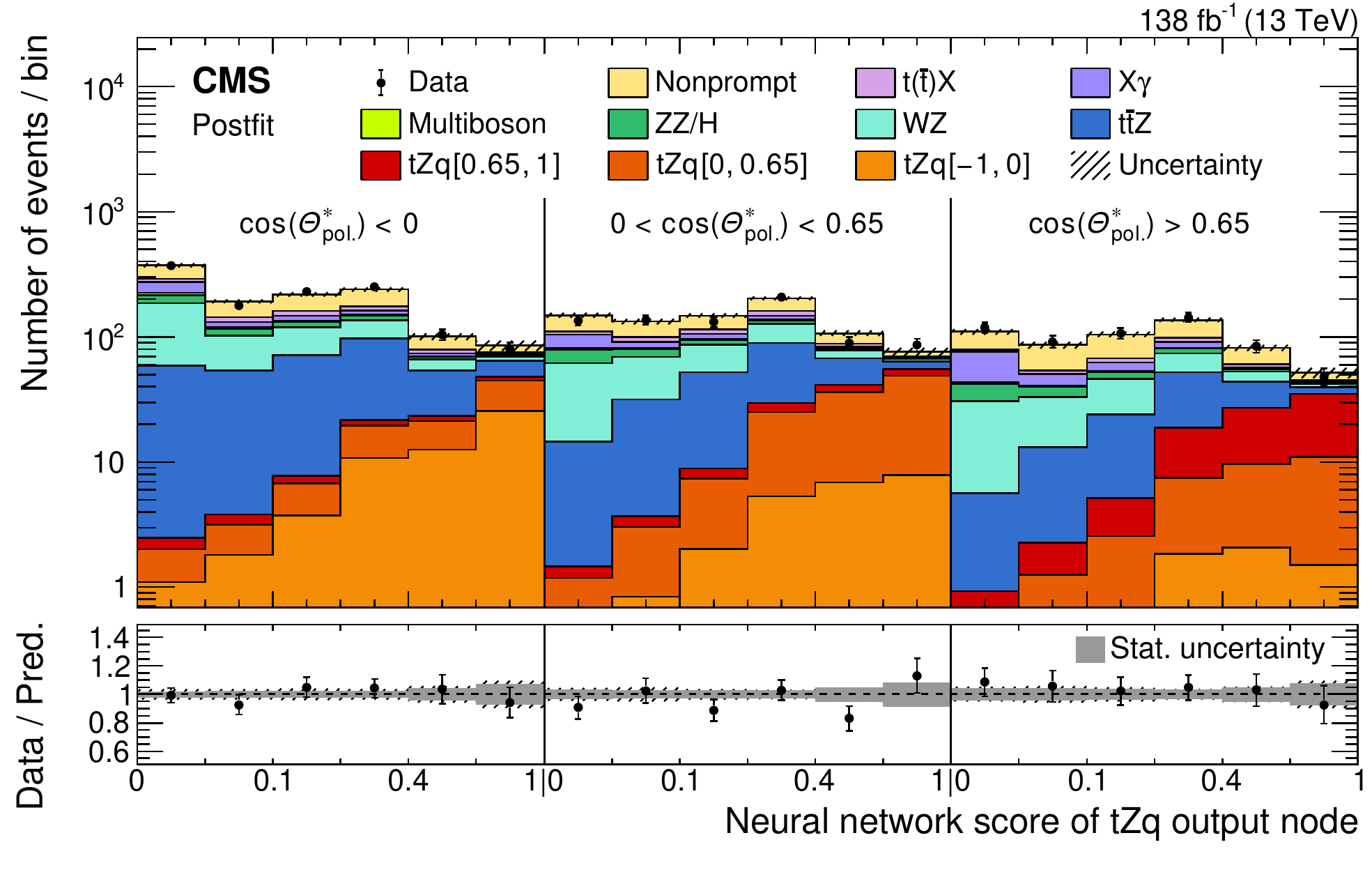}
\caption{
Prefit (upper) and postfit (lower) distributions of the neural network score from the \tZq output node for events in the signal region with fewer than four jets, used for the measurement of the spin asymmetry from the \costheta distribution at the parton level.
The data are shown by the points and the predictions by the colored histograms.
The vertical lines on the points represent the statistical uncertainty in the data, and the hatched region the total uncertainty in the prediction.
The events are split into three subregions based on the value of \costheta measured at the detector level.
Three different \tZq templates, defined by the same intervals of
\costheta at parton level and shown in different shades of orange and red,
are used to model the contribution of each parton-level bin.
The lower panels show the ratio of the data to the sum of the predictions,
with the \grey band indicating the uncertainty from the finite number of MC events.
}
\label{fig:SRFitAsymmetry}
\end{figure}

\begin{figure}[htb!]
	\centering
	\includegraphics[width=0.45\textwidth]{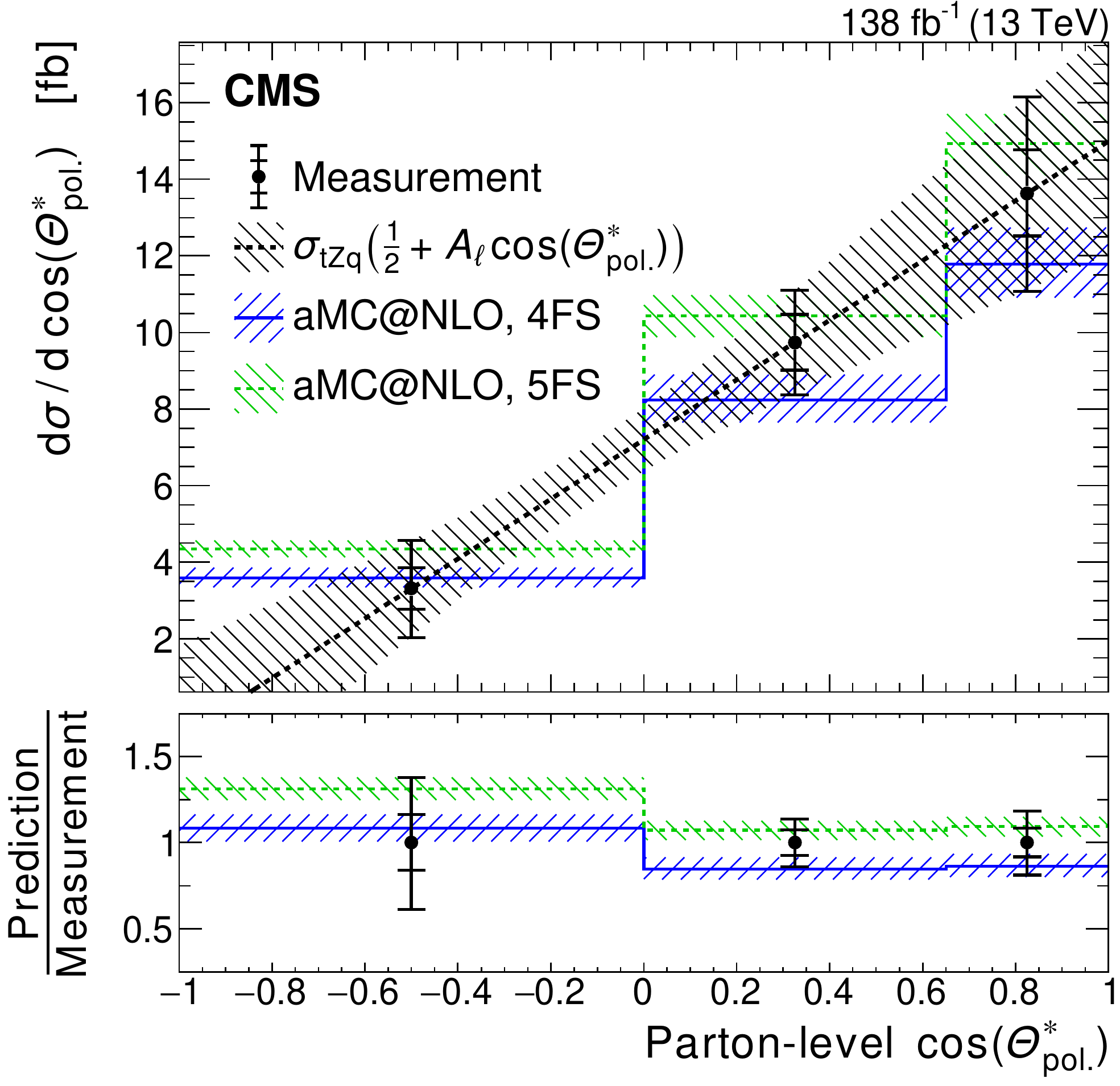}
	\includegraphics[width=0.45\textwidth]{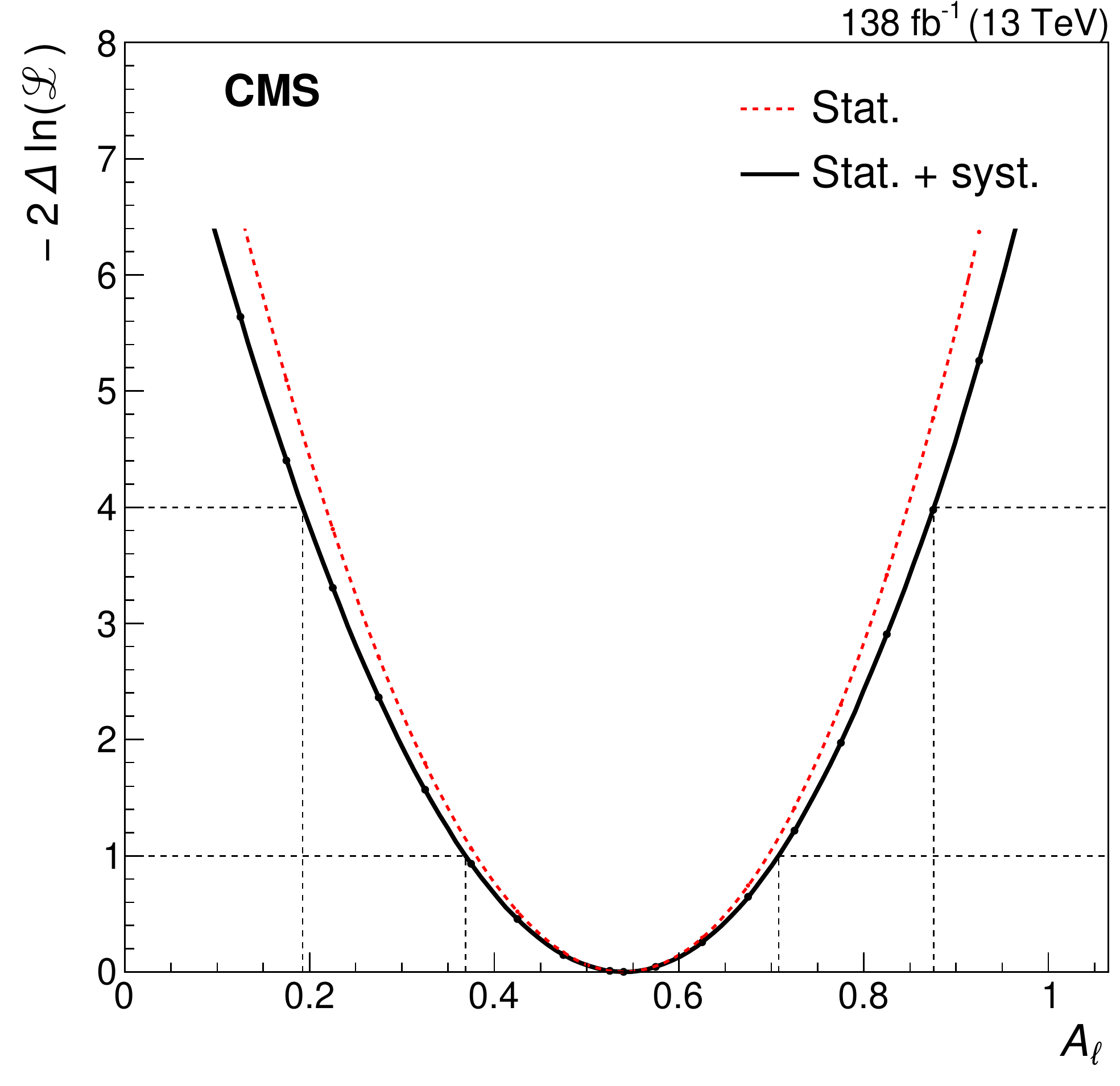}
	\caption{
    The left plot shows the measured absolute \costheta differential cross section at the parton level used in the extraction of the top quark spin asymmetry.
    The generator-level bins are parameterized according to Eq.~(\ref{eq:spin_asymmetry}), shown as a dashed black line in the plot, such that the spin asymmetry is directly used as a free parameter in the fit.
    The observed values of the generator-level bins are shown as black points with the inner and outer vertical bars giving the systematic and total uncertainties, respectively.
    The SM predictions for events simulated in the 5FS (dashed green line) and 4FS (solid blue line) are plotted as well.
    The hatched regions indicate the corresponding uncertainties, respectively.
    The lower panel displays the ratio of the MC prediction to the measurement.
    On the right, the negative log likelihood in the fit for the spin asymmetry \asym is shown when considering only the statistical uncertainties (dashed red line) or the combined statistical and systematic uncertainties (solid black line).
    The dotted black lines indicate the one (inner) and two (outer) standard deviation confidence intervals, respectively.
    }
	\label{fig:spinasymmetry}
\end{figure}

Additional material on the likelihood fit used for the extraction of the spin asymmetry is given.
The prefit and postfit distributions are shown in Fig.~\ref{fig:SRFitAsymmetry} where a good agreement with the data is visible.
The extracted parton-level bins and the likelihood as a function of the spin asymmetry are shown in Fig.~\ref{fig:spinasymmetry}.

\cleardoublepage \section{The CMS Collaboration \label{app:collab}}\begin{sloppypar}\hyphenpenalty=5000\widowpenalty=500\clubpenalty=5000\cmsinstitute{Yerevan~Physics~Institute, Yerevan, Armenia}
A.~Tumasyan
\cmsinstitute{Institut~f\"{u}r~Hochenergiephysik, Vienna, Austria}
W.~Adam\cmsorcid{0000-0001-9099-4341}, J.W.~Andrejkovic, T.~Bergauer\cmsorcid{0000-0002-5786-0293}, S.~Chatterjee\cmsorcid{0000-0003-2660-0349}, K.~Damanakis, M.~Dragicevic\cmsorcid{0000-0003-1967-6783}, A.~Escalante~Del~Valle\cmsorcid{0000-0002-9702-6359}, R.~Fr\"{u}hwirth\cmsAuthorMark{1}, M.~Jeitler\cmsAuthorMark{1}\cmsorcid{0000-0002-5141-9560}, N.~Krammer, L.~Lechner\cmsorcid{0000-0002-3065-1141}, D.~Liko, I.~Mikulec, P.~Paulitsch, F.M.~Pitters, J.~Schieck\cmsAuthorMark{1}\cmsorcid{0000-0002-1058-8093}, R.~Sch\"{o}fbeck\cmsorcid{0000-0002-2332-8784}, D.~Schwarz, S.~Templ\cmsorcid{0000-0003-3137-5692}, W.~Waltenberger\cmsorcid{0000-0002-6215-7228}, C.-E.~Wulz\cmsAuthorMark{1}\cmsorcid{0000-0001-9226-5812}
% ...
\vskip\cmsinstskip
% ...
1:  Also at TU~Wien, Wien, Austria\\
% ...
\end{sloppypar}
\end{document}